\documentclass[prd,nofootinbib,english]{revtex4}
\usepackage{amsmath}
\usepackage{amsfonts,color}
\usepackage{amssymb,float}
\usepackage[utf8]{inputenc}
\usepackage{color}
\allowdisplaybreaks
\usepackage[unicode=true,pdfusetitle,
 bookmarks=true,bookmarksnumbered=true,bookmarksopen=true,bookmarksopenlevel=3,
 breaklinks=false,pdfborder={0 0 1},backref=false,colorlinks=true,citecolor=blue]
 {hyperref}
\usepackage{enumitem,accents}

\usepackage{tikz}
\usetikzlibrary{shapes.geometric}
\usetikzlibrary{arrows.meta,arrows}

\usepackage[normalem]{ulem}

\newcommand{\lc}[1]{\accentset{\circ}{#1}}%Levi-Civita connection
\begin{document}
\title{Symmetric Teleparallel Horndeski Gravity}

\author{Sebastian Bahamonde$^{a}$}
\email{sbahamondebeltran@gmail.com, bahamonde.s.aa@m.titech.ac.jp}
%\affiliation{Department of Physics, Tokyo Institute of Technology 2-12-1 Ookayama, Meguro-ku, Tokyo 152-8551, Japan.}

\author{Georg Trenkler$^{b,c}$}
\email{trenkler@fzu.cz}
%\affiliation{CEICO, Institute of Physics of the Czech Academy of Sciences, Na Slovance 1999/2, 182 21, Prague 8, Czech Republic}
%\affiliation{Institute of Theoretical Physics, Faculty of Mathematics and Physics, Charles University, V Holešovičkách 2, 180 00 Prague 8, Czech Republic}

\author{Leonardo G. Trombetta$^{b}$}
\email{trombetta@fzu.cz}
%\affiliation{CEICO, Institute of Physics of the Czech Academy of Sciences, Na Slovance 1999/2, 182 21, Prague 8, Czech Republic}

\author{Masahide Yamaguchi$^{a}$}
\email{gucci@phys.titech.ac.jp}
%\affiliation{Department of Physics, Tokyo Institute of Technology 2-12-1 Ookayama, Meguro-ku, Tokyo 152-8551, Japan.}

\affiliation{${}^{a}$Department of Physics, Tokyo Institute of Technology 2-12-1 Ookayama, Meguro-ku, Tokyo 152-8551, Japan.\\
${}^{b}$CEICO, Institute of Physics of the Czech Academy of Sciences, Na Slovance 1999/2, 182 21, Prague 8, Czechia.\\
${}^{c}$Institute of Theoretical Physics, Faculty of Mathematics and Physics, Charles University, V Holešovičkách 2, 180 00 Prague 8, Czechia.}

\begin{abstract}
Horndeski gravity is the most general scalar-tensor theory with one scalar field leading to second-order Euler-Lagrange field equations for the metric and scalar field, and it is based on Riemannian geometry. In this paper, we formulate an analogue version of Horndeski gravity in a symmetric teleparallel geometry which assumes that both the curvature (general) and torsion are vanishing and gravity is only related to nonmetricity. Our setup requires that the Euler-Lagrange equations for not only metric and scalar field but also connection should be at most second order. We find that the theory can be always recast as a sum of the Riemannian Horndeski theory and new terms that are purely teleparallel. Due to the nature of nonmetricity, there are many more possible ways of constructing second-order theories of gravity. In this regard, up to some assumptions, we find the most general $k$-essence extension of Symmetric Teleparallel Horndeski gravity. We also formulate a novel theory containing higher-order derivatives acting on nonmetricity while still respecting the second-order conditions, which can be recast as an extension of Kinetic Gravity Braiding.  We finish our study by presenting the FLRW cosmological equations for our model. 
\end{abstract}

\maketitle
%\tableofcontents

\section{Introduction}
The recent observations of the accelerated expansion of the present Universe \cite{SupernovaCosmologyProject:1998vns,SupernovaSearchTeam:1998fmf} suggest that General Relativity (GR) needs to be modified on large scales (on infrared region) or unknown matter such as cosmological constant or dark energy needs to be introduced. In either case, a scalar-tensor theory as an extension of GR has been paid attention to since this new degree of freedom corresponding to the scalar field could be responsible for the modification of GR on large scales or the origin of dark energy. If a scalar degree of freedom is responsible for another (early stage) accelerated expansion of the Universe called inflation, such a scalar-tensor theory can accommodate inflation too. Then, the extensions of a scalar-tensor theory starting from the Einstein-Hilbert action plus a canonical scalar field were pursued. The generic function consisting of a scalar field and its canonical kinetic term was introduced in $k$-inflation~\cite{Armendariz-Picon:1999hyi} or $k$-essence \cite{Chiba:1999ka,Armendariz-Picon:2000nqq}. Further, the second-order derivative term of a scalar field was formulated in the context of the Galileon theory \cite{Nicolis:2008in}, and then dark energy (Kinetic gravity Braiding)~\cite{Deffayet:2010qz} and inflation (G-inflation) \cite{Kobayashi:2010cm} models have been proposed. In addition, the non-minimal coupling (including not only scalar field but also a kinetic term) to the Einstein tensor \cite{Germani:2010gm} as well as the Ricci scalar has been proposed. All of these terms are finally accommodated in the context of Generalized Galileon \cite{Deffayet:2011gz}. Further, that theory was shown to be equivalent \cite{Kobayashi:2011nu} to Horndeski gravity \cite{Horndeski:1974wa}, which was derived more than 50 years ago and is the most general scalar-tensor theory with one scalar field leading to second-order Euler-Lagrange field equations for the metric and scalar field.

Horndeski gravity was formulated in Riemannian geometry, in which the connection is fixed a priori as the Levi-Civita one and both the torsionless and metric compatibility conditions are imposed. However, if one regards gravity as a gauge theory, from the gauge theory viewpoint a connection, that is, a gauge field, could be an independent variable and its form and/or dynamics should be determined by taking the variation of an action with respect to it, rather than it being fixed a priori to be the Levi-Civita one. This interesting direction for modified gravity begins by considering a different geometry from the Riemannian one, which only contains curvature~\cite{Hehl:1976kj,Hehl:1994ue}.  One such possibility that recently has attained a lot of attention in the literature is a geometry endowed by torsion ($T^\lambda{}_{\mu\nu}=\Gamma^\lambda{}_{\nu\mu}-\Gamma^\lambda{}_{\mu\nu}$) or/and nonmetricity ($Q_{\alpha\mu\nu}=\nabla_\alpha g_{\mu \nu}\neq0$) with the general curvature $R^{\lambda}{}_{\mu\nu\alpha}$ being zero. Those geometries are labeled as teleparallel geometries and one can construct theories of gravity within that geometrical framework. One can formulate general teleparallel theories as in~\cite{BeltranJimenez:2019odq,Boehmer:2021aji}, but usually, in the literature, those theories are studied by either setting nonmetricity to zero~\cite{Bahamonde:2021gfp,Krssak:2018ywd,Aldrovandi:2013wha} (which are known as torsional teleparallel theories or metric teleparallel gravity) or by setting torsion to zero~\cite{Nester:1998mp,BeltranJimenez:2017tkd,Conroy:2017yln,BeltranJimenez:2018vdo} (which are known as symmetric teleparallel theories). In the following manuscript, we will concentrate on the sector where torsion is zero and construct theories of gravity only with nonmetricity. It turns out that in that framework, it is possible to formulate a theory that is equivalent to GR but expressed purely by nonmetricity, which is labeled as the symmetric teleparallel equivalent of GR (STEGR)~\cite{Nester:1998mp,BeltranJimenez:2019esp}. One can then modify the STEGR theory and formulate Symmetric Teleparallel (ST) modified gravity theories that in principle are different from the modified theories coming from GR~\cite{Heisenberg:2018vsk,CANTATA:2021ktz}. 

Following similar routes as in modified gravity based on GR, there are different ways of modifying STEGR. One difference is the fact that the connection now has some extra degrees of freedom (dof) that are independent of the metric. Although, the teleparallel condition, i.e., vanishing total curvature, implies that the form of the connection can be written in terms of a vector field which acts as a St\"uckelberg field associated with diffeomorphisms. Then, one can always choose a gauge (such as the coincident gauge) in which the connection vanishes, and then the theory is no longer diffeomorphism invariant~\cite{BeltranJimenez:2017tkd,BeltranJimenez:2022azb}. The simplest modification of STEGR is the so-called $f(Q)$ gravity where $Q$ is a particular contraction of nonmetricity tensor which differs by a boundary term with respect to the Ricci scalar $\lc{R}$ (constructed from the Levi-Civita connection)~\cite{BeltranJimenez:2017tkd}. That theory is different from $f(\lc{R})$ gravity and has some similarities to the torsional teleparallel gravity theory $f(T)$~\cite{Bengochea:2008gz,Ferraro:2008ey}. Several applications to that theory have been studied in the literature~\cite{BeltranJimenez:2019tme,Lazkoz:2019sjl,Harko:2018gxr,Mandal:2020buf,Khyllep:2021pcu,Barros:2020bgg,Frusciante:2021sio,Anagnostopoulos:2021ydo,Lu:2019hra,Soudi:2018dhv}. One can also formulate a theory by considering a linear combination of all the possible quadratic contractions of the nonmetricity tensor, which is known as Newer GR~\cite{BeltranJimenez:2017tkd,BeltranJimenez:2018vdo,Adak:2005cd}. 
Another route is to consider scalar fields minimally or nonminimally coupled to the nonmetricity scalars. In~\cite{Jarv:2018bgs,Runkla:2018xrv}, such a theory was formulated by considering non-minimal couplings between a scalar field and the nonmetricity scalar $Q$ of the form $G(\phi)Q$. Recently, it was found that there are scalarized black hole solutions for some particular coupling functions~\cite{Bahamonde:2022esv}. Similarly, one can extend those theories by considering derivatives of the nonmetricity tensor nonminimally coupled with a scalar field and, for example, find a particular subset of that theory which contains the Riemannian theory where the Riemannian Ricci scalar is nonminimally coupled as $F(\phi)\lc{R}$~\cite{Hohmann:2021ast}. 

The main goal of this manuscript is to formulate the most general (up to some assumptions) theory in ST gravity with one additional scalar field containing at most second-order Euler-Lagrange field equations. This means that we will formulate an analogue version to Horndeski gravity~\cite{Horndeski:1974wa} but now we modify the geometry to be teleparallel and torsionless as the starting point and find the resulting Lagrangian after imposing certain conditions. In~\cite{Bahamonde:2019shr}, the torsional teleparallel version of Horndeksi gravity was formulated and its form can be recast as a Lagrangian containing the Riemannian Horndeski gravity plus new contributions that are purely torsional teleparallel. It is well known that Riemannian Horndeski has been constrained for dark energy since the speed of gravity is almost equal to one~\cite{LIGOScientific:2017vwq,Copeland:2018yuh}. However, in the torsional teleparallel case, it was found that this theory allows for $G_5(\phi,X)=G_5(\phi)$ to be nontrivial and $G_4=G_4(\phi,X)$ while still having $c_T=1$, in contrast to the constraints derived in Riemannian Horndeski \cite{Baker:2017hug, Creminelli:2017sry, Ezquiaga:2017ekz, Sakstein:2017xjx}. This is possible due to the fact that torsional teleparallel corrections can be set in some ways to compensate the speed of tensor modes to travel at the speed of light~\cite{Bahamonde:2019ipm}. Further studies concerning gravitational wave polarizations~\cite{Bahamonde:2021dqn} and also Post-Newtonian parameters of the torsional Horndeski gravity theory~\cite{Bahamonde:2020cfv} give some evidence that such alternative description of gravity could be an interesting and new route to explore theories beyond the standard paradigm of the Riemannian case. Following that, we would like to find a theory in the ST formalism with the aim that in future works, we can perform similar studies to find if in this framework one can formulate theories that can alleviate or solve problems of GR.

This manuscript is organised as follows: In Sec.~\ref{sec:SymmetricTeleparallel} we give an introduction to ST gravity where the main ingredients of the underlying framework are defined. In Sec.~\ref{sec:HorndeskiTeleparallel} we give general guidelines for the construction of the ST Horndeski theory. Sec.~\ref{secIV} is devoted to providing a systematic construction of a subclass of ST extensions of Horndeski. To do that, first in Sec.~\ref{sec:NoHigherOrder} we construct the most general ST Horndeski theory with no higher-derivatives acting on purely teleparallel terms, which would be like considering that nonmetricity only modifies $L_2$ from the Riemannian Horndeski theory. Then, In Sec.~\ref{sec:STKGB} we construct a possible extension of that theory by considering teleparallel higher-order derivatives acting on nonmetricity but still, the theory constructed leads to at most second-order Euler-Lagrange field equations, which would be a ST extension of Kinetic Gravity Braiding/Cubic Horndeski by including nonmetricity at the linear level. In Sec.~\ref{FLRW} we discuss the flat Friedmann–Lemaître–Robertson–Walker (FLRW) cosmological equations for our formulated theory. Finally, we conclude our main results in Sec.~\ref{sec:Conclusions}. 

The notation of our paper considers the metric signature $(-+++)$, and uses natural units where $\hbar=c=1$. Furthermore, quantities constructed from the Levi-Civita connection (such as the Riemannian curvature) will be denoted with a overcircle (e.g. $\lc{R}^{\lambda}{}_{\mu\nu\beta}$, $\lc{\Gamma}^\alpha{}_{\mu\nu}$, etc.), whereas quantities without any symbol on top will be related to ST gravity.

\section{Introduction to Symmetric Teleparallel Gravity} \label{sec:SymmetricTeleparallel}

\subsection{Geometrical preliminaries}
In metric-affine geometry the metric $g_{\mu\nu}$ provides a notion of angles and distances, while the connection $\Gamma^{\lambda}{}_{\mu\rho}$ independently defines parallel transport and covariant derivatives $\nabla_{\mu}$~\cite{Hehl:1976kj,Hehl:1994ue}. ST spacetimes feature a flat (zero curvature), torsion-free but metric-incompatible connection, rendering the nonmetricity tensor the main geometrical quantity for the description of gravity~\cite{Nester:1998mp}.
It is per the definition given by the application of the covariant derivative onto the metric and therefore symmetric in the last two indices:
\begin{equation}\label{nonmetricity}
Q_{\lambda\mu\nu}\equiv\nabla_{\lambda}g_{\mu\nu}=\partial_{\lambda}g_{\mu\nu}-\Gamma^{\rho}{}_{\lambda\mu}g_{\rho\nu}-\Gamma^{\rho}{}_{\lambda\nu}g_{\mu\rho}\,.
\end{equation}  
The nonmetricity tensor can be split into its irreducible decomposition under the group of global Lorentz transformations as~\cite{McCrea:1992wa}
\begin{equation}
Q_{\lambda\mu\nu}=g_{\mu\nu}W_{\lambda}+{\nearrow\!\!\!\!\!\!\!Q}_{\lambda\mu\nu}\,,
\end{equation}
where the first piece is the so-called Weyl part defined as
\begin{align}
    W_{\mu}&=\frac{1}{4}\,Q_{\mu\nu}\,^{\nu}\,,
    \end{align}
and the second part is referred to as its traceless part defined as
\begin{equation}\label{irreducibletracelessnonmetricity}
    {\nearrow\!\!\!\!\!\!\!Q}_{\lambda\mu\nu}=g_{\lambda(\mu}\Lambda_{\nu)}-\frac{1}{4}g_{\mu\nu}\Lambda_{\lambda}+\frac{1}{3}\varepsilon_{\lambda\rho\sigma(\mu}\Omega_{\nu)}\,^{\rho\sigma}+q_{\lambda\mu\nu}\,,
\end{equation}
where $\varepsilon_{\lambda\rho\mu\nu}$ is the Levi-Civita (density) tensor and
\begin{subequations}\label{irreducible-parts}
\begin{align}
    \Lambda_{\mu}&=\frac{4}{9}\left(Q^{\nu}\,_{\mu\nu}-W_{\mu}\right)\,,\\
    \Omega_{\lambda}\,^{\mu\nu}&=-\,\left[\varepsilon^{\mu\nu\rho\sigma}Q_{\rho\sigma\lambda}+\varepsilon^{\mu\nu\rho}\,_{\lambda}\left(\frac{3}{4}\Lambda_{\rho}-W_{\rho}\right)\right]\,,\\
    q_{\lambda\mu\nu}&=Q_{(\lambda\mu\nu)}-g_{(\mu\nu}W_{\lambda)}-\frac{3}{4}g_{(\mu\nu}\Lambda_{\lambda)}\,,
\end{align}
\end{subequations}
constitute a vector $  \Lambda_{\mu}$,  one totally  traceless pseudotensor $\Omega_{\lambda}\,^{\mu\nu}$ and a totally traceless tensor $q_{\lambda\mu\nu}$, respectively. Notice that the traceless part of nonmetricity is only traceless in its last two indices, i.e., ${\nearrow\!\!\!\!\!\!\!Q}^{\lambda}{}_{\lambda\nu}\neq 0$. It is also useful to define the following  totally traceless tensor
\begin{eqnarray}
{*\Omega}^{\alpha\mu\nu}=\epsilon^{\rho\sigma\mu\nu}\Omega^{\alpha}{}_{\rho\sigma}\,,
\end{eqnarray}
which will be used later for constructing our theory.

A ST connection is characterised by a vanishing torsion and curvature tensors:
\begin{align}
T^{\lambda}{}_{\mu\nu} &\equiv \Gamma^{\lambda}{}_{\mu\nu}-\Gamma^{\lambda}{}_{\nu\mu}=0\,,\label{torsion}\\
\label{curvature}
R^{\alpha}_{\;\;\rho\mu\nu} &\equiv \partial_{\mu}\Gamma^{\alpha}{}_{\nu\rho}-\partial_{\nu}\Gamma^{\alpha}{}_{\mu\rho}+\Gamma^{\alpha}{}_{\mu\beta}\Gamma^{\beta}{}_{\nu\rho}-\Gamma^{\alpha}{}_{\nu\beta}\Gamma^{\beta}{}_{\mu\rho}=0\,,
\end{align}
where Eq.~\eqref{curvature} will be referred to as teleparalellism condition. One can express the connection as
\begin{equation}\label{genconn}
\Gamma^{\lambda}{}_{\mu\nu}=\mathring{\Gamma}^{\lambda}{}_{\mu\nu}+L^{\lambda}{}_{\mu\nu}(Q)\,,
\end{equation}  
where
\begin{align}\label{disformation}
   L^{\lambda}{}_{\mu\nu}(Q)=\frac{1}{2}Q^{\lambda}{}_{\mu\nu}-Q^{\;\;\;\lambda}_{(\mu\;\;\nu)}\,,
\end{align}
is the disformation tensor and
\begin{equation}\label{Christoffel}
\mathring{\Gamma}^{\lambda}{}_{\mu\nu}= \frac{1}{2}g^{\lambda\beta}\left(\partial_{\mu}g_{\beta\nu}+\partial_{\nu}g_{\mu\beta}-\partial_{\beta}g_{\mu\nu}\right) 
\end{equation}
is the Christoffel symbol associated with the Levi-Civita connection completely defined through the metric. Therefore, the torsionless connection consists of $40$ independent components. By replacing the above decomposition of the connection~\eqref{genconn} in the definition of the general curvature~\eqref{curvature} and imposing the teleparallel condition (curvatureless case), one can decompose the curvature tensor as
\begin{equation}\label{RiemannSplit}
    R^{\sigma}\,_{\rho\mu\nu}= \lc{R}^{\sigma}\,_{\rho\mu\nu}-\lc{\nabla}_{\nu}L^{\sigma}\,_{\mu\rho}+\lc{\nabla}_{\mu}L^{\sigma}\,_{\nu\rho}-L^{\sigma}\,_{\nu\lambda}L^{\lambda}\,_{\mu\rho}+L^{\sigma}\,_{\mu\lambda}L^{\lambda}\,_{\nu\rho}=0\,,
\end{equation}
which tells us that the conditions \eqref{torsion} and \eqref{curvature} allow us to relate the Riemann tensor associated with the Levi-Civita connection $\lc{R}^{\sigma}\,_{\rho\mu\nu}$ (denoted by a $\circ$ on top) to the nonmetricity tensor via quadratic contractions of the disformation tensor and its Levi-Civita covariant derivatives.

The teleparallelism condition \eqref{curvature} yields for the flat connection (zero curvature)
\begin{align}
  \Gamma^{\alpha}{}_{\mu\nu}=(\Lambda^{-1})^{\alpha}_{\;\;\lambda}\partial_{\mu}\Lambda^{\lambda}{}_{\nu}\,,
\end{align}
with $\Lambda \in GL(4,\mathbb{R})$, as in torsional teleparallel gravity. The absence of torsion \eqref{torsion} allows writing the connection as 
\begin{align}\label{CGRGamma}
    \Gamma^{\alpha}{}_{\mu\nu}=\frac{\partial x^{\alpha}}{\partial \xi^{\lambda}}\partial_{\mu}\partial_{\nu}\xi^{\lambda}\,,
\end{align}
where we have parametrized $\Lambda^{\alpha}_{\;\;\mu}=\partial_{\mu}\xi^{\alpha}$ in terms of the auxiliary field $\xi^{\alpha}$ associated to  diffeomorphisms (as a St\"uckelberg field). Thus, after imposing the torsionless and curvatureless conditions, the maximum number of dof in the nonmetricity part of the connection (disformation tensor) goes down from 40 dof to a maximum of 4 independent dof (that can be expressed via the vector $\xi^\mu$). 

It is worth mentioning that for a given ST theory coupled to extra fields $\Phi^i$ whose action is $S(g_{\mu\nu},\Gamma^\lambda{}_{\mu\nu},\Phi^i)$, the corresponding field equations of the theory are obtained by varying it with respect to all the fields, which in this case corresponds to taking variations with respect to the metric, the torsionless and flat connection, and all the extra fields. 

A remarkable property of ST gravity is that one can completely get rid of the connection by performing a coordinate transformation
\begin{align}\label{coincidentgauge}
   \xi^{\mu}=x^{\mu}\quad\Rightarrow\quad \Gamma^{\alpha}{}_{\mu\nu}=0 \quad\Rightarrow\quad \nabla_{\mu}=\partial_{\mu}\,,
\end{align}
which is referred to as fixing coincident gauge. Although we will not adopt this gauge fixing in the following sections, it is worth mentioning that it can simplify certain computations, at the price of full diffeomorphism invariance being given up and hence physics depending on the coordinates used to describe it. Other computations, however, for example in spherical symmetry, become more cumbersome in the coincident gauge, as it was shown in~\cite{Bahamonde:2022zgj}.

\subsection{Two formulations of gravity}
With the definitions of curvature and nonmetricity at hand, we are now ready to provide two equivalent geometric formulations of GR. In fact, there exists a third description utilizing torsion, such that these three formulations constitute the well-known geometric trinity of GR \cite{BeltranJimenez:2019esp}.

\subsubsection{GR à la Einstein} \label{GREinstein}
GR à la Einstein is formulated on a Riemannian spacetime, so the fundamental dynamical object is the metric $g_{\mu\nu}$ and gravity is ascribed to curvature, i.e. $R^{\alpha}_{\;\;\rho\mu\nu}=\mathring{R}^{\alpha}_{\;\;\rho\mu\nu}\neq 0$, $T^{\lambda}{}_{\mu\nu} = 0$ and $Q_{\lambda\mu\nu} = 0$. The action is given by the Einstein-Hilbert action
\begin{align}\label{GRaction}
    S=\frac{M_{\text{Pl}}^{2}}{2}\int d^{4}x\sqrt{-g}\mathring{R}\,,
\end{align}
where $M_{\text{Pl}}$ is the reduced Planck Mass.

In the Palatini formalism, the connection is assumed to be independent of the metric. It is remarkable that even when starting from the Einstein-Hilbert action in this formalism, i.e.,
\begin{align}
    S_{\text{Palatini}}=\frac{M_{\text{Pl}}^{2}}{2}\int d^{4}x\sqrt{-g}g^{\mu\nu}R_{\mu\nu}(\Gamma)\,,\label{GRpalatini}
\end{align}
where $g$ and $\Gamma$ are treated as independent fields, one still obtains Einstein's field equations. This can be seen in the following way. For a torsion free connection, the variation with respect to the connection enforces metric compatibility of the connection and hence uniquely fixes $\Gamma^{\alpha}{}_{\mu\nu}=\mathring{\Gamma}^{\alpha}{}_{\mu\nu}$, and then variations with respect to the metric gives the usual Einstein's equations. Therefore, one has $R=\mathring{R}$ in GR \`a la Einstein giving \begin{equation}
    \underbrace{10}_{g_{\mu\nu}}-\underbrace{2\times4}_{\text{diffs}}=2
\end{equation} 
propagating dof.

\subsubsection{Symmetric Teleparallel Equivalent of GR (STEGR)}\label{STEGR}
As mentioned above, ST gravity \cite{Nester:1998mp} is formulated on a  flat manifold without torsion, so the fundamental dynamical object is the metric $g_{\mu\nu}$ and gravity is ascribed to nonmetricity, i.e. $R^{\alpha}_{\;\;\rho\mu\nu}=0$, $T^{\lambda}{}_{\mu\nu} = 0$ and $Q_{\lambda\mu\nu} \neq 0$. Notice that the general curvature is zero but, in general, the Levi-Civita one $\lc{R}^{\alpha}_{\;\;\rho\mu\nu}$ is non-vanishing. This will be the framework, in which the rest of the paper is formulated in and hence is the most interesting for us.

One can construct a generic symmetric quadratic action described by~\cite{BeltranJimenez:2017tkd,BeltranJimenez:2018vdo,Adak:2005cd}
\begin{align}
    S_{\rm Newer\, GR}=\frac{M_{\text{Pl}}^{2}}{2}\int d^{4}x\sqrt{-g}\, \mathbb{Q}\,,\label{newerGR}
\end{align}
with 
\begin{align}\label{nonmetquadratic}
    \mathbb{Q}=c_{1}Q_{\alpha}^{\;\;\mu\nu}Q^{\alpha}_{\;\;\mu\nu}+c_{2}Q_{\alpha}^{\;\;\mu\nu}Q^{\;\;\alpha}_{\mu\;\;\nu}+c_{3}Q_{\mu}Q^{\mu}+c_{4}\tilde{Q}_{\mu}\tilde{Q}^{\mu}+c_{5}\tilde{Q}_{\mu}Q^{\mu}\,,
\end{align}
which consists of all five independent parity-preserving (because of the symmetry properties of nonmetricity) contractions quadratic in the nonmetricity, where $Q^{\mu}=Q^{\mu\alpha}_{\;\;\;\;\alpha}=4W^{\mu}$, $\tilde{Q}^{\mu}=Q_{\alpha}^{\;\;\alpha\mu}=W^{\mu}+\frac{9}{4}\Lambda^{\mu}$ and $c_{i}$ are arbitrary constants. These quadratic contractions can be equivalently written in terms of the irreducible pieces of the nonmetricity \eqref{irreducible-parts}
\begin{eqnarray} \label{quadratic-Q-invariants} 
Q_1=W^\mu W_\mu\,,\quad Q_2=\Lambda_\mu \Lambda^\mu\,,\quad Q_3=W_\mu \Lambda^\mu\,,\quad Q_4={*\Omega}_{\alpha\mu\nu}{*\Omega}^{\alpha\mu\nu}\,,\quad Q_5=q_{\lambda\mu\nu}q^{\lambda\mu\nu}\,,
\end{eqnarray}
such that
\begin{eqnarray}
\mathbb{Q}=\sum_{i=1}^{5}\tilde{c}_i Q_i\label{NewerGR}=L_{\rm Newer GR}\,.
\end{eqnarray}
Keeping the values of $c_{i}$ (or equivalently $\tilde{c}_{i}$) generic, one obtains a ST theory which is sometimes called ``Newer GR". It should be noted that $Q_4$ can be also defined as $\Omega_{\alpha\mu\nu}\Omega^{\alpha\mu\nu}$ since that scalar is parity preserving. However, for constructing more general theories, it is convenient to use the tensor ${*\Omega}_{\alpha\mu\nu}$ as one of the building blocks instead of $\Omega_{\alpha\mu\nu}$ for constructing parity preserving theories of gravity. 

Choosing the parameters $c_{i}$ in a very specific way, namely 
\begin{align}\label{constans}
    c_{1}=\frac{1}{4}\,,\quad c_{2}=-\frac{1}{2}\,,\quad c_{3}=-\frac{1}{4}\,,\quad c_{4}=0\,,\quad c_{5}=\frac{1}{2}\,,
\end{align}
or
\begin{eqnarray}
\tilde{c}_1=- \frac{3}{2}\,,\quad \tilde{c}_2= \frac{9}{32}\,,\quad \tilde{c}_3= \frac{9}{4}\,,\quad \tilde{c}_4=\frac{1}{24}\,,\quad \tilde{c}_5=- \frac{1}{4}\,,
\end{eqnarray}
the scalar constructed in~\eqref{nonmetquadratic} and \eqref{NewerGR}
becomes the so-called nonmetricity scalar 
\begin{eqnarray}\label{nonmetricityscalar}
Q=-\frac{3}{2}Q_1+\frac{9}{32}Q_2+\frac{9}{4}Q_3+\frac{1}{24}Q_4-\frac{1}{4}Q_5=\,\frac{1}{4}\,Q_{\lambda\mu\nu}Q^{\lambda\mu\nu}-\frac{1}{2}\,Q_{\lambda\mu\nu}Q^{\mu\nu\lambda}-\frac{1}{4}\,Q_{\mu\nu}{}^{\nu}Q^{\mu\beta}{}_\beta+\frac{1}{2}\,Q_{\mu\nu}{}^{\nu}Q_{\alpha\mu}{}^{\alpha}\,,
\end{eqnarray}
giving rise to the STEGR theory when appearing linearly in the Lagrangian. From the splitting \eqref{RiemannSplit}, one can construct the Ricci scalar and obtain that
\begin{align} \label{R and Q}
    0=R=\mathring{R}+Q+\mathring{\nabla}{}_{\mu}(Q^{\mu}-\tilde{Q}^{\mu}) \quad\iff\quad \lc{R}=-Q+\lc{\nabla}_{\mu}(\tilde{Q}^{\mu}-Q^{\mu}) :=-Q+B_Q\,,
\end{align}
where $\mathring{\nabla}{}_{\alpha}$ is the covariant derivative associated to the Levi-Civita connection, $Q$ being the nonmetricity scalar defined as~\eqref{nonmetricityscalar} and $B_Q$ a boundary term. The above equation shows that the Levi-Civita Ricci scalar differs by a boundary term from the nonmetricity scalar. This means that the field equations arising from an action constructed from~\eqref{nonmetricityscalar} give the same equations as Einstein's field equations. Also, one finds that the STEGR theory has 2 propagating dof in agreement with GR, which is particularly easy to see in the coincident gauge~\cite{BeltranJimenez:2019esp,DAmbrosio:2020nqu}.
 
 One can further do a similar computation in the case of torsional teleparallel gravity where only torsion is different from zero. In that case, again, one can formulate a theory with the same dynamics as GR but torsion is responsible of gravity (see~\cite{Bahamonde:2021gfp} on such torsional theories). Thus, the geometrical richness of a general manifold allows us to describe GR in three equivalent ways~\cite{BeltranJimenez:2019tjy}. 
 
 One outstanding property of STEGR in coincident gauge is that its action becomes
\begin{align}
    \left.S_{\text{STEGR}}\right\vert_{\Gamma=0}=\frac{M_{\rm pl}^{2}}{2}\int d^{4}x\sqrt{-g}g^{\mu\nu}\left(\lc{\Gamma}^{\alpha}{}_{\beta\mu}\lc{\Gamma}^{\beta}{}_{\nu\alpha}-\lc{\Gamma}^{\alpha}{}_{\beta\alpha} \lc{\Gamma}^{\beta}{}_{\mu\nu} \right)\,,
\end{align}
which is sometimes called the Einstein action (or Gamma squared action). In the remainder of this text, we work in the framework of ST gravity without fixing the coincident gauge.

\section{Symmetric Teleparallel Horndeski Theory}
\label{sec:HorndeskiTeleparallel}

In the Riemannian formulation of gravity, the most general way to couple a scalar field with the gravitational degrees of freedom with second-order Euler-Lagrange field equations is given by the well-known Horndeski Lagrangian. While containing arbitrary functions of the scalar field $\phi$ and its kinetic term $X=-\tfrac{1}{2} \partial^\mu \phi\partial_\mu \phi$, this theory can only involve the Riemannian curvature tensor $\lc{R}^{\mu}{}_{\nu\rho\sigma}$ and the second (covariant) derivatives of the scalar field $\lc{\nabla}_{\mu}\lc{\nabla}_{\nu}\phi$ in a very restricted form, such that the Euler-Lagrange field equations remain second order. In particular, the curvature invariants are not allowed to appear freely via arbitrary functions since they already contain second derivatives of the metric, and would in general lead to higher-order equations.

As described in the previous section, in the ST formulation of gravity, nonmetricity is responsible of gravity. Crucially, the nonmetricity tensor $Q_{\lambda\mu\nu}$ only contains first derviatives of the metric $g_{\mu\nu}$, and no derivatives of the connection $\Gamma^{\alpha}{}_{\mu\nu}$. For this reason, the nonmetricity invariants are allowed to enter the Lagrangian unrestricted through arbitrary functions. Moreover, one might expect a Galileon-like higher-derivative structure to also be possible involving derivatives of the nonmetricity, i.e. second derivatives of the metric. All of this leads to a much more general form for a theory coupling a scalar field with gravity with up to second-order Euler-Lagrange field equations \emph{\`a la} Horndeski. In this section, we will make the first steps towards the construction of such ST Horndeski Theory.

%-------------------------------------------------------------------------------

\subsection{Covariantization prescription and nonminimal coupling}

In order to consider Horndeski interactions in a ST formulation of gravity, we need a method of covariantizing scalar fields from tangent spaces to general manifolds. In pseudo-Riemannian manifolds (where only curvature is present), the Levi-Civita connection provides a unique prescription for how to promote locally Lorentz invariant objects to fully covariant ones, namely through the procedure
\begin{align}
\eta_{\mu\nu}\,&\rightarrow\,g_{\mu\nu}\,,\nonumber\\
\partial_{\mu}\,&\rightarrow\,\mathring{\nabla}{}_{\mu}\,,\label{GR_equiv_prin}
\end{align}
where the Minkowski metric is lifted to the general metric on the manifold, and partial derivatives are promoted to covariant derivatives defined through the Christoffel symbols describing corrections arising in parallel transport.

In ST manifolds (where only nonmetricity is present), Eq.~\eqref{GR_equiv_prin} is no longer the unique possibility, but one may also covariantize as follows
\begin{align}
\eta_{\mu\nu}\,&\rightarrow\,g_{\mu\nu}\,,\nonumber\\
\partial_{\mu}\,&\rightarrow\,\nabla{}_{\mu}\,,\label{equiv_prin}
\end{align}
using the ST connection instead of the Levi-Civita one. This alternative prescription includes a nonminimal coupling to nonmetricity, as it can be seen by means of Eq.~\eqref{genconn} which always allows us to perform the following split for an arbitrary tensor $P^{\alpha_1 \dots}{}_{\beta_1 \dots}$,
\begin{align} \label{two-nablas}
\nabla_\mu P^{\alpha_1 \dots}{}_{\beta_1 \dots} = \mathring{\nabla} _\mu P^{\alpha_1 \dots}{}_{\beta_1 \dots} &+ L^{\alpha_1}{}_{\mu\nu}(Q) \, P^{\nu \dots}{}_{\beta_1 \dots} + \dots - L^{\nu}{}_{\mu\beta_1}(Q) \, P^{\alpha_1 \dots}{}_{\nu \dots} - \dots \, .
\end{align}
It is therefore expected that the physics will be affected by the choice of covariantization prescription, and it will be important when adding matter to the theory.

However, when constructing a theory in the ST framework that couples a scalar field with gravity in the most general way, as we intend to do here, this becomes just a parametrization choice provided one allows for both the scalar field $\phi$ and nonmetricity tensor $Q_{\lambda\mu\nu}$ to enter in the Lagrangian with complete freedom. This is precisely what Eq.~\eqref{two-nablas} is telling us, we can either use $\nabla_\mu$ or $\mathring{\nabla}_\mu$, together with $Q_{\lambda\mu\nu}$ in the construction of such a theory.

In what follows we choose to work with $\mathring{\nabla}_\mu$ and $Q_{\lambda\mu\nu}$ as building blocks, since it will allow us to make easier contact with known results from the Riemannian formulation. For the same reason, we will also include $\mathring{R}^{\alpha}{}_{\beta\mu\nu}$ as an independent building block, without necessarily splitting it as in Eq.~\eqref{RiemannSplit}.

\subsection{Variations and field equations}
To calculate the equations of motion, we make use of the principle of stationary action. As mentioned in the previous section, in the case of ST gravity, variation with respect to any external fields (scalars, vectors, etc.) as well as with respect to the metric is performed in the standard way.

Since the connection is treated as an independent variable in metric affine theories, we will clarify again here how to correctly vary with respect to it while simultaneously ensuring that the teleparallelism condition \eqref{curvature} is always satisfied. To achieve this, there are in principle two ways. The first way is to take variations with respect to the flat connection (not the full) which can be done by replacing the  connection as~\eqref{CGRGamma} in the studied action $S=S(g_{\mu\nu},\Gamma^\lambda{}_{\mu\nu},\Phi^i)=S(g_{\mu\nu},\xi^\alpha,\Phi^i)$ and varying it with respect to the vector field $\xi^{\alpha}$. Alternatively, a second method can be used which relies on the language of Lagrange multipliers in the action which enforces curvature and torsion to vanish and vary with respect to a generic connection $\Gamma^{\alpha}{}_{\mu\nu}$. Those Lagrange multipliers are added to the action as $S=S(g_{\mu\nu},\Gamma^\lambda{}_{\mu\nu},\Phi^i)+\lambda_{\alpha}{}^{\beta\mu\nu}R^\alpha{}_{\beta\mu\nu}+\lambda_\alpha{}^{\mu\nu}T^\alpha{}_{\mu\nu}$. Then, one performs variations with respect to the full connection and the Lagrange multipliers, which would give in the end the same variations as starting from the action and varying with respect to the flat connection (i.e. varying with respect to $\xi^{\alpha}$). Both ways are equivalent and eventually should give the same results in ST theories.

In practice, it turns out that in all the explicit examples discussed in the following sections, variation with respect to the connection can trivially never give rise to higher-order derivative terms. This is due, for example, to the fact that the terms appearing in the action are at most linear in derivatives of $Q_{\lambda\mu\nu}$ and second-order derivatives of $\phi$ in the contractions presented in section \ref{sec:STKGB}. In general, this fact is not always true, however, such that one needs to be more careful.

\subsection{Conditions on a Symmetric Teleparallel Horndeski Theory\label{conditions}}
Let us now impose some conditions to formulate our theory. As a first step we enumerate the conditions that allow formulating a ST Horndeski Theory:
\begin{enumerate}
 \item The Euler-Lagrange field equations for the dynamical degrees of freedom, namely the scalar field $\phi$, the metric $g_{\mu\nu}$ and the teleparallel connection $\Gamma^{\alpha}{}_{\mu\nu}$ shall all be at most second order in their derivatives.
 \item The Lagrangian must be parity preserving.
 \item The Lagrangian shall contain at most quadratic contractions of the nonmetricity tensor $Q_{\lambda\mu\nu}$.
\end{enumerate}
The first condition ensures the non-existence of Ostrogradsky ghosts associated with higher-derivatives terms in the theory. Furthermore, the first two of the above conditions are analogues of the ones defining the Riemannian Horndeski theory in the curvature formulation. Notice however that in the ST formulation the connection $\Gamma^{\alpha}{}_{\mu\nu}$ is treated as an independent degree of freedom, and as such it is allowed to have second-order Euler-Lagrange field equations on its own, provided it satisfies the teleparallel condition, that is, the vanishing of the total curvature. Therefore in this formulation, the number of propagating degrees of freedom will generically be larger. The third condition is instead of a different nature, not being present in the Riemannian case. Its adoption follows closely the Torsional Teleparallel Horndeski Theory formulation \cite{Bahamonde:2019shr}, and it is required mostly for technical reasons in order to limit the otherwise infinite tower of invariants that can be constructed with both the nonmetricity tensor and derivatives of the scalar field which could in principle be included. Further, due to the geometrical nature of Symmetric Teleparallel gravity (nonmetricity has first derivatives of the metric and curvature already has second derivatives of the metric), there are actually infinite scalars that one can define with only those properties. That means that the theory implicitly described by Eq.~\eqref{PC-rule} contains infinite terms. In this regard, condition three is an additional condition  that we imposed to write down a theory with a finite Lagrangian. One can also think from the effective field theory point of view and understand this condition as the first correction of STEGR and then, the higher-order contractions of nonmetricity would have a smaller contribution.

Let us emphasize here that the ST framework is different from considering the case where both the curvature and nonmetricity are non-vanishing. As mentioned in previous sections, the teleparallel ($R^\alpha{}_{\mu\nu\beta}=0$) and torsionless ($T^\lambda{}_{\mu \nu}=0$) conditions restrict the independent connection $\Gamma^\lambda{}_{\mu\nu}$ to be of the form~\eqref{CGRGamma}, and hence, it carries a maximum of $4$ dof (from the vector $\xi^\lambda$). Summing up those maximum dof coming from the connection with the maximum dof coming from the metric (which is $6$ dof) and the extra scalar field (which is $1$ dof) would give us that the maximum number of dof for our theory would be $6+4+1=11$ dof. This argument only holds due to the fact that we will construct theories which are at most second-order in all of the fields $(g_{\mu\nu},\Gamma^\lambda{}_{\mu\nu},\phi)$. Another equivalent way of doing that counting is to assume the coincident gauge and set the connection to zero. That gauge choice would break diffeomorphishs invariance and due to that, the metric would now have a maximum of $6+4$ dof. This counting is consistent with the fact that these two formalisms, i.e., working with the flat connection~\eqref{CGRGamma} and the metric or working solely with the metric in the coicident gauge, are equivalent. Yet another equivalent way of counting the maximum number of dof coming from the torsionless and flat connection is the following. A general connection has $4^3 = 64$ components, which satisfy second-order Euler-Lagrange field equations and then require 128 initial conditions. The teleparallel condition $R^\alpha{}_{\mu\nu\beta}=0$ implies 96 equations ($=4\times4\times6$, since a general Riemann tensor for a generic connection is only antisymmetric in the last two indices), while the torsionless condition $T^\lambda{}_{\mu \nu}=0$ implies another 24 equations ($4\times6$). Therefore, we are left with 8 independent initial conditions for the connection, which implies 4 dof that are related to the connection.

An easy way to formulate the theory is to first consider the most general scalar-theory in Minkowski spacetime with a second-order equation of motion and then uplift it to include gravity by following the Levi-Civita covariantization prescription Eq. \eqref{GR_equiv_prin}. This is so far exactly the same procedure that leads to the Generalized Galileons which is equivalent to the Riemannian Horndeksi theory in the curvature formulation, precisely because of this coupling prescription. Therefore, so far, we know that Riemannian Horndeski is contained as a subcase provided one uses $\mathring{\nabla}{}_{\mu}$ instead of $\nabla_{\mu}$:
\begin{subequations}
\begin{align}
\mathring{L}{}_2 &= G_2(\phi,X)\,, \label{LC_Horn_Comp1}\\
\mathring{L}{}_3 &= -G_3(\phi,X)\mathring{\Box}\phi\,, \label{LC_Horn_Comp2}\\
\mathring{L}{}_4 &= G_4(\phi,X)\lc{R} + G_{4,X}(\phi,X)\left[(\mathring{\Box}\phi )^2 - \mathring{\nabla}{}_{\mu}\mathring{\nabla}{}_{\nu}\phi\mathring{\nabla}{}^{\mu}\mathring{\nabla}{}^{\nu}\phi\right]\,, \label{LC_Horn_Comp3}\\
\mathring{L}{}_5 &= G_5(\phi,X)\mathring{G}{}_{\mu\nu}\mathring{\nabla}{}^{\mu}\mathring{\nabla}{}^{\nu}\phi - \frac{1}{6}G_{5,X}(\phi,X)\left[(\mathring{\Box}\phi )^3 + 2\mathring{\nabla}{}_{\nu}\mathring{\nabla}{}_{\mu}\phi\mathring{\nabla}{}^{\nu}\mathring{\nabla}{}^{\lambda}\phi\mathring{\nabla}{}_{\lambda}\mathring{\nabla}{}^{\mu}\phi - 3\mathring{\Box}\phi \mathring{\nabla}{}_{\mu}\mathring{\nabla}{}_{\nu}\phi\mathring{\nabla}{}^{\mu}\mathring{\nabla}{}^{\nu}\phi\right]\,, \label{LC_Horn_Comp4}
\end{align}
\end{subequations}
where  $X=-\frac{1}{2}\mathring{\nabla}{}_{\mu}\phi\mathring{\nabla}{}^{\mu}\phi$, and the full Lagrangian will be
\begin{equation}
\mathring{L}=\sum_{i=1}^5 c_i \mathring{L}_i\,.\label{horn}
\end{equation}
Recall that the overcircles denote objects associated to the Levi-Civita connection. These Riemannian quantities, such as the Levi-Civita Ricci scalar can be rewritten in terms of  nonmetricity scalars as $\lc{R}=-Q+B_Q$ (see Eq.~\eqref{R and Q}) and the Levi-Civita Einstein tensor as (see Eq.~\eqref{RiemannSplit})
\begin{align}\label{einstein}
    \lc{G}_{\mu\nu}=2\nabla_\lambda P^\lambda{}_{\mu\nu}-\frac{1}{2}Qg_{\mu\nu}+P_{\rho\mu\nu}Q^{\rho\sigma}{}_{\sigma}+P_{\nu\rho\sigma}Q_\mu{}^{\rho\sigma}-2P_{\rho\sigma\mu}Q^{\rho\sigma}{}_{\nu},
\end{align}
with the so-called superpotential being equal to
\begin{eqnarray}
     P^\alpha{}_{\mu\nu}=-\,\frac{1}{4}Q^{\alpha}{}_{\mu\nu}+\frac{1}{2}Q_{(\mu}{}^\alpha{}_{\nu)}+\frac{1}{4}g_{\mu\nu}Q^\alpha-\frac{1}{4}(g_{\mu\nu}Q_{\beta}{}^{\alpha\beta}+\delta^\alpha{}_{(\mu}Q_{\nu)\beta}\,^{\beta})\,.
\end{eqnarray}
Note that the covariant derivative appearing in the above equation is with respect to the ST connection.
The explicit nonminimal couplings with the Ricci scalar $\mathring{R}$ and Einstein tensor $\mathring{G}{}_{\mu\nu}$ in $\mathring{L}{}_4$ and $\mathring{L}{}_5$, respectively, are required in order to balance terms with higher-than-second derivatives in the Euler-Lagrange field equations coming from the Galilean-like structure of second covariant (Levi-Civita) derivatives of the scalar field. Notice that the structure of the second derivatives is fixed, while the $G_i$ functions might only depend on invariants with at most first derivatives.

Then, due to the covariantization procedure, the Riemannian Horndeski theory also appears naturally in our construction since by upgrading the Galileon fields in Minkowski spacetime to the case where gravity is switched on, the theory coincides with the Riemannian Horndeski theory. However, in ST gravity, the procedure from Galileon fields in Minkowski spacetime to the gravity case does not coincide with the most general theory with second-order Euler-Lagrange field equations. Let us here now consider a systematic way of constructing the theory in the ST framework.

As discussed above, in the Riemannian formulation this is the end of the line, since we have run out of invariants with up to first derivatives of the dynamical fields to construct. This is however not the case in the ST formulation, and there are two ways in which this can be extended:
\begin{enumerate}[label=(\alph*)]
\item Including a dependence in the $G_i$ functions on new invariants that now can be constructed with the nonmetricity tensor $Q_{\lambda\mu\nu}$ and $\mathring{\nabla}{}_{\mu} \phi$, which contain at most first derivatives of the dynamical fields.
\item Adding new higher-derivative terms with $\mathring{\nabla}{}_{\alpha} Q_{\lambda\mu\nu}$, which contain second derivatives of the metric in novel ways and likely requiring a precise tuning analogous to the Galileon structure. 
\end{enumerate}
These two directions of generalisation are not completely independent, as for example, following the first one by adding a new dependence to the $G_i$ functions, can generically give rise to new terms with higher-than-second derivatives in the Euler-Lagrange field equations that need to be balanced out by including terms of the form described in the second option. Much like it happens in Riemannian Horndeski, we expect that one must include all the terms that have the same number of second derivatives such that they can balance each other and prevent the appearance of higher-order terms in the Euler-Lagrange field equations. For this reason, it is useful to rely on a power counting scheme to organise the operators in indivisible classes that must be considered together. With this in mind, we propose the full ST Horndeski Lagrangian to follow a power counting schematically of the form
\begin{eqnarray} \label{PC-rule}
L_i = \sum_{\substack{N_\phi,N_Q \geq 0 \\ N_\phi \geq n \geq 0}} C_{N_\phi,N_Q,n} \, \left[ \sum_{\substack{m,r,l \geq 0 \\ m + 2r + l = i-2}}  A^{(n,N_\phi,N_Q)}_{m,r,l} \,\, \phi^{N_\phi-n-m} (\partial \phi)^n \, (\mathring{\nabla} \mathring{\nabla} \phi)^{m} \, \mathring{\mathcal{R}}^r \, \mathcal{Q}^{N_Q-l} \, (\mathring{\nabla} \mathcal{Q})^{l} \right]\,,
\end{eqnarray}
where
\begin{equation}
    \mathcal{Q}=(W^\mu,\Lambda^\mu,q^{\mu\nu\rho},{*\Omega}^{\alpha\mu\nu})\,,
\end{equation}
generically represent the irreducible nonmetricity building blocks, which count as first derivatives (they contain $\partial g$), $\mathring{\mathcal{R}}$ stands schematically for the Riemannian curvature tensor $\mathring{R}^{\mu}{}_{\nu\rho\sigma}$, and the $C_{N_\phi,N_Q,n}$ and $A^{(n,N_\phi,N_Q)}_{m,r,l}$ coefficients should have appropriate mass dimensions. In the above Lagrangian we allow at most second derivatives acting on the scalar field and the metric. As already mentioned, due to Eq.~\eqref{RiemannSplit} the terms $\mathring{\mathcal{R}}$ (that only depend on the metric) can be always written as $\mathcal{Q}^p$ and $(\mathring{\nabla} \mathcal{Q})$ terms and then it would be sufficient to write the above general Lagrangian without the $\mathring{\mathcal{R}}$ term. However, by including it, one could get the corresponding Riemannian Horndeski case easily and due to that, we will keep it as a building block. This is a consistent choice that entails no loss of generality, but where the new terms are organised in a way that puts Riemannian Horndeski and the Levi-Civita covariantization procedure in the forefront.

The full Lagrangian then will be
\begin{equation}
L=\sum_{i=2} c_i L_i\,,\label{stele-horn}
\end{equation}
where in principle the summation can extend up to arbitrary $i$, but after some point, the Lagrangians will become trivial, i.e. total derivatives.

The above Lagrangians are constructed by considering nested summations. The inner sum in Eq. \eqref{PC-rule}, between brackets, is over the different ways of distributing the factors containing second derivatives, $\mathring{\nabla} \mathring{\nabla} \phi$, $\mathring{\mathcal{R}}$, and $\mathring{\nabla} \mathcal{Q}$, such that they add up to exactly $i-2$. Notice that here $\mathring{\mathcal{R}}$ enjoys a special counting inspired by the one from Riemannian Horndeski. As mentioned above, all of these terms must be considered together as an indivisible set (same $n$, $N_\phi$, $N_Q$), with relative coefficients $A^{(n,N_\phi,N_Q)}_{m,r,l}$ that are not free (except for an overall factor), but rather must be fixed by the requirement of second-order Euler-Lagrange field equations. This must be verified against variations with respect to all the fields, $\phi$, $g_{\mu\nu}$ and $\Gamma^{\alpha}{}_{\mu\nu}$. Notice however that due to the nature of nonmetricity, Eq.~\eqref{nonmetricity}, variations with respect to the connection automatically satisfy this for $i \leq 3$.

Importantly, while the proposed counting scheme cannot tell us anything at this stage about the proper nontrivial structure needed to achieve second-order Euler-Lagrange field equations, there cannot be contributions from other sets (different $n$, $N_\phi$, $N_Q$) as their powers of fields and derivatives would not match, enabling each class to be studied separately in the search for such a structure. The added benefit of this arrangement is that once each class manifestly leads to second-order Euler-Lagrange field equations, there is complete freedom in how to combine them. This freedom is realised through the outer sum in Eq. \eqref{PC-rule} where the $C_{N_\phi,N_Q,n}$ coefficients are unconstrained, and it expresses the generic dependence of the Lagrangian functions $G_i(\phi,\partial\phi,\mathcal{Q})$ associated to each $L_i$. Notice that once the proper structure has been found, these functions are actually not necessarily constrained to have a series expansion as apparently implied by Eq.~\eqref{PC-rule}. Finally, all of these $L_i$ Lagrangians can also be combined freely as in Eq. \eqref{stele-horn} to form the full ST Horndeski Lagrangian $L$.

\subsection{Known special cases}

The above proposed counting scheme is inspired by the Riemannian Horndeski one, with the purpose of acting as a guiding principle to organise the operators in classes that independently lead to second-order Euler-Lagrange field equations. It will be useful as long as we can describe known cases in a simple way. 

\subsubsection{Quadratic Symmetric Teleparallel gravity without a scalar field}

Let us consider a theory containing only nonmetricity $\mathcal{Q}$ ($N_\phi = 0$), and no higher-order terms ($i=2$),
\begin{eqnarray}
L = \sum_{N_Q \geq 0} C_{N_Q} \, \mathcal{Q}^{N_Q} \equiv f(\{\mathcal{Q}\})\,.
\end{eqnarray}
In order to cast this in a useful way, following our third condition above we further restrict the Lagrangian to be of the form
\begin{eqnarray} \label{f-of-Qs}
L = f(Q_1,Q_2,Q_3,Q_4,Q_5)\,,
\end{eqnarray}
with an arbitrary function depending on the five possible scalars constructed from the nonmetricity tensor up to quadratic order (see Eq.~\eqref{quadratic-Q-invariants}). This theory contains Newer GR in the particular case in which $f= \mathbb{Q}$ (see~\eqref{nonmetquadratic}), and therefore it also contains STEGR. Since the nonmetricity scalar $Q$ appears naturally in ST gravity and this quantity is related to the five irreducible scalars as~\eqref{nonmetricityscalar}, it is convenient to rewrite the above Lagrangian in terms of $Q$ instead of one of the other five scalars, such as $L=f(Q,Q_1,Q_2,Q_3,Q_4)$. That theory is of course equivalent to the above Lagrangian but both the $f(Q)$ and STEGR case can be easily obtained from it in a simpler way. Due to this, we will use this kind of parametrisation for constructing the theory when we add a scalar field.

\subsubsection{Riemannian Horndeski Theory}

The full Riemannian Horndeski Lagrangian is recovered by excluding terms with explicit nonmetricity ($N_Q = 0$) in Eq. \eqref{stele-horn}, but otherwise keeping everything else. Schematically we have
\begin{equation}
\mathring{L}_i = \sum_{\substack{N_\phi \geq n \geq 0}} C_{N_\phi,n} \, \left[ \sum_{\substack{m,r \geq 0 \\ m + 2r = i-2}}  A^{(n,N_\phi)}_{m,r} \,\, \phi^{N_\phi-n-m} (\partial \phi)^n \, (\mathring{\nabla} \mathring{\nabla} \phi)^{m} \, \mathring{\mathcal{R}}^r \right]\,,    
\end{equation}
where now $2 \leq i \leq 5$. It is easy to see that this is consistent with the Lagrangians in Eq. \eqref{horn}. The well-known Galilean-like structure amounts to fixing the $A^{(n,N_\phi)}_{m,r}$ coefficients, while the arbitrariness of the $G_i(\phi, X)$ functions is given by the $C_{N_\phi,n}$ coefficients, which are totally free. Thus, the Riemannian Horndeski gravity theory is contained in our formulated theory.

\subsubsection{Generalized Proca action with the Weyl part of nonmetricity and no scalar field} \label{sec:Proca}

One extension that we can immediately construct without much effort comes in the form of a Generalized Proca Lagrangian with only the Weyl part of nonmetricity, i.e. $W^\mu$, and no scalar field. In terms of the power-counting this is the\emph{ orthogonal} option to Riemannian Horndeski, namely, we let $N_Q$ run while keeping $N_\phi = 0$,
\begin{eqnarray} \label{proca1}
L_i = \sum_{N_Q \geq 0} C_{N_Q} \, \left[ \sum_{\substack{r,l \geq 0 \\ 2r + l = i-2}}  A^{(N_Q)}_{r,l} \,\, \mathring{\mathcal{R}}^r \, W^{N_Q-l} \, (\mathring{\nabla} W)^{l} \right]\,,
\end{eqnarray}
and then we sum over $i$,
\begin{equation} \label{proca2}
L=\sum_{i=2}^6 c_i L_i\,.
\end{equation}
Note that restricting the construction to only use $\mathcal{Q} = \{W^\mu \}$ is only consistent because of the special form of the Weyl vector $W^\mu$ which, in regard to variations with respect to the metric $g_{\mu\nu}$, it has the structure analogous to the gradient of a scalar field
\begin{equation}
    \delta W_\mu \supset \frac{1}{4} g^{\alpha\beta} \partial_\mu \left(\delta g_{\alpha\beta}\right) \to - \frac{1}{4} \delta g_{\alpha\beta} g^{\alpha\beta} \, \partial_\mu( \dots )\,,
\end{equation}
where in the first expression we are not showing terms without derivatives acting on the variation, and the last expression is obtained by integrating by parts and disregarding terms that are not dangerous in the sense of leading to higher-order Euler-Lagrange field equations. Therefore, for the purpose of tracking the cancellation of higher-derivative terms in the metric equations, we can make the replacement $g^{\alpha\beta} \delta g_{\alpha\beta} \to \delta \Psi$, or equivalently write $W_\mu = \frac{1}{4} \partial_\mu \Psi$. Notice that this is not possible in general, and in particular it is not valid for $\Lambda^\mu$. Variations with respect to the flat connection lead instead to the same number of derivatives as those performed with respect to $W^{\mu}$, so it is sufficient to only track the latter ones for ensuring their second-order nature. For these reasons, a theory following Eqs. \eqref{proca1} and \eqref{proca2} can be made to lead to second-order Euler-Lagrange field equations on their own by arranging the Lagrangian in the well-known form~\cite{Heisenberg:2014rta}
\begin{subequations}
\begin{eqnarray} \label{proca3}
    L_2 &=& G_2(Q_1, \mathring{F}_{\mu\nu},\mathring{\tilde{F}}^{\mu\nu}) \, , \\
    L_3 &=& G_3(Q_1) \mathring{\nabla}_\mu W^\mu \, ,\\
    L_4 &=& G_4(Q_1) \mathring{R} - 2 G_{4,Q_1} \left[ (\mathring{\nabla}_\mu W^\mu)^2 - \mathring{\nabla}_\rho W_\sigma \mathring{\nabla}^\sigma W^\rho \right] \, ,\\
    L_5 &=& G_5(Q_1) \mathring{G}_{\mu\nu} \mathring{\nabla}^\mu W^\nu + \frac{1}{3} G_{5,Q_1} \Bigl[ (\mathring{\nabla}_\mu W^\mu)^3 \notag  
     + 2 \mathring{\nabla}_\rho W_\sigma \mathring{\nabla}^\gamma W^\rho \mathring{\nabla}^\sigma W_\gamma - 3 (\mathring{\nabla}_\mu W^\mu) \mathring{\nabla}_\rho W_\sigma \mathring{\nabla}^\sigma W^\rho \Bigr] \notag \\
    && - g_5(Q_1) \mathring{\tilde{F}}^{\alpha\mu} \mathring{\tilde{F}}^{\beta}{}_{\mu} \mathring{\nabla}_\alpha W_\beta \, ,\\
    L_6 &=& G_6(Q_1) \mathring{\mathcal{L}}^{\mu\nu\alpha\beta} \mathring{\nabla}_\mu W_\nu \mathring{\nabla}_\alpha W_\beta 
    - G_{6,Q_1} \mathring{\tilde{F}}^{\alpha\beta} \mathring{\tilde{F}}^{\mu\nu} \mathring{\nabla}_\alpha W_\mu \mathring{\nabla}_\beta W_\nu \,,
\end{eqnarray}
\end{subequations}
where $\mathring{\mathcal{L}}^{\mu\nu\alpha\beta}$ is the double dual Riemann tensor, while $\mathring{F}^{\mu\nu} = \mathring{\nabla}^\mu W^\nu - \mathring{\nabla}^\nu W^\mu$ and $\mathring{\tilde{F}}^{\mu\nu}$ its dual. The Lagrangian $L_2$ is constructed from an arbitrary function that depends on $Q_1$ and all the possible scalars constructed from $F_{\mu\nu}$ and its dual respecting the U(1) symmetry. It should be noted that the theories that we are constructing in this manuscript are parity preserving, so that, any combination in $L_2$ leading to parity-violating invariants would not be part of our theory (such as for example $\mathring{F}^{\mu\nu}\mathring{\tilde{F}}_{\mu\nu}$). Then, the above equation will produce the same field equations as the generalized Proca gravity theory but now the vector field would be associated with the Weyl part of nonmetricity and its vector field norm with $Q_1$. Although the equations would be mathematically equivalent, the physical interpretation of the above model could be different due to the fact that the Weyl part of nonmetricity is invariant under the dilation group. Furthermore, since we are working in the framework of teleparallel gravity (and not metric-affine gravity in general), the form of the vector $W_\mu$ will be constrained under the condition of having a vanishing curvature. Extending this construction to include all of the irreducible components of nonmetricity in full generality is highly nontrivial and beyond the scope of this paper.

\section{Systematic construction of a subclass of Symmetric Teleparallel extensions of Horndeski}\label{secIV}

Constructing the ST Horndeski, that is, the most general second-order Euler-Lagrange field equation theory containing a scalar field $\phi$ and nonmetricity $Q_{\lambda\mu\nu}$ as described by the power counting scheme devised in the previous section, is a very daunting task. On the road to constructing such a theory, it is reasonable to exploit such a scheme by systematically considering each of the $L_i$ Lagrangians one by one, with increasing complexity. 

\subsection{Most general action for a scalar and nonmetricity with no purely teleparallel Higher-Derivative terms}\label{sec:NoHigherOrder}

We begin by formulating the most general theory with a scalar field and nonmetricity with no purely teleparallel higher-derivative terms. According to Eq. \eqref{PC-rule}, this would be $L_2$, which is guaranteed to remain second-order in general since it is at most first-order in derivatives. The simplified power counting in this scenario is simply
\begin{eqnarray} \label{L2-PC}
L_2 = \sum_{N_\phi, N_Q \geq 0} \sum_{n=0}^{N_\phi} c_{N_\phi,N_Q,n} \, \phi^{N_\phi-n} (\partial \phi)^n \mathcal{Q}^{N_Q} \equiv L_2(\phi, \partial\phi, \{\mathcal{Q}\})\,,
\end{eqnarray}
however, similarly to the example given before for the quadratic ST gravity, we need to specify a closed set of variables on which $L_2$ above can depend if we hope to express it in a practically useful way. 

In order to account for the possible dependencies that can enter in Eq. \eqref{L2-PC}, it is useful to classify the invariant quantities that can be built with $Q_{\lambda\mu\nu}$ and the scalar field $\phi$ that contain up to first derivatives. For this purpose, we use the irreducible decomposition of the nonmetricity tensor, defined in Eqs.~\eqref{irreducible-parts}.

First, we consider contractions of nonmetricity up to quadratic order without a scalar field. These were already given in Eq. \eqref{quadratic-Q-invariants}. As mentioned in the previous section, the first part of the theory (without the scalar field and $i=2$) can be written as an arbitrary function of $f(Q,Q_1,Q_2,Q_3,Q_4)$. Since $Q$ is a scalar that appears in STEGR and it is a linear combination of $Q_i$, we will instead consider $Q$, $Q_1$, $Q_2$, $Q_3$, $Q_4$ as a basis of quadratic invariants that can freely enter as dependencies in $L_2$. On top of this, the trivial way to incorporate the scalar field is by purely scalar invariants $\phi$ and $X$, just as in typical Riemannian Horndeski.

Then there are the invariants that can be built from nontrivial contractions of the nonmetricity tensor and the gradient of the scalar field. We can classify them in terms of the number $n$ of powers of $\phi_{;\mu} = \partial_\mu \phi$. With one single factor of $\phi_{;\mu}$, i.e. $n=1$, we have only two possibilities,
\begin{subequations} \label{invariants-n1}
\begin{eqnarray} 
I_1&=&W^\mu \phi_{;\mu}\,,\\
I_2&=&\Lambda^\mu \phi_{;\mu}\,,
\end{eqnarray}
\end{subequations}
both with $N_Q = 1$. With $n=2$ factors of $\phi_{;\mu}$ instead we need to use at least two factors of nonmetricity, i.e. $N_Q \geq 2$, but on the other hand we cannot have more than two if we want to stop at quadratic order (while remaining nontrivial, that is, not reducible to products of simpler invariants). Therefore, we have the following eight invariants:
\begin{subequations} \label{invariants-n2}
\begin{eqnarray}
J_1&=&q_{\lambda\mu \nu}W^\lambda \phi^{;\mu}\phi^{;\nu}\,,\\
J_2&=&q_{\lambda\mu \nu}\Lambda^\lambda \phi^{;\mu}\phi^{;\nu}\,,\\
J_3&=&
{*\Omega}_{\mu}{}^{\nu\sigma}W_\sigma \phi^{;\mu}\phi_{;\nu}\,,\\
J_{4}&=&
{*\Omega}_{\mu}{}^{\nu\sigma}\Lambda_\sigma \phi^{;\mu}\phi_{;\nu}\,,\\
J_{5}&=&q_{\lambda\alpha \mu}q^{\lambda\alpha}{}_\nu \phi^{;\mu}\phi^{;\nu}
\,,\\
J_{6}&=&q_{\lambda\alpha \mu}{*\Omega}^{\lambda\alpha}{}_\nu \phi^{;\mu}\phi^{;\nu}\,\\
J_{7}&=& {*\Omega}_{\lambda\alpha \mu}{*\Omega}^{\lambda\alpha}{}_\nu \phi^{;\mu}\phi^{;\nu}
\,,\\
J_{8}&=& {*\Omega}_{\lambda\alpha \mu}{*\Omega}_\nu{}^{\lambda\alpha} \phi^{;\mu}\phi^{;\nu}
\,.
\end{eqnarray}
\end{subequations}
For $n=3$ we have a single invariant up to quadratic contractions:
\begin{eqnarray} \label{invariants-n3}
J_{9}&=&q_{\lambda\mu\nu}\phi^{;\lambda}\phi^{;\mu}\phi^{;\nu}\,,
\end{eqnarray}
which is actually $N_Q = 1$. Finally, there are three extra invariants with $n=4$ factors of $\phi_{;\mu}$ and quadratic in nonmetricity, $N_Q = 2$,
\begin{subequations} \label{invariants-n4}
\begin{eqnarray}
J_{10}&=&q_{\mu\nu\sigma }q_{\lambda\alpha}{}^\sigma\phi^{;\mu}\phi^{;\nu}\phi^{;\lambda}\phi^{;\alpha}\,,\\
J_{11}&=&q_{\mu\nu\sigma} {*\Omega}_{\lambda\alpha}{}^\sigma\phi^{;\mu}\phi^{;\nu}\phi^{;\lambda}\phi^{;\alpha}\,,\\
J_{12}&=&{*\Omega}_{\mu\nu\sigma }{*\Omega}_{\lambda\alpha}{}^\sigma\phi^{;\mu}\phi^{;\nu}\phi^{;\lambda}\phi^{;\alpha}\,.
\end{eqnarray}
\end{subequations}
In the construction of all the above invariants, we have taken into consideration the symmetry properties of the irreducible components of the nonmetricity tensor. Also, we have taken the condition of having a parity-preserving theory.

Hence, the most general $L_2$ we can construct under our assumptions is 
\begin{equation}
   L_2 = \tilde{G}_{\rm ST}(Q_i,\phi,X,I_i,J_i)\,,
\end{equation}
where we have introduced the notation \begin{eqnarray}\label{QQQQQ}
    Q_i:=\{Q,Q_1,Q_2,Q_3,Q_4\}\,,\quad I_i:=\{I_1,I_2\}\,,\quad J_i:=\{J_1,J_2,J_3,J_4,J_5,J_6,J_7,J_8,J_9,J_{10},J_{11},J_{12}\}\,.
\end{eqnarray}
Furthermore, the above Lagrangian is the ST generalization of Horndeski's $\mathring{L}_2$ Lagrangian, and it also contains the theory of Eq. \eqref{f-of-Qs}. Equivalently, we can separate it into the Riemannian Horndeski part plus a purely ST contribution
\begin{equation}
 L_2 = \mathring{L}_2 + L_{\rm 2-ST}= G_2(\phi, X) + G_{\rm ST}(Q_i,\phi,X,I_i,J_i) \,.
\end{equation}
It can be of course combined with any other Lagrangian that stands as independently leading to second-order Euler-Lagrange field equations. If we only consider a theory containing no higher-derivative terms acting on nonmetricity (a theory that cannot be recast as a purely Riemannian higher-order theory), then, the others $L_i$ will only correspond to set $N_Q=0$ which would give us Horndeski gravity. This means that the most general ST Horndeski gravity theory without higher-order derivatives acting on purely teleparallel terms would be given by the above Lagrangian plus the Riemannian Horndeski Lagrangian $\mathring{L}$, namely,
\begin{equation}\label{Lagrangian}
L= \mathring{L} + L_{\rm 2-ST} = \sum_{i=2}^5 \mathring{L}_i + G_{\rm ST}(Q_i,\phi,X,I_i,J_i) \,,
\end{equation}
as long as one remembers that the Riemannian Horndeski $G_i$ functions are only dependent on $\phi$ and $X$, and only $G_{\rm ST}$ here contains explicit dependence on nonmetricity. Extending this dependence to $i > 2$ would require the inclusion of higher-order terms involving nonmetricity (that cannot be recast as purely Riemannian gravity), as described by our power-counting scheme of Eq. \eqref{PC-rule}. Doing this, in general, is beyond the scope of this work, but we will examine an example in the following section and find an extension of $\mathring{L}_3$.

\subsection{Symmetric Teleparallel Kinetic Gravity Braiding (STKGB) up to linear order in nonmetricity}\label{sec:STKGB}

So far we have discussed a ST extension of Horndeski that does not involve higher-derivative terms with nonmetricity. The previously formulated theory contains higher-order derivatives in nonmetricity but only the particular combinations such that those terms can be purely written as a Riemannian gravity theory (i.e., $L_3,L_4$ and $L_5$ from Horndeski gravity). In terms of the general power-counting scheme of Eq. \eqref{PC-rule}, this means staying at $i=2$, and letting $N_Q$ run (with the restriction of having at most quadratic contractions of nonmetricity), or, allowing arbitrary $i$ but then forcing $N_Q = 0$, which is nothing else than the Riemannian Horndeski Lagrangian. Combining these two possibilities lead to the Lagrangian in Eq. \eqref{Lagrangian}, which is the most general $L_2$. Then, in this section we would like to explore the possibility of going beyond, and, while we will not formulate here the most general theory contained in Eq. \eqref{PC-rule}, we will at least show that there is a path forward toward the construction of nontrivial extensions.

In this section, we will focus on the extension form of $L_3$ by considering possible higher-order derivatives acting on the nonmetricity tensor (that cannot be recast as purely Riemannian terms). The power counting scheme of Eq. \eqref{PC-rule} for $i=3$ specialises to
\begin{eqnarray} \label{PC-KGB}
L_3 = \sum_{\substack{N_\phi,N_Q \geq 0 \\ N_\phi \geq n \geq 0}} C_{N_\phi,N_Q,n} \, \left[ \sum_{\substack{m,l \geq 0 \\ m + l = 1}}  A^{(n,N_\phi,N_Q)}_{m,l} \,\, \phi^{N_\phi-n-m} (\partial \phi)^n \, (\mathring{\nabla} \mathring{\nabla} \phi)^{m} \, \mathcal{Q}^{N_Q-l} \, (\mathring{\nabla} \mathcal{Q})^{l} \right]\,,
\end{eqnarray}
where the condition $m + 2r + l = 1$ forces $r = 0$, i.e. no factor of the Riemannian curvature tensor is present. This condition also implies that the only terms allowed are those with exactly one factor of second derivatives (either $\mathring{\nabla} \mathring{\nabla} \phi$ or $\mathring{\nabla} \mathcal{Q}$). These are expected properties of a ST generalization of Cubic Horndeski/Kinetic Gravity Braiding. 

The above expression is still very general in that it allows both the scalar and nonmetricity to enter arbitrarily (generic $N_\phi$ and $N_Q$). Since in this paper we are interested in extending Riemannian Horndeski by the inclusion of nonmetricity, we take the approach to be general in the $\phi$ and $X$ dependence and instead incorporate $Q_{\alpha\mu\nu}$ gradually. Therefore, we start by noting as already discussed in Sec. \ref{sec:HorndeskiTeleparallel} that the full Riemannian Cubic Horndeski/Kinetic Gravity Braiding Lagrangian is contained in this scheme simply as the $N_Q = 0$ terms of the outer sum, in which case the inner sum collapses to a single type of term with the second derivatives acting on $\phi$,
\begin{eqnarray} \label{KGB}
\mathring{L}_3 = L_3(N_Q = 0) &=& \sum_{\substack{N_\phi \geq 0 \\ N_\phi \geq n \geq 0}} C_{N_\phi,N_Q=0,n} \, \left[ A^{(n,N_\phi,N_Q=0)}_{1,0} \,\, \phi^{N_\phi-n-1} (\partial \phi)^n \, (\mathring{\nabla} \mathring{\nabla} \phi) \right] \notag \\
&=&- G_3(\phi, X) \, \mathring{\Box} \phi \, ,
\end{eqnarray}
where the other potential term that could have been included $\phi^{;\mu} \phi^{;\nu}\, \mathring{\nabla}_\mu \mathring{\nabla}_\nu \phi$ has been removed by integration by parts. This Lagrangian is known to lead to second-order Euler-Lagrange field equations. The next step is to include nonmetricity in the simplest way, that is linearly ($N_Q = 1$). As we will now show, this will prove to be already somewhat involved. For $N_Q = 1$ the above schematic Lagrangian reduces to
\begin{eqnarray}
L_3(N_Q = 1) &=& \sum_{\substack{N_\phi \geq 0 \\ N_\phi \geq n \geq 0}} C_{N_\phi,N_Q=1,n} \, \left[ A^{(n,N_\phi,N_Q=1)}_{1,0} \,\, \phi^{N_\phi-n-1} (\partial \phi)^n \, (\mathring{\nabla} \mathring{\nabla} \phi) \, \mathcal{Q} + A^{(n,N_\phi,N_Q=1)}_{0,1} \phi^{N_\phi-n} (\partial \phi)^n \, (\mathring{\nabla} \mathcal{Q}) \right] \notag \\
&=& \sum_{a} \left[ \tilde{G}^{(a)}_3(\phi, X) \tilde{\mathcal{O}}_a + F^{(a)}_3(\phi, X) \hat{\mathcal{O}}_a \right] \, ,
\end{eqnarray}
where in the first expression we have explicitly developed the inner sum of Eq. \eqref{PC-KGB} into the two types of terms described before, and the remaining summation over $n$ and $N_\phi$ implements the generic dependence on $\phi$ and $X$. On the second equality we resummed that dependence into generic functions $\tilde{G}^{(a)}_3(\phi, X)$ and $F^{(a)}_3(\phi, X)$ which multiply invariants $\tilde{\mathcal{O}}_a \sim \mathring{\nabla} \mathring{\nabla} \phi$ of the first kind ($m=1,l=0$) and $\hat{\mathcal{O}}_a \sim \mathring{\nabla} \mathcal{Q}$ of the second kind ($m=0,l=1$), respectively. These are invariants linear in $Q_{\alpha\mu\nu}$ constructed with $\phi^{;\rho}$ and one $\mathring{\nabla}_\lambda$ properly positioned. The summation over \``$a$" loosely stands for summing over all the possible ways of contracting indices to build such invariants\footnote{Notice that there need not be the same number of $\tilde{\mathcal{O}}_a$ and $\hat{\mathcal{O}}_a$ scalars.}.

We proceed first to construct all the invariants with no derivatives acting on $Q_{\lambda\alpha\beta}$, and one factor of $\mathring{\nabla}_\mu \mathring{\nabla}_\nu \phi$, namely the ones we denote $\tilde{\mathcal{O}}_a$. In what follows, it will be convenient to work directly with $Q_{\alpha\mu\nu}$ instead of its irreducible components $W^\mu$, $\Lambda^\mu$, $\Omega_{\alpha\mu\nu}$, and $q_{\alpha\mu\nu}$, as the variations with respect to the metric take a simpler form, making it easier to find the appropriate structure that ensures second-order Euler-Lagrange field equations. We provide some details about the variations in Appendix \ref{app:variations}. One can, later on, reexpress the Lagrangians in terms of the irreducible components if desired. There are twelve $\tilde{\mathcal{O}}_a$ invariants, which in increasing order in factors of $\phi^{;\rho}$'s are
\begin{subequations} \label{B-operators}
\begin{eqnarray}
\tilde{\mathcal{O}}_{1} &=& \phi^{;\mu} Q_{\mu\nu}{}^{\nu} \, \mathring{\Box} \phi \, ,\\
\tilde{\mathcal{O}}_{2} &=& \phi^{;\mu} Q_{\nu\mu}{}^{\nu} \, \mathring{\Box} \phi \, ,\\
\tilde{\mathcal{O}}_{3} &=& \phi_{;\alpha} Q_{\beta\mu}{}^{\mu} \, \mathring{\nabla}^\alpha \mathring{\nabla}^\beta \phi \, ,\\
\tilde{\mathcal{O}}_{4} &=& \phi_{;\alpha} Q_{\mu\beta}{}^{\mu} \, \mathring{\nabla}^\alpha \mathring{\nabla}^\beta \phi \, ,\\
\tilde{\mathcal{O}}_{5} &=& \phi^{;\mu} Q_{\alpha\mu\beta} \, \mathring{\nabla}^\alpha \mathring{\nabla}^\beta \phi \, ,\\
\tilde{\mathcal{O}}_{6} &=& \phi^{;\mu} Q_{\mu\alpha\beta} \, \mathring{\nabla}^\alpha \mathring{\nabla}^\beta \phi \, ,\\
\tilde{\mathcal{O}}_{7} &=& \phi^{;\mu} \phi^{;\nu} \phi^{;\alpha} Q_{\mu\nu\alpha} \, \mathring{\Box} \phi \, ,\\
\tilde{\mathcal{O}}_{8} &=& \phi^{;\mu} \phi^{;\nu} \phi^{;\alpha} Q_{\nu\beta}{}^{\beta} \, \mathring{\nabla}_\mu \mathring{\nabla}_\alpha \phi \, ,\\
\tilde{\mathcal{O}}_{9} &=& \phi^{;\mu} \phi^{;\nu} \phi^{;\alpha} Q_{\beta\nu}{}^{\beta} \, \mathring{\nabla}_\mu \mathring{\nabla}_\alpha \phi \, ,\\
\tilde{\mathcal{O}}_{10} &=& \phi^{;\mu} \phi^{;\nu} \phi^{;\alpha} Q_{\mu\nu}{}^{\beta} \, \mathring{\nabla}_\alpha \mathring{\nabla}_\beta \phi \, ,\\
\tilde{\mathcal{O}}_{11} &=& \phi^{;\mu} \phi^{;\nu} \phi^{;\alpha} Q^{\beta}{}_{\mu\nu}\, \mathring{\nabla}_\alpha \mathring{\nabla}_\beta \phi \, ,\\
\tilde{\mathcal{O}}_{12} &=& \phi^{;\mu} \phi^{;\nu} \phi^{;\alpha} \phi^{;\rho} \phi^{;\sigma} Q_{\mu\nu\alpha} \, \mathring{\nabla}_\rho  \mathring{\nabla}_\sigma \phi \, ,
\end{eqnarray}
\end{subequations}
where the symmetry property $Q_{\alpha\mu\nu} = Q_{\alpha\nu\mu}$ has been taken into account. As explained above, on top of this set of invariants one can also construct another different set $\hat{\mathcal{O}}_a$, with $\mathring{\nabla}_\mu$ acting on $Q_{\lambda\alpha\beta}$ instead and then only factors of $\phi^{;\rho}$, i.e. no second derivatives of $\phi$. However, at the level we are working on here linear in nonmetricity ($N_Q = 1$) and with only one factor with second derivatives in total ($i=3$), it is clear that the $\hat{\mathcal{O}}_a$ can in fact be expressed in terms of the $\tilde{\mathcal{O}}_a$ by integration by parts, as we can always move the derivative acting on $Q_{\lambda\alpha\beta}$ onto some factor of $\phi$. Then, one can absorb this redundancy in a redefinition of the $\tilde{G}_3^{(a)}(\phi, X)$, which are so far generic. For completeness, we list the seven $\hat{\mathcal{O}}_a$ invariants in Appendix \ref{app:int-by-parts} and show their relations with those in Eqs. \eqref{B-operators}. This equivalence involves discarding terms that do not contain second derivatives and hence belong to $L_2$ rather than $L_3$.

We can now construct Lagrangians that lead to second-order Euler-Lagrange field equations by carefully examining the variations with respect to both the scalar field $\phi$, the metric $g_{\mu\nu}$ and the connection $\Gamma^\rho{}_{\mu\nu}$ of the invariants in Eqs. \eqref{B-operators}, allowing them to be multiplied by arbitrary functions of $\phi$ and $X$. We find five independent pair-wise combinations for which any terms that would lead to higher-than-second-order Euler-Lagrange field equations are canceled. These are
\begin{subequations}\label{O-operators-combined}
\begin{eqnarray}
\tilde{L}_3^{(1)} &=& \tilde{G}_{3}^{(1)}(\phi, X) ( \tilde{\mathcal{O}}_3 - \tilde{\mathcal{O}}_1 ) =2\tilde{G}_{3}^{(1)}(\phi, X)\phi_{;\alpha}L^{\mu}{}_{\beta\mu}\left[\mathring{\nabla}^\alpha \mathring{\nabla}^\beta\phi-g^{\alpha\beta}\mathring{\Box}\phi\right]\, , \\
    \tilde{L}_3^{(2)} &=& \tilde{G}_{3}^{(2)}(\phi, X) ( \tilde{\mathcal{O}}_4 - \tilde{\mathcal{O}}_6 )=\tilde{G}_{3}^{(2)}(\phi, X) \left(g_{\alpha\beta,\mu}-\Gamma^{\lambda}{}_{\mu\alpha}g_{\lambda\beta}-\Gamma^{\lambda}{}_{\mu\beta}g_{\alpha\lambda}\right)\left[\phi_{;\rho}g^{\mu\alpha}\mathring{\nabla}^\rho \mathring{\nabla}^\beta\phi-\phi^{;\mu}\mathring{\nabla}^\alpha \mathring{\nabla}^\beta\phi\right]\, , \\
    \tilde{L}_3^{(3)} &=& \tilde{G}_{3}^{(3)}(\phi, X) ( \tilde{\mathcal{O}}_5 - \tilde{\mathcal{O}}_2 ) =\tilde{G}_{3}^{(3)}(\phi, X) \phi^{;\mu}\left(g_{\mu\beta,\alpha}-\Gamma^{\lambda}{}_{\alpha\mu}g_{\lambda\beta}-\Gamma^{\lambda}{}_{\alpha\beta}g_{\mu\lambda}\right)\left[\mathring{\nabla}^\alpha \mathring{\nabla}^\beta\phi-g^{\alpha\beta}\mathring{\Box}\phi\right] \, , \\
   \tilde{L}_3^{(4)} &=& \tilde{G}_{3}^{(4)}(\phi, X) ( \tilde{\mathcal{O}}_{10} - \tilde{\mathcal{O}}_9 ) = \tilde{G}_{3}^{(4)}(\phi, X) \phi^{;\alpha}\phi^{;\nu}\left(g_{\nu\rho,\mu}-\Gamma^{\lambda}{}_{\mu\nu}g_{\lambda\rho}-\Gamma^{\lambda}{}_{\mu\rho}g_{\nu\lambda}\right)\left[\phi^{;\mu}g^{\beta\rho}-\phi^{;\beta}g^{\rho\mu}\right]\mathring{\nabla}_\alpha \mathring{\nabla}_\beta\phi \, , \\
    \tilde{L}_3^{(5)} &=& \tilde{G}_{3}^{(5)}(\phi, X) ( \tilde{\mathcal{O}}_{11} - \tilde{\mathcal{O}}_7 ) = \tilde{G}_{3}^{(5)}(\phi, X)\phi^{;\rho}\phi^{;\nu}\left(g_{\nu\rho,\mu}-\Gamma^{\lambda}{}_{\mu\nu}g_{\lambda\rho}-\Gamma^{\lambda}{}_{\mu\rho}g_{\nu\lambda}\right)\left[\phi^{;\alpha}g^{\mu\beta}\mathring{\nabla}_\alpha \mathring{\nabla}_\beta\phi-\phi^{;\mu}\mathring{\Box}\phi\right]  \, , 
\end{eqnarray}
\end{subequations}
where we explicitly expressed the nonmetricity tensor in terms of the metric and the connection.

Notice that there is a pair missing, as there are terms in $\tilde{\mathcal{O}}_{8}$ and $\tilde{\mathcal{O}}_{12}$ that cannot be canceled away. Therefore, their coefficients must be set to zero. It is worth noting that in $L_3$, the variation with respect to the connection $\Gamma^\alpha{}_{\mu\nu}$ will always lead automatically to second-order equations due to it appearing in the above construction without any derivatives acting on it (see the definition of nonmetricity~\eqref{nonmetricity}). 

A linear combination of these Lagrangians gives the most general $L_3$ with $N_Q = 1$,
\begin{equation}
    L_3(N_Q=1) = \sum_{a=1}^5 \tilde{L}_3^{(a)} \, .
\end{equation}
Finally, we can expand the Lagrangians in terms of the irreducible components of nonmetricity using the alternative basis of twelve invariants
\begin{subequations} \label{B-operators-irrep}
\begin{eqnarray}
\mathcal{O}_{W1} &=& W_{\mu} \phi^{;\mu} \, \mathring{\Box} \phi \, , \\
\mathcal{O}_{W2} &=& W_{\alpha} \phi_{;\beta} \, \mathring{\nabla}^\alpha \mathring{\nabla}^\beta \phi \, , \\
\mathcal{O}_{W3} &=& W_{\mu} \phi^{;\mu} \phi^{;\alpha} \phi^{;\beta} \, \mathring{\nabla}_\alpha \mathring{\nabla}_\beta \phi \, , \\
\mathcal{O}_{\Lambda1} &=& \Lambda_\mu \phi^{;\mu} \, \mathring{\Box} \phi \, , \\
\mathcal{O}_{\Lambda2} &=& \Lambda_{\alpha} \phi_{;\beta} \, \mathring{\nabla}^\alpha \mathring{\nabla}^\beta \phi \, , \\
\mathcal{O}_{\Lambda3} &=& \Lambda_\mu \phi^{;\mu} \phi^{;\alpha} \phi^{;\beta}\, \mathring{\nabla}_\alpha \mathring{\nabla}_\beta \phi \, , \\
\mathcal{O}_{\Omega1} &=& {*\Omega}_{\alpha\beta\mu} \phi^{;\mu} \, \mathring{\nabla}^\alpha \mathring{\nabla}^\beta \phi \, , \\
\mathcal{O}_{\Omega2} &=& {*\Omega}_{\alpha\beta\mu} \phi^{;\alpha} \phi^{;\beta} \phi_{;\nu} \, \mathring{\nabla}^\mu \mathring{\nabla}^\nu \phi \, , \\
\mathcal{O}_{q1} &=& q_{\alpha\beta\mu} \phi^{;\mu} \mathring{\nabla}^\alpha \mathring{\nabla}^\beta \phi \, , \\
\mathcal{O}_{q2} &=& q_{\alpha\beta\mu} \phi^{;\mu} \phi^{;\alpha} \phi^{;\beta} \, \mathring{\Box} \phi \, , \\
\mathcal{O}_{q3} &=& q_{\alpha\beta\mu} \phi^{;\alpha} \phi^{;\beta} \phi_{;\nu} \, \mathring{\nabla}^\mu \mathring{\nabla}^\nu \phi \, , \\
\mathcal{O}_{q4} &=& q_{\alpha\beta\mu} \phi^{;\mu} \phi^{;\alpha} \phi^{;\beta} \phi^{;\rho} \phi^{;\sigma} \, \mathring{\nabla}_\rho \mathring{\nabla}_\sigma \phi \, , 
\end{eqnarray}
\end{subequations}
where the relation between the two bases is given in Appendix \ref{app:operators-irrep}. We can also take linear combinations of these Lagrangians to decouple them as much as possible. We find
\begin{eqnarray}\label{L3one}
L_3^{(1)}&=&(\mathcal{O}_{W2} - \mathcal{O}_{W1})G_3^{(1)}(\phi,X):=\mathcal{O}_1G_3^{(1)}(\phi,X) \,,\\
L_3^{(2)}&=&\left(5\mathcal{O}_{\Lambda 1}+\mathcal{O}_{\Lambda 2}-4\mathcal{O}_{q1}\right)G_3^{(2)}(\phi,X):=\mathcal{O}_2 G_3^{(2)}(\phi,X) \,,\\
L_3^{(3)}&=&\Big(3(\mathcal{O}_{\Lambda 2}-\mathcal{O}_{\Lambda 1})+\mathcal{O}_{\Omega 1 }\Big)G_3^{(3)}(\phi,X):=\mathcal{O}_3G_3^{(3)}(\phi,X) \,,\\
L_3^{(4)}&=&\Big(3X(\mathcal{O}_{\Lambda 2}-\mathcal{O}_{\Lambda 1})+2(3\mathcal{O}_{\Lambda 3}+\mathcal{O}_{q2})-6 \mathcal{O}_{q 3}\Big)G_3^{(4)}(\phi,X):= \mathcal{O}_4 G_3^{(4)}(\phi,X)\,,\\
L_3^{(5)}&=&\Big(3X(\mathcal{O}_{\Lambda 2}+\mathcal{O}_{\Lambda 1})+2(3\mathcal{O}_{\Lambda 3}-\mathcal{O}_{q 2})-\mathcal{O}_{\Omega 2}\Big)G_3^{(5)}(\phi,X):=\mathcal{O}_5G_3^{(5)}(\phi,X) \,.\label{L3five}
\end{eqnarray}
Notice that in contrast with the Riemannian Cubic Horndeski/Kinetic Gravity Braiding, Eq. \eqref{KGB}, here it is not possible to have any of these scalars independently give second-order Euler-Lagrange field equations, but rather it is necessary to combine them. This is a novel property of the ST extension. 

To conclude this section, we state explicitly its main result, namely, a Lagrangian with $i=3$ and $N_Q \leq 1$, which we call Symmetric Teleparallel Kinetic Gravity Brading (STKGB) which would be constructed by adding the Riemannian Kinetic Gravity Brading term (i.e., $G_5=0,G_4=1$) with the above teleparallel contribution, namely, 
\begin{eqnarray}
    L_{\rm STKGB}&=&\lc{R}+\mathring{L}_{2}+\mathring{L}_{3}+\sum_{a=1}^5 L_3^{(a)}\\
    &=&\lc{R}+G_2(\phi,X)-G_3(\phi,X)\mathring{\Box}\phi+\mathcal{O}_1G_3^{(1)}(\phi,X)+\mathcal{O}_2G_3^{(2)}(\phi,X)+\mathcal{O}_3G_3^{(3)}(\phi,X)\nonumber\\
    &&+\mathcal{O}_4G_3^{(4)}(\phi,X)+\mathcal{O}_5G_3^{(5)}(\phi,X)\,,\label{STKGB}
\end{eqnarray}
which contains $\mathring{L}_{3}$ and $\mathring{L}_{2}$ as in the Riemannian Kinetic Gravity Braiding/Cubic Horndeski Lagrangian. In principle, the above systematic procedure can be carried on to higher $N_Q$ in order to incorporate nonmetricity in a more general way into $L_3$. The steps are clear: first, use the power-counting scheme of Eq. \eqref{PC-KGB} to identify the relevant kinds of operators at a given $N_Q$, then construct all the possible invariants with the appropriate factors of $Q_{\lambda\alpha\beta}$, $\phi^{;\rho}$ and $\mathcal{\nabla}_\mu$ (one in the case of $L_3$). Compute the variations with respect to $\phi$, $g_{\mu\nu}$ and $\Gamma^{\alpha}{}_{\mu\nu}$ (the latter not being necessary for $L_3$) of the operators built with the invariants multiplied by arbitrary functions of $\phi$ and $X$. One needs only to track terms that can lead to higher-order Euler-Lagrange field equations. Finally, find the proper combinations of these operators that ensure the cancellation of such higher-order terms for all Euler-Lagrange field equations.

It is clear though that this approach becomes increasingly complicated as the number of possible invariants that can be constructed increases very quickly with $N_Q$. The same happened even in $L_2$, where we had to limit the number of invariants to those containing up to quadratic contractions of the nonmetricity tensor $Q_{\lambda\alpha\beta}$. Following the same prescription here, it would only be necessary to go up to quadratic invariants for $L_3$ as well, of which there are many more. In contrast to the $L_2$ case, however, here one needs to then compute the variations of all of them while multiplied by arbitrary functions of $\phi$ and $X$.

Potentially one could also include a more general dependency on nonmetricity by including the invariants from Eqs. \eqref{quadratic-Q-invariants}, \eqref{invariants-n1}, \eqref{invariants-n2}, \eqref{invariants-n3} and \eqref{invariants-n4} in the arbitrary functions as well. This equates to letting $N_Q$ be arbitrary.

\subsection{Towards the most general Symmetric Teleparallel Horndeski theory}\label{HO}

The full ST Horndeski should include $L_i$ with $i>3$ as well. According to the general power-counting scheme of Eq. \eqref{PC-rule}, this implies terms with two or more factors of second derivatives (that cannot be recast as just Riemannian higher-order terms). For example, terms of the following form are to be expected in $L_4$ at linear order in nonmetricity,
\begin{equation}
    \mathcal{Q} (\mathring{\nabla} \mathring{\nabla} \phi )^2 \, , \qquad  (\mathring{\nabla} \mathring{\nabla} \phi ) \mathring{\nabla} \mathcal{Q} \, .
\end{equation}
On top of the obvious increase in complexity compared to the case discussed in the previous section due to the sheer number of possible invariants than can be built, there is also a novel ingredient. In contrast to $L_3$, here it is not guaranteed that the equation for the connection $\Gamma^{\rho}{}_{\mu\nu}$ is automatically second-order, nor that the connection itself never appears with more than two derivatives acting on it. Take for example the second term above, we can find the connection with one derivative acting on it inside the  $\mathring{\nabla} \mathcal{Q}$ factor. Upon varying with respect to it, this will generate terms that go like $\mathring{\nabla} \mathring{\nabla} \mathring{\nabla} \phi$. Similarly, the scalar field equation can now have terms like $\mathring{\nabla} \mathring{\nabla} \mathring{\nabla} \Gamma$ as well. Special cases like the Generalized Proca theory with $W^\mu$ discussed in Sec.~\ref{sec:Proca} are known to be safe in this respect due to the special properties of the Weyl component of nonmetricity, so it needed not be discussed in detail there, but generally, the connection and its field equations can no longer be neglected in the procedure. These facts make the construction of $L_i$ with $i>3$ highly nontrivial when nonmetricity is included. We leave such efforts for future work.

\section{Flat FLRW cosmology in a subclass of Symmetric Teleparallel Horndeski gravity}\label{FLRW}
As an immediate application of the above construction, in this section, we will study flat FLRW cosmology for a subclass of ST Horndeski which we consider to be constructed from the Riemannian Horndeski plus the Teleparallel contributions that are related to $L_2$ and $L_3$, namely
\begin{eqnarray}
L_{\rm ST-Horn}&=&\sum_{k=4}^5 \mathring{L}_k + G_{\rm ST}(Q_i,\phi,X,I_i,J_i)+ L_{\rm STKGB}\,.\label{telehor}
\end{eqnarray}
As said before, in ST gravity, the connection and the metric are independent. This means that if one assumes that the metric follows certain symmetries, it is not guaranteed that the connection (and then the field equations) will also respect those symmetries. The simplest way to work in these theories is to consider that both the connection (or nonmetricity tensor) and the metric follow the same symmetries. This can be achieved by choosing the invariance under the same set of Killing vector fields $Z_\zeta$, with $\zeta=\{1,..,m\}$, which gives the following conditions
\begin{align}\label{LieD_mag}
(\mathcal{L}_{Z_\zeta}g)_{\mu\nu}=0\,,\qquad
(\mathcal{L}_{Z_\zeta}\Gamma)^{\lambda}\,_{\mu\nu}=0\,.
\end{align}
The last equation implies that also the nonmetricity tensor satisfies $(\mathcal{L}_{Z_\zeta}Q)_{\alpha\mu\nu}=0$. 

If one solves the condition~\eqref{LieD_mag} for an isotropic and homogeneous spacetime,  the metric would be described by the FLRW metric that can be written in spherical coordinates $(t,r,\vartheta,\varphi)$ for its flat case as
\begin{eqnarray}
ds^2=-N(t)^2dt^2+a(t)^2(dr^2+r^2d\vartheta^2+r^2\sin^2\vartheta d\varphi^2)\,.
\end{eqnarray}
For the connection part, it is convenient to decompose the metric as
\begin{equation}
    g_{\mu\nu}=-n_{\mu}n_{\nu} + h_{\mu\nu}\,,\quad n_\mu=(-N,0,0,0)\,,
\end{equation}
and then, we can write the nonmetricity tensor satisfying~\eqref{LieD_mag} as~\cite{Hohmann:2021ast}
\begin{equation}\label{eq:cosmononmet}
Q_{\rho\mu\nu} = 2F_1n_{\rho}n_{\mu}n_{\nu} + 2F_2n_{\rho}h_{\mu\nu} + 2F_3h_{\rho(\mu}n_{\nu)}\,,
\end{equation}
with $F_i=F_i(t)$ which ensures that the connection does not contain torsion and depending on the values of the functions, one can also ensure the flat curvature condition. This condition can be established in three different ways which give three different branches. It is worth mentioning that the above nonmetricity tensor respects cosmological symmetries and it is easy to notice that always its pseudotensor part vanishes ($\Omega_{\lambda\mu\nu}=0$). This means that for cosmology at the background level, we have that
\begin{eqnarray}
Q_4=J_3=J_4=J_6=J_7=J_8=J_{11}=J_{12}=0\,.
\end{eqnarray}

Hereafter, we will introduce the Hubble parameter as
\begin{equation}
H = \frac{\partial_t a}{ N a }=\frac{\dot{a}}{Na}\,,
\end{equation}
where dots are a differentiation with respect to the time coordinate\footnote{Note that in~\cite{Hohmann:2021ast}, dots are differentiation with respect to the conformal time}. 

We can further add extra matter content to our Lagrangian. Since the connection is independent of the metric in ST gravity, we can add an arbitrary matter Lagrangian $L_{\rm m}$ that not only depends on the metric but also on the connection, namely
\begin{eqnarray}\label{Laggg}
L=L_{\rm ST-Horn}+L_{\rm m}(g_{\mu\nu},\Gamma^\lambda{}_{\mu\nu})\,,
\end{eqnarray}
with $L_{\rm ST-Horn}$ given by Eq.~\eqref{telehor}. This choice would introduce a new matter source that is associated with assuming that the matter sector is also coupled to the connection and amounts to a specific choice of covariantization prescription as in Eq.~\eqref{equiv_prin}. Thus, by taking variations with respect to the metric and the connection, we define the energy-momentum tensor and the hypermomentum tensor as
\begin{eqnarray}
T^{\mu\nu}=\frac{2}{\sqrt{-g}}\frac{\delta(\sqrt{-g}L_{\rm m})}{\delta g_{\mu\nu}}\,,\quad \Delta_{\alpha}{}^{\mu\nu}=\frac{-2}{\sqrt{-g}}\frac{\delta(\sqrt{-g}L_{\rm m})}{\delta \Gamma^\alpha{}_{\mu\nu}}\,,
\end{eqnarray}
which for our case we would choose them in a way that both respect the cosmological symmetries. Note again here that the variation of the matter sector with respect to the connection is established with respect to the flat one (not the full connection).

For the energy-momentum tensor, we assume the standard perfect fluid given by
\begin{align}
    T_{\mu\nu}=\rho(t)n_\mu u_\nu +p(t)h_{\mu\nu}\,,
\end{align}
where $\rho$ and $p$ are the energy density and pressure of the fluid. For the hypermomentum tensor we use the following one:~\cite{Iosifidis:2020gth}
\begin{equation}
    	\Delta_{\alpha\mu\nu}=\phi(t) h_{\mu\alpha}n_{\nu}+\chi(t) h_{\nu\alpha}n_{\mu}+\psi(t) n_{\alpha}h_{\mu\nu}+\omega(t) n_{\alpha}n_{\mu} n_{\nu}+\zeta(t)\epsilon_{\alpha\mu\nu\kappa}n^{\kappa} \label{hyper}\,,
\end{equation}
which is compatible with cosmological symmetries. In general, this quantity is related to the intrinsic spin (related to torsion) and the intrinsic dilations and shears of nonmetricity. For a theory constructed only with the nonmetricity tensor, $	\Delta_{[\alpha\mu]\nu}=0$, meaning that $\psi=\chi$ and then the hypermomentum would be
\begin{equation}
    	\Delta_{\alpha\mu\nu}=\phi(t) h_{\mu\alpha}n_{\nu}+2\psi(t) n_{(\alpha}h_{\mu)\nu}+\omega(t) n_{\alpha}n_{\mu} n_{\nu}+\zeta(t)\epsilon_{\alpha\mu\nu\kappa}n^{\kappa} \label{hyper}\,.
\end{equation}
By demanding that both $L_{\rm ST-Horn}$ and $L_{\rm m}$ are invariant under diffeomorphisms separately, we can arrive at the following energy-momentum-hypermomentum conservation law~\cite{Hohmann:2021ast}:
\begin{eqnarray}
\sqrt{-g}\lc{\nabla}_\nu T_\mu{}^\nu=\nabla_\nu \nabla_\rho (\Delta_{\mu}{}^{\nu\rho}\sqrt{-g})\,.\label{conservationlaw}
\end{eqnarray}
It should be noted that the covariant derivatives on the right-hand side of the above equation are computed with respect to the teleparallel connection. For ST gravity, we find that the above equation is given by
\begin{eqnarray}\label{conservationlaw2}
[\dot{\rho} + 3H(\rho + p)]n_{\mu}&=&\big\{\ddot{\omega} + (6H + F_1)\dot{\omega} + 3[\dot{H} + H(3H + F_1)]\omega
\nonumber\\
&&+ 3(H + F_2 - F_3)[\dot{\psi} + (3H + 2F_1 + 2F_2)\psi] + 3(H + F_2)[\dot{\phi} + \dot{\psi} + 3H(\phi + \psi)]\big\}n_{\mu}\,.
\end{eqnarray}

The cosmological equations would be related to varying the scale factor $a(t)$, the lapse $N(t)$ (the two FLRW equations), the scalar field $\phi(t)$, and finally, the extra dof coming from the connection component (which is related to the functions $F_i(t)$). However, one can eliminate one of the equations by using the above energy-momentum-hypermomentum conservation law. In the next sections, instead of presenting the connection equation, we will instead present the conservation laws.

The cosmological equations can be written as a combination of the Riemannian Horndeski part and the ST gravity part. Since there are three different ways of obtaining a flat curvature, this means that there will be three different field equations coming from the different branches.

The first FLRW field equation obtained by varying with respect to the lapse function can be then written as (see~\cite{Kobayashi:2011nu} for the Riemannian Horndeski contribution)
\begin{equation}
-\frac{1}{2}\Big[{\cal E}_{\rm ST}+\sum_{i=2}^5 {\cal E}_i\Big] =\kappa^2\rho\,,\label{FRW1}
\end{equation}
where
\begin{eqnarray}
{\cal E}_2&=&2XG_{2,X}-G_2\,,\label{F1b}\\
{\cal E}_3&=&6X\dot\phi HG_{3,X}-2XG_{3,\phi}\,,
\\
{\cal E}_4&=&-6H^2G_4+24H^2X(G_{4,X}+XG_{4,XX})
-12HX\dot\phi G_{4,\phi X}-6H\dot\phi G_{4,\phi }\,, \label{F1c}
\\
{\cal E}_5&=&2H^3X\dot\phi\left(5G_{5,X}+2XG_{5,XX}\right)\label{F1d}
-6H^2X\left(3G_{5,\phi}+2XG_{5,\phi X}\right)\,,
\end{eqnarray}
where $G_{2,X}=\partial G_2/\partial X$, $G_{5,XX}=\partial^2 G_5/\partial X^2$ and so on; therefore commas denote differentiation. 

Now, if one varies the action with respect to the scale factor $a(t)$ one gets the following set of equations:
\begin{eqnarray}
\frac{1}{2}\Big[{\cal P}_{\rm ST}+\sum_{i=2}^5{\cal P}_i\Big]=-\kappa^2 p\,,  \label{FRW2}
\end{eqnarray}
where
\begin{eqnarray}
{\cal P}_2&=&G_2\,,\\
{\cal P}_3&=&-2X\left(G_{3,\phi}+\ddot\phi G_{3,X} \right) \,,
\\
{\cal P}_4&=&2\left(3H^2+2\dot H\right) G_4
-12 H^2 XG_{4,X}-4H\dot X G_{4,X}
-8\dot HXG_{4,X}-8HX\dot X G_{4,XX}
\nonumber\\&&
+2\left(\ddot\phi+2H\dot\phi\right) G_{4,\phi}
+4XG_{4,\phi\phi}
+4X\left(\ddot\phi-2H\dot\phi\right) G_{4,\phi X}\,,
\\
{\cal P}_5&=&-2X\left(2H^3\dot\phi+2H\dot H\dot\phi+3H^2\ddot\phi\right)G_{5,X}
-4H^2X^2\ddot\phi G_{5,XX}
\nonumber\\&&
+4HX\left(\dot X-HX\right)G_{5,\phi X}
+2\left[2\frac{d}{dt}\left(HX\right)+3H^2X\right]G_{5,\phi}
+4HX\dot\phi G_{5,\phi\phi}\,.
\end{eqnarray}

Furthermore, the variations with respect to the scalar field give us
\begin{eqnarray}
\frac{1}{a^3}\frac{d}{dt}\Big[a^3 (\mathcal{J}+\mathcal{J}_{\rm ST})\Big]=P_{\phi}+P_{\phi \rm ST}\,,\label{FRW3}
\end{eqnarray}
where
\begin{eqnarray}
\mathcal{J}&=&\dot\phi G_{2,X}+6HXG_{3,X}-2\dot\phi G_{3,\phi}
+6H^2\dot\phi\left(G_{4,X}+2XG_{4,XX}\right)-12HXG_{4,\phi X}
\nonumber\\&&
+2H^3X\left(3G_{5,X}+2XG_{5,XX}\right)
-6H^2\dot\phi\left(G_{5,\phi}+XG_{5,\phi X}\right)\,,\\
P_{\phi}&=&G_{2,\phi}-2X\left(G_{3,\phi\phi}+\ddot\phi G_{3,\phi X}\right)
+6\left(2H^2+\dot H\right)G_{4,\phi}
+6H\left(\dot X+2HX\right)G_{4,\phi X}
\nonumber\\&&
-6H^2XG_{5,\phi\phi}+2H^3X\dot\phi G_{5,\phi X}\,.
\end{eqnarray}

The terms ${\cal E}_{\rm ST}, {\cal P}_{\rm ST}$ and $P_{\phi\rm ST}$ are the ST contribution that will be calculated for each branch in the next sections. It should be noted that the variation with respect to the connection only has a contribution from the ST part and then, the hypermomentum tensor would only contribute in the connection equation that we will not present in the following sections since this equation can be omitted by considering~\eqref{conservationlaw}. The cosmological equations for the ST Horndeski pieces are involved. Then, we will show the equations by splitting them into all the contributions as follows:
\begin{eqnarray}
{\cal E}_{\rm ST}&=& -G_{\rm ST}+2 X G_{\rm ST,X}+{\cal E}_{\rm Q_i}+{\cal E}_{\rm I_i}+{\cal E}_{\rm J_i}+{\cal E}_{\rm G_{3}^{(i)}}\,,\label{FRW1Tele}\\
{\cal P}_{\rm ST}&=& G_{\rm ST}+{\cal P}_{\rm Q_i}+{\cal P}_{\rm I_i}+{\cal P}_{\rm J_i}+{\cal P}_{\rm G_{3}^{(i)}}\,,\label{FRW2Tele}\\
\mathcal{J}_{\rm ST}&=& \dot{\phi}G_{\rm ST,X}+\mathcal{J}_{\rm Q_i}+\mathcal{J}_{\rm I_i}+\mathcal{J}_{\rm J_i}+\mathcal{J}_{\rm G_{3}^{(i)}}\,,\label{FRW3aTele}\\
P_{\phi \rm ST}&=& G_{\rm ST,\phi}+P_{\phi \rm Q_i}+P_{\phi \rm I_i}+P_{\phi \rm J_i}+P_{\phi \rm G_{3}^{(i)}}\,.\label{FRW3bTele}
\end{eqnarray}
Note again that in our convention, commas denote derivatives. The first terms in the above equation correspond to the $\{\phi,X\}$ derivative contributions which are independent on the connection branch, and then, they have the same form for all cosmological branches. The other terms are split in contributions from $Q_i,I_i, J_i$ as~\eqref{QQQQQ} and $G_3^{(i)}$ denoting the STKGB contributions. For all the branches, the scalar field contribution coming from $G_{\rm ST}$ is the same and it is given by
\begin{eqnarray}
\mathcal{J}_{\rm Q_i}&=&\mathcal{J}_{\rm I_i}=\mathcal{J}_{\rm J_i}=P_{\phi\rm Q_i}=0\,,\\
P_{\phi\rm I_i}&=&\frac{I_1}{\dot{\phi}}G_{\rm ST,I_1}+\frac{I_2}{\dot{\phi}}G_{\rm ST,I_2}\,,\\
P_{\phi\rm J_i}&=&2\frac{J_1}{\dot{\phi}}G_{\rm ST,J_1}+2\frac{J_2}{\dot{\phi}}G_{\rm ST,J_2}+2\frac{J_5}{\dot{\phi}}G_{\rm ST,J_5}+3\frac{J_9}{\dot{\phi}}G_{\rm ST,J_9}+4\frac{J_{10}}{\dot{\phi}}G_{\rm ST,J_{10}}\,,
\end{eqnarray}
where the invariants take different forms for the branches but the form of the equations can be written in the same way. The contribution from STKGB has a different behavior for the different branches. Hereafter, we will show the cosmological equations for each branch but since the equations coming from $J_i$ and STKGB are cumbersome, we will show them in Appendix~\ref{appendixcosmo}. Note that those branches are obtained by taking the nonmetricity tensor as~\eqref{eq:cosmononmet} and then by setting $F_i$ in a way that satisfies the teleparallel condition, that is, vanishing general curvature.

\subsubsection{Branch 1: $F_1 = K\,,F_2 = -H,\,
F_3 = 0$}\label{branch1}
The first branch which gives us a vanishing general curvature is obtained when the functions appearing in the nonmetricity tensor~\eqref{eq:cosmononmet} become
\begin{equation}
F_1 = K\,, \quad
F_2 = -H\,, \quad
F_3 = 0\,,
\end{equation}
with $K=K(t)$. The function $K$ is related to an extra degree of freedom coming from the connection (or nonmetricity). The form of the scalars related to the extension of $L_2$ can be written in terms of the following two scalars:
\begin{eqnarray}
Q&=&6H^2\,,\quad Q_1=-\frac{1}{4}(3H+K)^2\,,
\end{eqnarray}
from where one finds that
\begin{eqnarray} 
Q_2&=&-\frac{16}{81}\left(-3 \sqrt{-Q_1}+  \sqrt{6Q}\right)^2\,,\quad Q_3= \sqrt{Q_1Q_2}\,,\\
I_1&=&-\sqrt{-2XQ_1}\,,\quad  I_2=\sqrt{-2XQ_2}\,,\quad J_1=-3X\sqrt{Q_1Q_2}\quad J_2=-3XQ_2\,,\quad J_5=-6XQ_2\,,\\ 
J_9&=&-3X\sqrt{-2X Q_2}\,,\quad J_{10}=9X^2Q_2\,,\quad Q_4=J_3=J_4=J_6=J_7=J_8=J_{11}=J_{12}=0\,.
\end{eqnarray}
The scalars appearing in the STKGB theory (see~\eqref{L3one}-\eqref{L3five}),  become
\begin{eqnarray}
\mathcal{O}_1&=&-\frac{3 H (3 H+K) \dot{\phi}^2}{2 N^2}\,,\quad 
\mathcal{O}_2=-\frac{7}{3}\mathcal{O}_3=-\frac{7}{9X}\mathcal{O}_4=\frac{7}{9X}\mathcal{O}_5=-\frac{14 H (H-K) \dot{\phi}^2}{N^2}\,.
\end{eqnarray} 
Since some of the STKGB scalars are related by some factors or by $X$, it is then convenient to introduce the following function
\begin{eqnarray}
\bar{G}_{3}^{(2)}(\phi,X)=\frac{9}{7} X (G_{3}^{(5)}(\phi,X)- G_{3}^{(4)}(\phi,X))-\frac{3}{7} G_{3}^{(3)}(\phi,X)+G_{3}^{(2)}(\phi,X)\,.
\end{eqnarray}
Then, the cosmological equations coming from STKGB would depend only on $G_{3}^{(1)}(\phi,X)$ and $\bar{G}_{3}^{(2)}(\phi,X)$.

The contributions from ST gravity appearing in the first cosmological equation~\eqref{FRW1} with~\eqref{FRW1Tele} are
\begin{eqnarray}
{\cal E}_{\rm Q_i}&=&\frac{1}{18} H^2 (216 G_{\rm ST,Q}-81 G_{\rm ST,Q_1}-16 G_{\rm ST,Q_2}+36 G_{\rm ST,Q_3})-\frac{1}{18} H K (27 G_{\rm ST,Q_1}-16 G_{\rm ST,Q_2}+12 G_{\rm ST,Q_3})\,,\\
{\cal E}_{\rm I_i}&=&-\frac{\sqrt{X/2}}{3} (3 G_{\rm ST,I_1} (6 H+K)+4 G_{\rm ST,I_2} (K-2 H))\,,
\end{eqnarray}
while the corresponding contributions form the second FLRW~\eqref{FRW2} with~\eqref{FRW2Tele} become
\begin{eqnarray}
{\cal P}_{\rm Q_i}&=&\frac{1}{54} \Big[H (3 H (-216 G_{\rm ST,Q}+81 G_{\rm ST,Q_1}+16 G_{\rm ST,Q_2}-36 G_{\rm ST,Q_3})+81 \dot{G}_{\rm ST,Q_1}+16 \dot{G}_{\rm ST,Q_2}\nonumber\\
&&-36 (\dot{G}_{\rm ST,Q_3}+6 \dot{G}_{\rm ST,Q}))+\dot{H} (-216 G_{\rm ST,Q}+81 G_{\rm ST,Q_1}+16 G_{\rm ST,Q_2}-36 G_{\rm ST,Q_3})\Big]\nonumber\\
&&+\frac{1}{54} K \Big[3 H (27 G_{\rm ST,Q_1}-16 G_{\rm ST,Q_2}+12 G_{\rm ST,Q_3})+27 \dot{G}_{\rm ST,Q_1}-16 \dot{G}_{\rm ST,Q_2}+12 \dot{G}_{\rm ST,Q_3}\Big]\nonumber\\
&&+\dot{K} \left(\frac{G_{\rm ST,Q_1}}{2}-\frac{8 G_{\rm ST,Q_2}}{27}+\frac{2 G_{\rm ST,Q_3}}{9}\right)\,,\\
{\cal P}_{\rm I_i}&=&\frac{\sqrt{X/2} }{9}\Big[3 H (9 G_{\rm ST,I_1}-4 G_{\rm ST,I_2})+9 \dot{G}_{\rm ST,I_1}-4 \dot{G}_{\rm ST,I_2}\Big]+\frac{1}{18} (9 G_{\rm ST,I_1}-4 G_{\rm ST,I_2}) \ddot{\phi}\,.
\end{eqnarray}
It should be noted again that the contributions coming from $J_i$ and $G_{3}^{(i)}$ are written in Appendix~\ref{appendixa}. 

Finally, the energy-momentum-hypermomentum conservation law for this branch~\eqref{conservationlaw2} is reduced to
\begin{eqnarray}
[\dot{\rho} + 3H(\rho + p)]n_{\mu}&=&
\left\{\ddot{\omega} + (6H + K)\dot{\omega} + 3[\dot{H} + H(3H + K)]\omega\right\}n_{\mu}
\end{eqnarray}
Clearly, when $G_i=G_{3}^{(1)}=\bar{G}_{3}^{(2)}=0$ and $G_{\rm ST}= -f(Q)$,  the equations coincide with the flat FLRW equations for $f(Q)$ gravity reported in~\cite{Hohmann:2021ast,BeltranJimenez:2019tme}. 

%%%
\subsubsection{Branch 2: $F_1 = 2H + \frac{1}{K N}\frac{dK}{dt},\, F_2 = -H\,, F_3 = K$ }\label{branch2}
The second branch giving a zero curvature is obtained by setting the functions
\begin{equation}
F_1 = 2H + \frac{\dot{K}}{K N}\,, \quad
F_2 = -H\,, \quad
F_3 = K\,,
\end{equation}
where $K=K(t)$ is an additional degree of freedom that comes from nonmetricity.
In this branch, we find that all the teleparallel scalars depend on $N,H,K,dK/dt$ and $X$. The form of the scalars appearing in $G_{\rm ST}$ can be rewritten in terms of three scalars, which can be chosen to be
\begin{eqnarray}
Q&=&9 H K+6 H^2+\frac{3 \dot{K}}{N}\,,\quad Q_1=-\frac{1}{4 N^2}\left(5 H N+\frac{\dot{K}}{K}\right)^2\,,\quad Q_2=-\frac{4 }{9 N^2}\left(H N+\frac{\dot{K}}{K}-2 K N\right)^2\,,
\end{eqnarray}
giving us
\begin{eqnarray}
Q_3&=&-\sqrt{Q_1Q_2}\,,\quad I_1=-\sqrt{-2XQ_1}\,,\quad I_2=-\sqrt{-2XQ_2}\,,\\
J_1&=&\frac{1}{2}X\sqrt{-Q_1(48Q_1-72Q_3-9Q_2+32Q)}\,,\quad J_2=\frac{1}{2}X\sqrt{-Q_2(48Q_1-72Q_3-9Q_2+32Q)}\\
J_5&=&-\frac{2J_1^2}{3XQ_1}\,,\quad J_9=X\sqrt{3J_5}\,,\quad J_{10}=-\frac{3}{2}X J_5\,,\\
Q_4&=&J_{3}=J_{4}=J_{6}=J_{7}=J_{8}=J_{11}=J_{12}=0\,.
\end{eqnarray}
For this branch, the scalars appearing in the STKG theory are not related as in the previous branch. They behave as
\begin{eqnarray}
\mathcal{O}_1&=&-\frac{3 H X }{K}\left(5 H K+\dot{K}\right)\,,\\
\mathcal{O}_2&=&H \left(\frac{28 X \dot{K}}{K}-32 K X\right)+28 H^2 X-12 \sqrt{2X} K  \ddot{\phi}\,,\\
\mathcal{O}_3&=&-\frac{12 H X }{K}\left(H K+\dot{K}-2 K^2\right)\,,\\
\mathcal{O}_4&=&-\frac{36 H X^2 \dot{K}}{K}-36 H^2 X^2+24 \sqrt{2} K X^{3/2} \ddot{\phi}\,,\\
\mathcal{O}_5&=&\frac{36 H X^2 \dot{K}}{K}+36 H^2 X^2+12 \sqrt{2} K X^{3/2} \ddot{\phi}\,.
\end{eqnarray}
Thus, the corrections coming to the ST contributions in the flat FLRW equations for the first FLRW equation \eqref{FRW1} with \eqref{FRW1Tele} are 
\begin{eqnarray}
{\cal E}_{\rm Q_i}&=&\frac{1}{18} \Big[H^2 (216 G_{\rm ST,Q}-225 G_{\rm ST,Q_1}-16 G_{\rm ST,Q_2}-60 G_{\rm ST,Q_3})+2 \dot{K} (27 G_{\rm ST,Q}+16 G_{\rm ST,Q_2}+6 G_{\rm ST,Q_3})\Big]\nonumber\\
&&+\frac{1}{9} H K (81 G_{\rm ST,Q}+16 G_{\rm ST,Q_2}+30 G_{\rm ST,Q_3})-\frac{H \dot{K} }{9 K}(45 G_{\rm ST,Q_1}+16 G_{\rm ST,Q_2}+36 G_{\rm ST,Q_3})\nonumber\\
&&-\frac{\dot{K}^2 (9 G_{\rm ST,Q_1}+16 G_{\rm ST,Q_2}+12 G_{\rm ST,Q_3})}{18 K^2}\,,\\
{\cal E}_{\rm I_i}&=&-\frac{\sqrt{2X} }{3 K}\Big[3 G_{\rm ST,I_1} \left(5 H K+\dot{K}\right)+4 G_{\rm ST,I_2} \left(H K+\dot{K}-K^2\right)\Big]\,,
\end{eqnarray}
while for the second FLRW equation~\eqref{FRW2} with \eqref{FRW2Tele} we find
\begin{eqnarray}
{\cal P}_{\rm Q_i}&=&\frac{1}{54} \Big[H \Big(3 H (-216 G_{\rm ST,Q}+225 G_{\rm ST,Q_1}+16 G_{\rm ST,Q_2}+60 G_{\rm ST,Q_3})+225 \dot{G}_{\rm ST,Q_1}+16 \dot{G}_{\rm ST,Q_2}+60 \dot{G}_{\rm ST,Q_3}\nonumber\\
&&-216 \dot{G}_{\rm ST,Q}\Big)+\dot{H} (-216 G_{\rm ST,Q}+225 G_{\rm ST,Q_1}+16 G_{\rm ST,Q_2}+60 G_{\rm ST,Q_3})\Big]\nonumber\\
&&+\frac{1}{27} K \Big[-3 \Big(H (81 G_{\rm ST,Q}+16 G_{\rm ST,Q_2}+30 G_{\rm ST,Q_3})+10 \dot{G}_{\rm ST,Q_3}+27 \dot{G}_{\rm ST,Q}\Big)-16 \dot{G}_{\rm ST,Q_2}\Big]\nonumber\\
&&-\dot{K} \Big[3 G_{\rm ST,Q}+\frac{2}{27} (8 G_{\rm ST,Q_2}+15 G_{\rm ST,Q_3})\Big]+\frac{1}{K}\Big[\frac{1}{54} \dot{K} (3 H (45 G_{\rm ST,Q_1}+16 G_{\rm ST,Q_2}+36 G_{\rm ST,Q_3})+45 \dot{G}_{\rm ST,Q_1}\nonumber\\
&&+16 \dot{G}_{\rm ST,Q_2}+36 \dot{G}_{\rm ST,Q_3})+\frac{1}{54} \ddot{K} (45 G_{\rm ST,Q_1}+16 G_{\rm ST,Q_2}+36 G_{\rm ST,Q_3})\Big]\nonumber\\
&&-\frac{\dot{K}^2 }{K^2}\Big[\frac{1}{6} 5 G_{\rm ST,Q_1}+\frac{2}{27} (4 G_{\rm ST,Q_2}+9 G_{\rm ST,Q_3})\Big]\,,\\
{\cal P}_{\rm I_i}&=&\frac{\sqrt{X/2} }{9}\Big[3 H (15 G_{\rm ST,I_1}+4 G_{\rm ST,I_2})+15 \dot{G}_{\rm ST,I_1}+4 \dot{G}_{\rm ST,I_2}\Big]+\frac{1}{18} (15 G_{\rm ST,I_1}+4 G_{\rm ST,I_2}) \ddot{\phi}\,.
\end{eqnarray}
The contributions from $J_i$ and $G_3^{(i)}$ are written in Appendix~\ref{appendixsecondbranch}.
Finally, the conservation equation~\eqref{conservationlaw2} leads to
\begin{eqnarray}
[\dot{\rho} + 3H(\rho + p)]n_{\mu}&=&\Big\{\ddot{\omega} + \Big(8H + \frac{\dot{K}}{K}\Big)\dot{\omega} + 3[\dot{H} + H\Big(5H + \frac{\dot{K}}{K}\Big)]\omega - 3K[\dot{\psi} + \Big(5H + 2\frac{\dot{K}}{K}\Big)\psi]\Big\}n_{\mu}\,.
\end{eqnarray}
In principle, one would need to solve the above equation for $K$ to determine the form of the extra dof of nonmetricity and then use this in the FLRW equations and the modified Klein-Gordon one. This set of cosmological equations are more involved than in the previous branch. Note that again the equations coincide with the ones reported for the case of $f(Q)$ gravity or Newer GR in their respective limit~\cite{Hohmann:2021ast}.

%%%
\subsubsection{Branch 3: $F_1 = -K - \frac{1}{K}\frac{dK}{dt}\,, F_2 = K - H\,,F_3 = K $}\label{branch3}
The last branch satisfying the condition of having a flat curvature is obtained when
\begin{equation}
F_1 = -K - \frac{\dot{K}}{K}\,, \quad
F_2 = K - H\,, \quad
F_3 = K\,,
\end{equation}
where again $K(t)$ is an additional degree of freedom related to nonmetricity. Similarly, as in the previous branch, we obtain that the scalars coming from $G_{\rm ST}$ are
\begin{eqnarray}
Q&=&-9 H K+6 H^2-\frac{3 \dot{K}}{N}\,,\quad Q_1=-\frac{\left(-3 H N+\frac{\dot{K}}{K}+4 K N\right)^2}{4 N^2}\,,\quad Q_2=-\frac{4 \left(H N+\frac{\dot{K}}{K}+2 K N\right)^2}{9 N^2}\,,
\end{eqnarray}
from which one can reconstruct all the other ones:
\begin{eqnarray}
Q_3&=&-\sqrt{Q_1Q_2}\,,\quad I_1=\sqrt{-2XQ_1}\,,\quad I_2=\sqrt{-2XQ_2}\,,\\
J_1&=&\frac{1}{2}X\sqrt{-Q_1(48Q_1-72Q_3-9Q_2+32Q)}\,,\quad J_2=\frac{1}{2}X\sqrt{-Q_2(48Q_1-72Q_3-9Q_2+32Q)}\\
J_5&=&-\frac{2J_1^2}{3XQ_1}\,,\quad J_9=-X\sqrt{3J_5}\,,\quad J_{10}=-\frac{3}{2}X J_5\,,\\
Q_4&=&J_{3}=J_{4}=J_{6}=J_{7}=J_{8}=J_{11}=J_{12}=0\,.
\end{eqnarray}
One can notice that the relationships between the scalars have the same form as the previous branch but with different signs in $I_1,I_2$ and $J_9$. 

The scalars appearing in the STKG theory have the same form:
\begin{eqnarray}
\mathcal{O}_1&=&-\frac{3 H X }{K}\left(3 H K-\dot{K}-4 K^2\right)\,,\\
\mathcal{O}_2&=&-\frac{4 H X }{K}\left(7 \dot{K}+8 K^2\right)-28 H^2 X-12 \sqrt{2X} K  \ddot{\phi}\,,\\
\mathcal{O}_3&=&\frac{12 H X }{K}\left(H K+\dot{K}+2 K^2\right)\,,\\
\mathcal{O}_4&=&\frac{36 H X^2 \dot{K}}{K}+36 H^2 X^2+24 \sqrt{2} K X^{3/2} \ddot{\phi}\,,\\
\mathcal{O}_5&=&-\frac{36 H X^2 \dot{K}}{K}-36 H^2 X^2+12 \sqrt{2} K X^{3/2} \ddot{\phi}\,,
\end{eqnarray}
where again we notice that they have a similar structure as Branch 2.

Thus, the teleparallel contribution to the FLRW equations (\eqref{FRW1} with~\eqref{FRW1Tele} and \eqref{FRW2} with~\eqref{FRW2Tele}) for the third branch become
\begin{eqnarray}
{\cal E}_{\rm Q_i}&=&\frac{1}{18} \Big[H^2 (216 G_{\rm ST,Q}-81 G_{\rm ST,Q_1}-16 G_{\rm ST,Q_2}+36 G_{\rm ST,Q_3})-2 \dot{K} (27 G_{\rm ST,Q}+2 (9 G_{\rm ST,Q_1}+8 G_{\rm ST,Q_2}+9 G_{\rm ST,Q_3}))\Big]\nonumber\\
&&+\frac{1}{9} H K (-81 G_{\rm ST,Q}+54 G_{\rm ST,Q_1}-16 G_{\rm ST,Q_2}+6 G_{\rm ST,Q_3})+\frac{H \dot{K} }{9 K}\Big[27 G_{\rm ST,Q_1}-16 G_{\rm ST,Q_2}+12 G_{\rm ST,Q_3}\Big]\nonumber\\
&&-\frac{\dot{K}^2}{18 K^2} \Big[9 G_{\rm ST,Q_1}+16 G_{\rm ST,Q_2}+12 G_{\rm ST,Q_3}\Big]\,,\\
{\cal E}_{\rm I_i}&=&\frac{\sqrt{2X}  }{3 K}\Big[G_{\rm ST,I_1} \left(-9 H K+3 \dot{K}+6 K^2\right)+4 G_{\rm ST,I_2} \left(H K+\dot{K}+K^2\right)\Big]\,,\\
{\cal P}_{\rm Q_i}&=&\frac{1}{54} \Big[H \Big(3 H (-216 G_{\rm ST,Q}+81 G_{\rm ST,Q_1}+16 G_{\rm ST,Q_2}-36 G_{\rm ST,Q_3})+81 \dot{G}_{\rm ST,Q_{1}}+16 \dot{G}_{\rm ST,Q_{2}}-36 (\dot{G}_{\rm ST,Q_{3}}+6 \dot{G}_{\rm ST,Q})\Big)\nonumber\\
&&+\dot{H} (-216 G_{\rm ST,Q}+81 G_{\rm ST,Q_1}+16 G_{\rm ST,Q_2}-36 G_{\rm ST,Q_3})\Big]+\frac{1}{27} K \Big[3 H (81 G_{\rm ST,Q}-54 G_{\rm ST,Q_1}+16 G_{\rm ST,Q_2}-6 G_{\rm ST,Q_3})\nonumber\\
&&-54 \dot{G}_{\rm ST,Q_{1}}+16 \dot{G}_{\rm ST,Q_{2}}-6 \dot{G}_{\rm ST,Q_{3}}+81 \dot{G}_{\rm ST,Q}\Big]+\dot{K} \Big[3 G_{\rm ST,Q}-2 G_{\rm ST,Q_1}+\frac{16 G_{\rm ST,Q_2}}{27}-\frac{2 G_{\rm ST,Q_3}}{9}\Big]\nonumber\\
&&+\frac{1}{K}\Big[\frac{1}{54} \dot{K} \Big(-3 H (27 G_{\rm ST,Q_1}-16 G_{\rm ST,Q_2}+12 G_{\rm ST,Q_3})-27 \dot{G}_{\rm ST,Q_{1}}+16 \dot{G}_{\rm ST,Q_{2}}-12 \dot{G}_{\rm ST,Q_{3}}\Big)\nonumber\\
&&-\frac{1}{54} \ddot{K} (27 G_{\rm ST,Q_1}-16 G_{\rm ST,Q_2}+12 G_{\rm ST,Q_3})\Big]+\frac{\dot{K}^2 }{K^2}\Big[\frac{G_{\rm ST,Q_1}}{2}-\frac{8 G_{\rm ST,Q_2}}{27}+\frac{2 G_{\rm ST,Q_3}}{9}\Big]\,,\\
{\cal P}_{\rm I_i}&=&\frac{\sqrt{X/2} }{9 }\Big[3 H (9 G_{\rm ST,I_1}-4 G_{\rm ST,I_2})+9 \dot{G}_{\rm ST,I_1}-4 \dot{G}_{\rm ST,I_2}\Big]+\frac{1}{18} (9 G_{\rm ST,I_1}-4 G_{\rm ST,I_2}) \ddot{\phi}\,,
\end{eqnarray}
where again we have displayed the cosmological contributions coming from $J_i$ and $G_3^{(i)}$ in the Appendix~\ref{appendixthirdbranch}.

The conservation equation~\eqref{conservationlaw2} for this branch becomes \begin{eqnarray}
[\dot{\rho} + 3H(\rho + p)]n_{\mu}&=&\Big\{\ddot{\omega} + \Big(6H - K - \frac{\dot{K}}{K}\Big)\dot{\omega} + 3\Big[\dot{H} + H\Big(3H - K - \frac{\dot{K}}{K}\Big)\Big]\omega + 3K[\dot{\phi} + \dot{\chi} + 3H(\phi + \chi)]\Big\}n_{\mu}\,,\nonumber\\
\end{eqnarray}
which again gives us an extra equation for $K$. One can mention that the form of the equations is very similar to the previous branch but there are some different signs appearing in the equations. Therefore, even though both branches look similar, the cosmological dynamics might give different descriptions of the Universe. The daunting task of analysing their properties and consequences in detail is beyond the scope of this paper and will be pursued in the future.

\section{Conclusions}
\label{sec:Conclusions}

Scalar-tensor theories have attracted a lot of attention since they are simple models that can explain observations such as dark energy or inflation. One of the most famous theories concerning them is Horndeski gravity which is the most general theory with one scalar field leading to second-order Euler-Lagrange equations, although the word ``most" is true under certain assumptions that sometimes are not explicitly said. One of them is the fact that Horndeski found his theory by assuming a manifold that only contains curvature, which is the Riemannian geometry. If one modifies the geometry as a starting point to construct theories of gravity, then the resulting theory would be different. As a family of theories, Riemannian Horndeski has a very varied phenomenology, but also highly pathological subclasses and severe observational constraints in some scenarios. Moreover, finding suitable UV completions has proven to be difficult~\cite{Serra:2022pzl}. For this reason, it is important to explore other possible extensions of GR by including scalar fields coupled to gravity. For this task, we focused on the construction of another unexplored route related to geometry based purely on nonmetricity and possible couplings with a scalar field.

For this reason, we formulated a theory of gravity within the same idea of Horndeski but in a torsionless and flat geometry (zero curvature and torsion) endowed with nonmetricity ($\nabla_{\alpha}g_{\mu\nu}\neq0$). In that geometry, nonmetricity is responsible for generating the gravitational interactions, and then, the metric and the flat connection are independent fields. Since the definition of the nonmetricity tensor contains only first derivatives of the metric, there is a larger array of possibilities that lead to second-order Euler-Lagrange field equations. As we have explained throughout this manuscript, the form of the Horndeski ST gravity Lagrangian can always be recast as the sum of the Riemannian Horndeski gravity Lagrangian plus a new additional piece which exists only due to nonmetricity. This means that even though the geometry assumed is different from the Riemannian case, still, the Riemannian Horndeski theory is obtained. Then, the generic implicit form of the theory can be written as Eq.~\eqref{PC-rule} with the unspecified form of the coefficients such that the theory respects the condition of being at most second-order.

In order to formulate the theory, in Sec.~\ref{sec:NoHigherOrder} we first concentrated on its simplest construction which is by considering that there are no higher-order derivatives acting on nonmetricity such that those terms cannot be recast as purely Riemannian contributions. By assuming that, it is possible to write down an explicit form for the resulting Lagrangian which is expressed in Eq.~\eqref{Lagrangian} and it has the form of Horndeski plus a new independent function $G_{\rm ST}$ which would correspond to the most general form of $L_2$ in a ST framework (within the condition of having at most quadratic contractions of nonmetricity). That theory is the analogous version of the torsional Horndeski gravity theory presented in~\cite{Bahamonde:2019shr}. It should be noted here that the number of invariants obtained in the torsional case for the extension of $L_2$ is smaller ($12$) compared to the nonmetricity teleparallel case ($21$). The reason for this is the fact that nonmetricity carries more dof than torsion and then, it is possible to write many more invariants from nonmetricity (and then couple them with a scalar field).

After formulating the teleparallel $L_2$ extension, we studied the case where higher-order derivatives can act not only on the metric independent components (which give Horndeski) but also in the nonmetricity sector which could lead to purely teleparallel higher-order invariant contributions. The general form of that general case can be schematically written as in Eq.~\eqref{PC-rule} and we showed two different theories which respect the second-order condition and have derivatives acting on the nonmetricity tensor. The systematic way of constructing them requires finding the correct counterbalance terms which cancel the higher-order derivatives after performing variations. The first example, which was presented in Sec.~\ref{sec:Proca}, with this property is the Generalized Proca action with only the Weyl part of the nonmetricity tensor. Due to the mathematical nature of this irreducible mode of nonmetricity, then, one can easily notice that the Generalized Proca action formulated in Ref.~\cite{Heisenberg:2014rta} can be obtained from choosing the vector field to be the Weyl part of nonmetricity $W_\mu$ and its vector field norm is $Q_1=W_\mu W^\mu$. Thus, the mathematical form of this Lagrangian is identical to the one presented in Ref.~\cite{Heisenberg:2014rta}. Secondly, we formulated a ST analogous version of the so-called Kinetic Gravity Braiding/Cubic Horndeski theory~\cite{Deffayet:2010qz} by finding the corresponding $L_3$ extensions coming from nonmetricity with couplings between derivatives of the scalar field and nonmetricity scalars which are linear in nonmetricity. The final form of that theory, which we labeled as Symmetric Teleparallel Kinetic Gravity Braiding theory was presented in Eq.~\eqref{STKGB}. Again, our theory contains the Riemannian Kinetic Braiding theory. In Sec.~\ref{HO}, we commented on a way of systematically finding possible extensions to other higher-order terms such as purely teleparallel $L_4$ or $L_5$ but we leave that construction for future works.

After formulating the theory, we presented the flat cosmological FLRW equations in Sec.~\ref{FLRW} by considering a theory constructed from the $L_2$ extension and the STKGB theory. To do this, we impose the condition that the flat connection satisfies the cosmological symmetries. That condition implies that there are three different branches of cosmological equations since the condition leading to a curvatureless manifold provides three different ways where the connection is homogeneous and isotropic. This means that there would be three different sets of equations depending on the branch of the connection. Those equations generalize previous studies where particular cases of our theory can be obtained, such as $f(Q)$ gravity, or Newer GR~\cite{Hohmann:2021ast}. One important aspect of those cosmological equations is the fact that the ST contributions are highly non-trivial even at the background level. In the torsional case~\cite{Bahamonde:2019shr}, the torsional teleparallel contributions from the cosmological equations can be written only in term of just four scalars $(\phi,X,T,I_2)$ but in our ST Horndeski case, the equations depend on many more invariants ($\phi,X,Q_i,I_i,J_1,J_2,J_5,J_9,J_{10}$). Furthermore, in the nonmetricity case, there are three sets of cosmological equations while in the torsion one, there is only one branch in flat FLRW. This suggests that the phenomenology of the cosmological equations for the ST Horndeski case would be richer than both the teleparallel torsional and Riemannian Horndeski case. It would be interesting then to study those cosmological equations to find out if our theory can accommodate the cosmological observations correctly.

\begin{figure}[H]\label{Tele_Map}
	\centering
\small{\begin{tikzpicture}
[auto,
decision/.style={diamond, draw=blue, thick, fill=blue!20,
	text width=8em,align=flush center,
	inner sep=1pt},
block/.style ={rectangle, draw=blue, thick, fill=blue!20,
	text width=9.5em,align=center, rounded corners,
	minimum height=3em},
	block2/.style ={rectangle, draw=blue, thick, fill=blue!20,
	text width=4em,align=center, rounded corners,
	minimum height=3em},
line/.style ={draw, thick, -latex',shorten >=2pt},
cloud/.style ={draw=red, thick, ellipse,text width=8em,fill=red!20,	text width=6em,align=center,	minimum height=3em},cloud2/.style ={draw=red, thick, ellipse,text width=8em,fill=red!20,	text width=3em,align=center,	minimum height=3em}]
\matrix [column sep=12mm,row sep=30mm]
{
\node [block] (expertnew2) {Generalized Proca gravity with $W_\mu$~\cite{Heisenberg:2014rta}};
& & \node [block] (expertnew) {Symmetric \\ Teleparallel Kinetic Gravity Braiding};
 	& 	\node [cloud] (expert0) {GR}; \\
			& \node [decision] (evaluate) {Symmetric Teleparallel Horndeski
	}
		;& & \node [cloud] (identify) {Horndeski\newline~\cite{Horndeski:1974wa,Kobayashi:2011nu}};\\[-21ex]
\\[7ex]
\node [block] (evaluate3) {Non-minimally Couplings between $\phi,X$ and $(Q_i,I_i,B_Q)$}; & & \node [block] (evaluate2) {$f(Q,Q_1,Q_2,Q_3,Q_4)$\\\cite{Dialektopoulos:2019mtr,Flathmann:2020zyj}}; & \\[-5ex]
	% row 5
	\node [block] (stop2) {Non-minimally Couplings between $\phi$ and $Q,I_i$~\cite{Hohmann:2021ast}};	& 
	& \node [block] (evaluate4) {$f(Q)$~\cite{BeltranJimenez:2017tkd}}; & \node [block] (evaluate5) {Newer General Relativity~\cite{BeltranJimenez:2017tkd}};\\[-11ex]
	\node [block] (stop3) {Non-minimally Couplings between $\phi$ and $Q$~\cite{Jarv:2018bgs,Runkla:2018xrv}}; &	&\node [block] (evaluate6) {STEGR~\cite{Nester:1998mp}};\\
};
\begin{scope}[-stealth,every path/.style]
	\path (identify) -- (1,1);
%%%%%%%%%%%%%%%%%%%%%%%%%%%%%%%%%%%%%%%%%%%%
%%%%%%%%%%%%%%%%%%%%%%%%%%%%%%%%%%%%%%%%%%%%
%%%%%%%%%%%%%%%%%%%%%%%%%%%%%%%%%%%%%%%%%%%%
\path (evaluate) edge node [below] {$G_{\rm ST}=0$}  (identify);
\path (evaluate) edge node [above] {Zero STG terms}  (identify);
\node[text width=3cm] at (-1.5,0) 
    {Zero STG Higher Order Terms};
\node[text width=3cm] at (-1.8,6.5) 
    {Non-zero STG Higher Order Terms};
\path (evaluate) edge node [above,sloped] {$\phi=0$} (evaluate2);
\path (evaluate) edge node [below,sloped] {$G_{\rm ST}=\tilde{G}_{\rm ST}(\phi,X,Q_i,I_i)+Q\tilde{G}_4(\phi)$} (evaluate3);
\path (evaluate) edge node [above,sloped] {$G_4=1+\tilde{G}_4(\phi),G_2=G_3=G_5=0$} (evaluate3);
\path (evaluate2) edge node [right] {$f=f(Q)$} (evaluate4);
\path (evaluate2) edge node [above,sloped] {$f=c_0Q+\sum_{i=1}^{4}c_i Q_i$} (evaluate5);
\path (evaluate5) edge node [above,sloped] {$c_i=0$} (evaluate6);
\path (evaluate4) edge node [right] {$f=Q$} (evaluate6);
\path (evaluate3) edge node [left] {$\tilde{G}_{\rm ST}=F_1(\phi)Q$} (stop2);
\path (evaluate3) edge node(A) [right] {\hspace{-0.1cm}$+F_2(\phi)I_1+F_3(\phi)I_2+X+V(\phi),$} (stop2);
\path (evaluate3) edge node [below of=A, node distance=0.4cm] {$\hspace{1.5cm}\tilde{G}_4=0$} (stop2);
\path (stop2) edge node [right] {$F_2=F_3=0$} (stop3);
\path (stop3) edge node [below] {$\phi=V(\phi)=0$} (evaluate6);
\path (identify) edge node [below,sloped] {$G_2=G_3=G_5=0, G_4=1$} (expert0);
\path (evaluate) edge node [below,sloped] {$L_{\rm STKGB}=\lc{R}+\mathring{L}_{2}+\mathring{L}_{3}+\sum\limits_{a=1}^5 L_3^{(a)}$} (expertnew);
\path (evaluate) edge node [above,sloped] {\hspace{1cm}(Eq.~\eqref{STKGB})} (expertnew);
\path (evaluate) edge node [above,sloped] {$L=\sum_{i=2}^6 c_i L_i$ (Eq.~\eqref{proca2}) } (expertnew2);
\path (evaluate) edge node [below,sloped] {$\phi=0$ and $  {\nearrow\!\!\!\!\!\!\!Q}_{\lambda\mu\nu}=0$ } (expertnew2);
	\node[cloud] at (4,6.5) (KB) {Kinetic Gravity Braiding~\cite{Deffayet:2010qz}};
		\path (expertnew) edge node [below,sloped] {\hspace{-0.2cm}$ L_3^{(a)}=0$ } (KB);
	\path (KB) edge node [below,sloped] {$G_3=0$ } (expert0);
		\path (identify) edge node [below,sloped] {$G_5=0$ } (KB);
			\path (identify) edge node [above,sloped] {$G_4=1$ } (KB);
					\node [cloud2] at (7.2,-9.25) (GR4) {GR};	
				\path (evaluate6) edge node [above] {Equivalent in} (GR4);
				\path (GR4) edge node [below] {field equations} (evaluate6);
	\end{scope}
	\end{tikzpicture}}
	\caption{Relationship between Symmetric Teleparallel Horndeski gravity and various theories are known in the literature. The blue blocks are theories that need Symmetric Teleparallel gravity and the red ones are purely Riemannian theories. }
	\label{fig}
\end{figure}
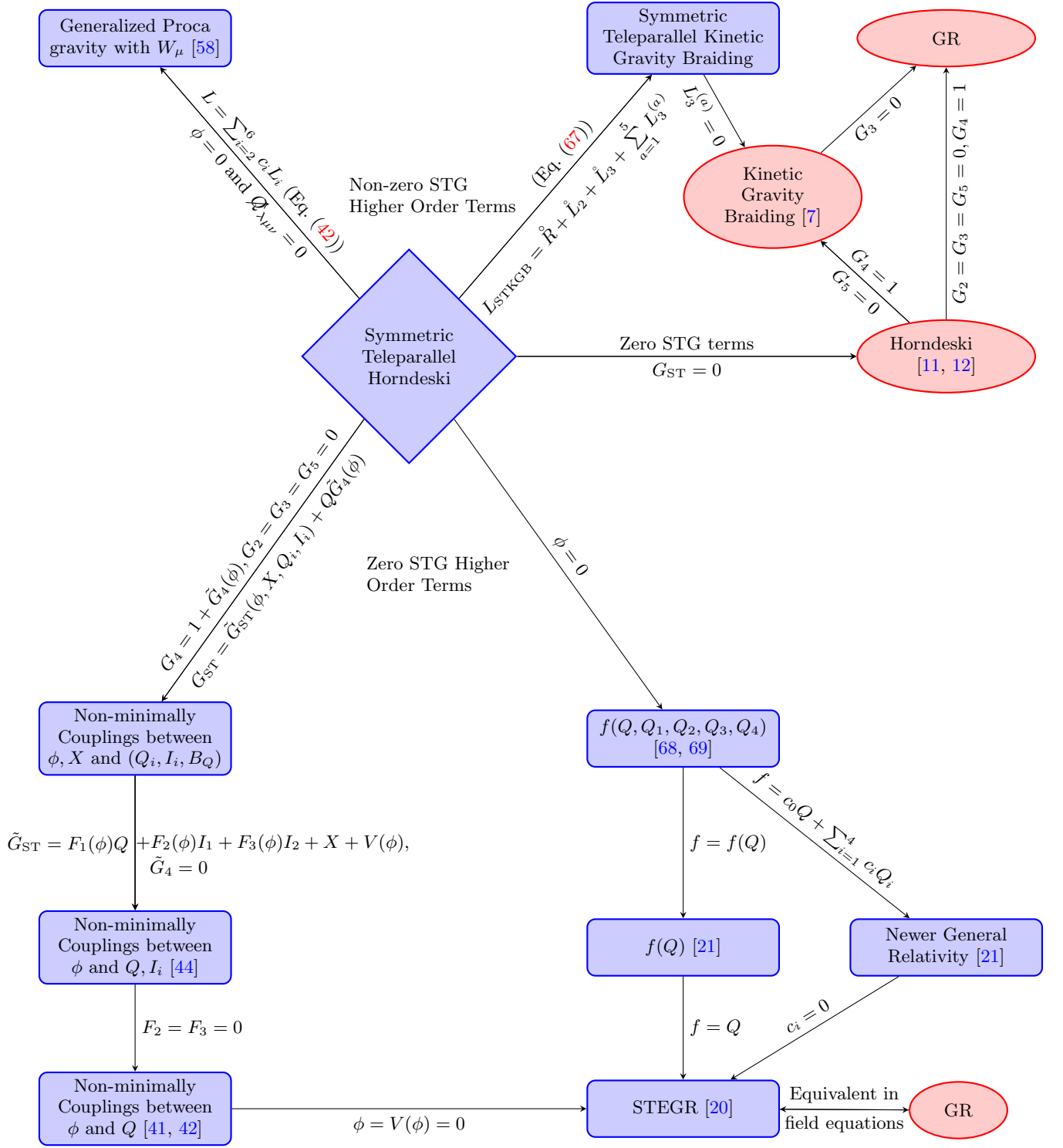

In Fig.~\ref{fig} we present a schematic representation of possible different theories which belong to our ST Horndeski gravity theory. In the lower corner of the figure, we showed the sector where there are no higher-order derivatives acting on nonmetricity, and the Riemannian Horndeski is switched off. Then, the theory is given by~\eqref{Lagrangian}. One can notice that several previous theories can be obtained from assuming several limits related to $G_{\rm ST}$. The theories presented in the figure are only some examples of known theories since the Lagrangian contains a much richer form that could give theories of gravity that have not been presented before. On the upper center part of the diagram, we present the two examples of theories containing derivatives acting on nonmetricity (that cannot be recast as purely Riemannian) and still respecting the conditions of having second-order Euler-Lagrange field equations. Finally, the upper right part of the corner represents the limit of the well-known Horndeski gravity theory. Recall that the last part of the diagram leading to GR would be equivalent to the lowest part of the diagram which leads to STEGR (since those two theories have the same equations of motion, i.e, the Einstein's field equations). It is worth mentioning that due to the nature of ST gravity, the maximum number of dof that Horndeski ST gravity can have is 11 dof. However, the propagating dof of the theory would depend on the theory chosen and a Hamiltonian analysis is needed to understand that. In addition, the theory constructed does not have Ostrogradsky ghosts but depending on the theory, one could have other types of instabilities (as it happens in Riemannian Horndeksi gravity). 

 One potential problem of our formulated theory is the possibility of having strongly coupled modes around FLRW and Minkowski since there are some known ST theories (such as $f(Q)$) that might suffer from that problem~\cite{BeltranJimenez:2019tme}. Furthermore, $f(Q)$ is also part of our theory, so that, if one insists on trying to avoid those issues, one would need to eliminate such dependence in our Lagrangian. After saying this, still, the number of propagating degrees of freedom in $f(Q)$ is under debate. For example, in~\cite{Hu:2022anq}, the authors found that $f(Q)$ has 8 dof while in~\cite{BeltranJimenez:2019tme} the authors claimed that the maximum number of dof is 6. The study of cosmology has been mainly devoted to understanding the first branch, so to our knowledge, there is still not a final conclusion regarding strongly coupled modes around FLRW for all the branches for $f(Q)$ gravity. Moreover, the analysis for more general theories (using Hamiltonian analysis and perturbation theory) has not been studied yet.

 Let us remark here that the cosmological equations of our theory have a much richer structure than in the Riemannian Horndeski case. The reason for this is the fact that our constructed theory contains the Riemannian Horndeski as a subset and new additional degrees of freedom related to nonmetricity appear. Furthermore, due to the nature of Symmetric Teleparallel geometry, there are three sets of cosmological field equations. In this regard, the structure of our theory has a richer cosmology than the standard Horndeski and the repercussions in cosmology need to be studied further in the future with great detail. One important theory present in our construction is $f(Q)$ gravity. Obviously, that theory does not appear in the Riemannian Horndeski and different studies~\cite{Khyllep:2021pcu,Barros:2020bgg,Anagnostopoulos:2021ydo,Ayuso:2020dcu} have shown that already this theory can explain dark energy purely with nonmetricity and the $\sigma_8$ is reduced within this framework. However, as explained above, one must take those results with caution due to the strong coupling problem. For that reason, our constructed theory which contains both $f(Q)$, Riemannian Horndeski, and new symmetric theories of gravity, can help one understand the role of nonmetricity in cosmology and to solve the strong coupling problem for this sector by allowing new degrees of freedom as considered in our theory.

Since our formulated theory contains Horndeski gravity, it is expected that the speed of tensor modes would be in general different from one. However, as the theory is more general, it is expected to have more possibilities satisfying the condition $c_T=1$ while still keeping non-trivial couplings in the Riemannian Horndeski sector given by $G_4$ or $G_5$, as well as to evade bounds from GW decay \cite{Creminelli:2018xsv, Creminelli:2019nok} and/or from GW induced instabilities \cite{Creminelli:2019kjy}. Actually, this argument was already proved in~\cite{Bahamonde:2019ipm} for the torsional Horndeski gravity case where it was found that that theory can still provide $c_T=1$ while having $G_4$ and $G_5$ being non-trivial.  As a future work, we would like to explore this property in our ST theory presented in this manuscript. It would be interesting to study the radiative stability of theories within this framework, extending known results from Riemannian Horndeski and Beyond Horndeski theories where a weakly broken Galilean symmetry ensures their nonrenormalization~\cite{Pirtskhalava:2015nla, Santoni:2018rrx}, as well as how the positivity bounds derived from the analyticity properties of the scattering amplitudes around Minkowski might be affected in the presence of the extra gravitational dof. There are also plenty of applications that one can further consider for the future such as scalarized black holes or studying the possibility of explaining the cosmological observations by our presented theory. Those studies will be pursued in the future to then analyse if our presented theory can solve the recent tensions in observational cosmology and inspect if our theory can be considered as a potentially viable extension of GR.

\newpage

\newpage

\newpage

\section*{Acknowledgements}
The authors would like to thank Ignacy Sawicki and Francesco Serra for useful comments. S.B. is supported by JSPS Postdoctoral Fellowships for Research in Japan and KAKENHI Grant-in-Aid for Scientific Research No. JP21F21789. The work of G.T. and L.G.T.~is supported by the Grant Agency of the Czech Republic, GACR grant 20-28525S. M.Y. acknowledges financial support from JSPS Grant-in-Aid for Scientific Research No. JP18K18764, JP21H01080, JP21H00069. The authors would like to thank the support by the Bilateral Czech-Japanese Mobility Plus Project JSPS-21-12 (JPJSBP120212502). G.T. and L.G.T. would like to thank the Tokyo Institute of Technology for their hospitality during the early stages of this work. S.B. and M.Y. would also like to thank the CEICO, Institute of Physics of the Czech Academy of Sciences, for their hospitality during the final stages of this work.

\bibliographystyle{utphys}
\bibliography{references}

\appendix

\section{Redundant invariants linear in second derivatives and in $Q_{\alpha\mu\nu}$} \label{app:int-by-parts}

These are all the possible invariants that can be constructed linear in $\mathring{\nabla}_{\lambda} Q_{\alpha\mu\nu}$ and then only factors of $\phi^{;\rho}$,
\begin{subequations} \label{A-operators}
\begin{eqnarray}
\hat{\mathcal{O}}_{1} &=& \mathring{\nabla}^{\mu} Q_{\mu\nu}{}^{\nu} = 4\mathring{\nabla}^{\mu} W_{\mu} \, ,\\
\hat{\mathcal{O}}_{2} &=& \mathring{\nabla}^{\mu} Q_{\nu\mu}{}^{\nu} \, ,\\
\hat{\mathcal{O}}_{3} &=& \phi^{;\mu} \phi^{;\nu} \mathring{\nabla}_{\mu} Q_{\nu\alpha}{}^{\alpha} \, ,\\
\hat{\mathcal{O}}_{4} &=& \phi^{;\mu} \phi^{;\nu} \mathring{\nabla}_{\mu} Q_{\alpha\nu}{}^{\alpha} \, ,\\
\hat{\mathcal{O}}_{5} &=& \phi^{;\mu} \phi^{;\nu} \mathring{\nabla}^{\alpha} Q_{\alpha\mu\nu} \, ,\\
\hat{\mathcal{O}}_{6} &=& \phi^{;\mu} \phi^{;\nu} \mathring{\nabla}^{\alpha} Q_{\mu\alpha\nu} \, ,\\
\hat{\mathcal{O}}_{7} &=& \phi^{;\mu} \phi^{;\nu} \phi^{;\alpha} \phi^{;\beta} \mathring{\nabla}_{\mu} Q_{\nu\alpha\beta} \, .
\end{eqnarray}
\end{subequations}
Multiplying these operators by generic functions of $\phi$ and $X$, they can all be expressed up to total derivatives in terms of the operators in Eqs. \eqref{B-operators} as follows, 
\begin{subequations}
\begin{eqnarray}
F_1 \, \hat{\mathcal{O}}_{1} &=& F_{1,X} \, \tilde{\mathcal{O}}_{3} - 4 F_{1,\phi} \, I_1 \, , \\
F_2 \, \hat{\mathcal{O}}_2 &=& F_{2,X} \, \tilde{\mathcal{O}}_4 - F_{2,\phi} \left(\frac{9}{4} I_2 + I_1 \right) \, , \\
F_{3} \, \hat{\mathcal{O}}_{3}&=& - F_{3}\left(\tilde{\mathcal{O}}_{1}+\tilde{\mathcal{O}}_{3}\right)+F_{3X}\tilde{\mathcal{O}}_{8}+8F_{3\phi}XI_{1} \, , \\ 
F_4 \, \hat{\mathcal{O}}_{4} &=& - F_4 \left( \tilde{\mathcal{O}}_{2} + \tilde{\mathcal{O}}_{4} \right) + F_{4X} \tilde{\mathcal{O}}_{9} + 2 F_{4,\phi} X \left(\frac{9}{4} I_2 + I_1 \right) \,  \\
F_{5} \, \hat{\mathcal{O}}_{5} &=&  -2F_{5} \, \tilde{\mathcal{O}}_{5}+F_{5X}\tilde{\mathcal{O}}_{11}- F_{5\phi} \, \phi^{;\alpha} \phi^{;\mu} \phi^{;\nu} Q_{\alpha\mu\nu}\\
F_{6} \, \hat{\mathcal{O}}_6 &=& - F_{6} \left( \tilde{\mathcal{O}}_5 + \tilde{\mathcal{O}}_6 \right) + F_{6,X} \tilde{\mathcal{O}}_{10} - F_{6,\phi} \, \phi^{;\alpha} \phi^{;\mu} \phi^{;\nu} Q_{\alpha\mu\nu} \, ,\\
F_{7} \, \hat{\mathcal{O}}_7 &=& - F_{7} \left( \tilde{\mathcal{O}}_7 + 2 \tilde{\mathcal{O}}_{10} + \tilde{\mathcal{O}}_{11} \right) + F_{7,X} \tilde{\mathcal{O}}_{12} + 2 F_{7,\phi} \, X \phi^{;\alpha} \phi^{;\mu} \phi^{;\nu} Q_{\alpha\mu\nu} \, .
\end{eqnarray}
\end{subequations}
In the context of the construction of $L_3$, we are allowed to discard the last term of each of these equations (the ones with $\phi$-derivatives of the $F_i$ functions), as they do not contain second derivatives and therefore belong in $L_2$.

\section{Variations} \label{app:variations}
Variations with respect to $\phi$ of the invariants in Eqs. \eqref{B-operators}, multiplied by generic functions of $\phi$ and $X$ are (only keeping terms with contain at least 3 derivatives)
\begin{subequations} \label{variations-phi}
\begin{eqnarray}
\delta_\phi\left(\tilde{G}_{3}^{(1)}\tilde{\mathcal{O}}_{1}\right) &\supset& \tilde{G}_{3X}^{(1)}\phi^{;\mu}\phi^{;\rho} Q_{\mu\nu}{}^{\nu}\left(\mathring{\nabla}_{\rho}\mathring{\Box}\phi-\mathring{\Box}\mathring{\nabla}_{\rho}\phi\right) +\tilde{G}_{3}^{(1)}\biggl[Q_{\mu\nu}{}^{\nu}\left(\mathring{\Box}\mathring{\nabla}^{\mu} \phi-\mathring{\nabla}^{\mu}\mathring{\Box} \phi\right)+\phi^{;\mu}\mathring{\Box} Q_{\mu\nu}{}^{\nu}\biggr] \, ,\\
\delta_\phi\left(\bar{G}_{3}^{(2)}\tilde{\mathcal{O}}_{2}\right) &\supset& \tilde{G}_{3X}^{(2)}\phi^{;\mu}\phi^{;\rho} Q_{\nu\mu}{}^{\nu}\left(\mathring{\nabla}_{\rho}\mathring{\Box}\phi-\mathring{\Box}\mathring{\nabla}_{\rho}\phi\right) +\bar{G}_{3}^{(2)}\biggl[Q_{\nu\mu}{}^{\nu}\left(\mathring{\Box}\mathring{\nabla}^{\mu} \phi-\mathring{\nabla}^{\mu}\mathring{\Box} \phi\right)+\phi^{;\mu}\mathring{\Box} Q_{\nu\mu}{}^{\nu}\biggr] \, ,\\
\delta_\phi\left(\tilde{G}_{3}^{(3)}\tilde{\mathcal{O}}_{3}\right) &\supset& \tilde{G}_{3X}^{(3)}\phi^{;\rho}\phi_{;\alpha}Q_{\beta\mu}{}^{\mu}\biggl[\mathring{\nabla}_{\rho} \mathring{\nabla}^\alpha\mathring{\nabla}^\beta\phi-\mathring{\nabla}^\alpha\mathring{\nabla}^\beta\mathring{\nabla}_{\rho} \phi\biggr]\nonumber\\
&&+\tilde{G}_{3}^{(3)}\biggl[Q_{\beta\mu}{}^{\mu}\left(\mathring{\nabla}^\beta\mathring{\Box} \phi-\mathring{\Box}\mathring{\nabla}^{\beta} \phi\right)+\phi_{;\alpha}\mathring{\nabla}^\beta \mathring{\nabla}^\alpha Q_{\beta\mu}{}^{\mu}\biggr]  \, ,\\
\delta_\phi\left(\tilde{G}_{3}^{(4)}\tilde{\mathcal{O}}_{4}\right) &\supset& \tilde{G}_{3X}^{(4)}\phi^{;\rho}\phi_{;\alpha}Q_{\mu\beta}{}^{\mu}\biggl[\mathring{\nabla}_{\rho} \mathring{\nabla}^\alpha\mathring{\nabla}^\beta\phi-\mathring{\nabla}^\alpha\mathring{\nabla}^\beta\mathring{\nabla}_{\rho} \phi\biggr]\nonumber\\
&&+\tilde{G}_{3}^{(4)}\biggl[Q_{\mu\beta}{}^{\mu}\left(\mathring{\nabla}^\beta\mathring{\Box} \phi-\mathring{\Box}\mathring{\nabla}^{\beta} \phi\right)+\phi_{;\alpha}\mathring{\nabla}^\beta \mathring{\nabla}^\alpha Q_{\mu\beta}{}^{\mu}\biggr]  \, ,\\
\delta_\phi\left(\tilde{G}_{3}^{(5)}\tilde{\mathcal{O}}_{5}\right) &\supset& \tilde{G}_{3X}^{(5)}\phi^{;\rho}\phi^{;\mu}Q_{\alpha\mu\beta}\biggl[\mathring{\nabla}_{\rho} \mathring{\nabla}^\alpha\mathring{\nabla}^\beta\phi-\mathring{\nabla}^\alpha\mathring{\nabla}^\beta\mathring{\nabla}_{\rho} \phi\biggr]\nonumber\\
&&+\tilde{G}_{3}^{(5)}\biggl[Q_{\alpha\mu\beta}\left( \mathring{\nabla}^\alpha\mathring{\nabla}^\beta\mathring{\nabla}^{\mu} \phi-\mathring{\nabla}^{\mu} \mathring{\nabla}^\alpha\mathring{\nabla}^\beta \phi\right)+\phi^{;\mu}\mathring{\nabla}^\beta \mathring{\nabla}^\alpha Q_{\alpha\mu\beta}\biggr]  \, ,\nonumber \\ \\
\delta_\phi\left(\tilde{G}_{3}^{(6)}\tilde{\mathcal{O}}_{6}\right) &\supset& \tilde{G}_{3X}^{(6)}\phi^{;\mu}\phi^{;\rho}Q_{\mu\alpha\beta}\left(\mathring{\nabla}_\rho\mathring{\nabla}^\alpha\mathring{\nabla}^\beta\phi-\mathring{\nabla}^\alpha\mathring{\nabla}^\beta\mathring{\nabla}_\rho\phi\right) \notag \\
&\quad&+\tilde{G}_{3}^{(6)}\biggl[\phi^{;\mu}\mathring{\nabla}^\alpha\mathring{\nabla}^\beta Q_{\mu\alpha\beta}+Q_{\mu\alpha\beta}\left(\mathring{\nabla}^\mu\mathring{\nabla}^\alpha\mathring{\nabla}^\beta\phi-\mathring{\nabla}^\alpha\mathring{\nabla}^\beta\mathring{\nabla}^\mu\phi\right)\biggr]  \, ,\\
\delta_\phi\left(\tilde{G}_{3}^{(7)}\tilde{\mathcal{O}}_{7}\right) &\supset& \tilde{G}_{3X}^{(7)}\phi^{;\mu}\phi^{;\nu}\phi^{;\rho}\phi^{;\alpha}Q_{\mu\nu\alpha}\left(\mathring{\nabla}_{\rho}\mathring{\Box} \phi-\mathring{\Box}\mathring{\nabla}_{\rho} \phi\right)+\tilde{G}_{3}^{(7)}\phi^{;\nu}\phi^{;\mu}\phi^{;\alpha}\mathring{\Box}Q_{\mu\nu\alpha} \notag\\
&\quad&+\tilde{G}_{3}^{(7)}\phi^{;\nu}Q_{\mu\nu\alpha}\biggl[2\phi^{;\mu}\left(\mathring{\Box}\mathring{\nabla}^\alpha\phi-\mathring{\nabla}^\alpha\mathring{\Box}\phi\right)+\phi^{;\alpha}\left(\mathring{\nabla}^\mu\mathring{\Box}\phi-\mathring{\Box}\mathring{\nabla}^\mu\phi\right)\biggr] \, ,\\
\delta_\phi\left(\tilde{G}_{3}^{(8)}\tilde{\mathcal{O}}_{8}\right) &\supset& \tilde{G}_{3X}^{(8)}\phi^{;\mu}\phi^{;\nu}\phi^{;\rho}\phi^{;\alpha}Q_{\nu\beta}{}^{\beta}\left(\mathring{\nabla}_{\rho}\mathring{\nabla}_\mu \mathring{\nabla}_\alpha \phi-\mathring{\nabla}_\mu \mathring{\nabla}_\alpha\mathring{\nabla}_{\rho} \phi\right)+\tilde{G}_{3}^{(8)}\phi^{;\alpha}\phi^{;\mu}\phi^{;\nu}\mathring{\nabla}_\mu \mathring{\nabla}_\alpha Q_{\nu\beta}{}^{\beta} \notag \\
&\quad&+\tilde{G}_{3}^{(8)}\phi^{;\alpha}Q_{\nu\beta}{}^{\beta}\biggl[2\phi^{;\nu}\left(\mathring{\nabla}_\alpha\mathring{\Box}\phi-\mathring{\Box}\mathring{\nabla}_\alpha\phi\right)+\phi^{;\mu}\left(\mathring{\nabla}_\mu \mathring{\nabla}_\alpha\mathring{\nabla}^\nu\phi-\mathring{\nabla}^\nu \mathring{\nabla}_\mu \mathring{\nabla}_\alpha\phi\right)\biggr] \, ,\\
\delta_\phi\left(\tilde{G}_{3}^{(9)}\tilde{\mathcal{O}}_{9}\right) &\supset& \tilde{G}_{3X}^{(9)}\phi^{;\mu}\phi^{;\nu}\phi^{;\rho}\phi^{;\alpha}Q_{\beta\nu}{}^{\beta}\left(\mathring{\nabla}_{\rho}\mathring{\nabla}_\mu \mathring{\nabla}_\alpha \phi-\mathring{\nabla}_\mu \mathring{\nabla}_\alpha\mathring{\nabla}_{\rho} \phi\right)+\tilde{G}_{3}^{(9)}\phi^{;\alpha}\phi^{;\mu}\phi^{;\nu}\mathring{\nabla}_\mu \mathring{\nabla}_\alpha Q_{\beta\nu}{}^{\beta} \notag \\
&\quad&+\tilde{G}_{3}^{(9)}\phi^{;\alpha}Q_{\beta\nu}{}^{\beta}\biggl[2\phi^{;\nu}\left(\mathring{\nabla}_\alpha\mathring{\Box}\phi-\mathring{\Box}\mathring{\nabla}_\alpha\phi\right)+\phi^{;\mu}\left(\mathring{\nabla}_\mu \mathring{\nabla}_\alpha\mathring{\nabla}^\nu\phi-\mathring{\nabla}^\nu \mathring{\nabla}_\mu \mathring{\nabla}_\alpha\phi\right)\biggr] \, ,\\
\delta_\phi\left(\tilde{G}_{3}^{(10)}\tilde{\mathcal{O}}_{10}\right) &\supset& \tilde{G}_{3X}^{(10)}\phi^{;\mu}\phi^{;\nu}\phi^{;\rho}\phi^{;\alpha}Q_{\mu\nu}{}^{\beta}\left(\mathring{\nabla}_{\rho}\mathring{\nabla}_\beta \mathring{\nabla}_\alpha \phi-\mathring{\nabla}_\beta \mathring{\nabla}_\alpha\mathring{\nabla}_{\rho} \phi\right)+\tilde{G}_{3}^{(10)}\phi^{;\alpha}\phi^{;\mu}\phi^{;\nu}\mathring{\nabla}_\beta \mathring{\nabla}_\alpha Q_{\mu\nu}{}^{\beta} \notag \\
&\quad&+\tilde{G}_{3}^{(10)}Q_{\mu\nu}{}^{\beta}\biggl[\phi^{;\alpha}\phi^{;\nu}\left(\mathring{\nabla}_\beta \mathring{\nabla}_\alpha\mathring{\nabla}^\mu\phi-\mathring{\nabla}^\mu \mathring{\nabla}_\beta \mathring{\nabla}_\alpha\phi\right)+\phi^{;\mu}\phi^{;\nu}\left(\mathring{\nabla}_\beta\mathring{\Box}\phi-\mathring{\Box}\mathring{\nabla}_\beta\phi\right)\notag\\
&\quad&\qquad\qquad\qquad+\phi^{;\alpha}\phi^{;\mu}\left(\mathring{\nabla}_\beta \mathring{\nabla}_\alpha\mathring{\nabla}^\nu\phi-\mathring{\nabla}^\nu \mathring{\nabla}_\beta \mathring{\nabla}_\alpha\phi\right)\biggr] \, ,\\
\delta_\phi\left(\tilde{G}_{3}^{(11)}\tilde{\mathcal{O}}_{11}\right) &\supset& \tilde{G}_{3X}^{(11)}\phi^{;\mu}\phi^{;\nu}\phi^{;\rho}\phi^{;\alpha}Q^{\beta}{}_{\mu\nu}\left(\mathring{\nabla}_{\rho}\mathring{\nabla}_\beta \mathring{\nabla}_\alpha \phi-\mathring{\nabla}_\beta \mathring{\nabla}_\alpha\mathring{\nabla}_{\rho} \phi\right)+\tilde{G}_{3}^{(11)}\phi^{;\alpha}\phi^{;\mu}\phi^{;\nu}\mathring{\nabla}_\beta \mathring{\nabla}_\alpha Q^{\beta}{}_{\mu\nu} \notag \\
&\quad&+\tilde{G}_{3}^{(11)}Q^{\beta}{}_{\mu\nu}\biggl[\phi^{;\alpha}\phi^{;\nu}\left(\mathring{\nabla}_\beta \mathring{\nabla}_\alpha\mathring{\nabla}^\mu\phi-\mathring{\nabla}^\mu \mathring{\nabla}_\beta \mathring{\nabla}_\alpha\phi\right)+\phi^{;\mu}\phi^{;\nu}\left(\mathring{\nabla}_\beta\mathring{\Box}\phi-\mathring{\Box}\mathring{\nabla}_\beta\phi\right)\notag\\
&\quad&\qquad\qquad\qquad+\phi^{;\alpha}\phi^{;\mu}\left(\mathring{\nabla}_\beta \mathring{\nabla}_\alpha\mathring{\nabla}^\nu\phi-\mathring{\nabla}^\nu \mathring{\nabla}_\beta \mathring{\nabla}_\alpha\phi\right)\biggr] \, , \\
\delta_\phi\left(\tilde{G}_{3}^{(12)}\tilde{\mathcal{O}}_{12}\right) &\supset& \tilde{G}_{3X}^{(12)}\phi^{;\mu}\phi^{;\nu}\phi^{;\rho}\phi^{;\sigma}\phi^{;\alpha}\phi^{;\beta}Q^{\mu\nu\alpha}\left(\mathring{\nabla}_{\beta}\mathring{\nabla}_\rho \mathring{\nabla}_\sigma \phi-\mathring{\nabla}_\rho \mathring{\nabla}_\sigma\mathring{\nabla}_{\sigma} \phi\right)+\tilde{G}_{3}^{(12)}\phi^{;\mu}\phi^{;\nu}\phi^{;\rho}\phi^{;\sigma}\phi^{;\alpha}\mathring{\nabla}_\rho \mathring{\nabla}_\sigma Q_{\mu\nu\alpha} \notag \\
&\quad&+2\tilde{G}_{3}^{(12)}Q_{\mu\nu\alpha}\phi^{;\mu}\phi^{;\nu}\phi^{;\sigma}\biggl[\phi^{;\rho}\left(\mathring{\nabla}_\rho \mathring{\nabla}_\sigma\mathring{\nabla}^\alpha\phi-\mathring{\nabla}^\alpha \mathring{\nabla}_\rho \mathring{\nabla}_\sigma\phi\right)+\phi^{;\alpha}\left(\mathring{\nabla}_\sigma\mathring{\Box}\phi-\mathring{\Box}\mathring{\nabla}_\sigma\phi\right)\biggr]\notag\\
&\quad&+\tilde{G}_{3}^{(12)}\phi^{;\nu}\phi^{;\rho}\phi^{;\sigma}\phi^{;\alpha}Q_{\mu\nu\alpha}\left(\mathring{\nabla}_\rho \mathring{\nabla}_\sigma\mathring{\nabla}^\mu\phi-\mathring{\nabla}^\mu \mathring{\nabla}_\rho \mathring{\nabla}_\sigma\phi\right) \, ,
\end{eqnarray}
\end{subequations}

Variations with respect to $g_{\mu\nu}$ are (only keeping terms with at least 3 derivatives)
\begin{subequations} \label{variations-g}
\begin{eqnarray}
\frac{\delta}{\delta g_{\mu\nu}}\left(\tilde{G}_{3}^{(1)}\tilde{\mathcal{O}}_{1}\right) &\supset& -\tilde{G}_{3}^{(1)}\phi^{;\alpha}g^{\mu\nu}\mathring{\Box}\mathring{\nabla}_\alpha\phi \, ,\\
\frac{\delta}{\delta g_{\mu\nu}}\left(\bar{G}_{3}^{(2)}\tilde{\mathcal{O}}_{2}\right) &\supset& -\bar{G}_{3}^{(2)}\phi^{;\mu}\mathring{\Box}\mathring{\nabla}^\nu\phi \, ,\\
\frac{\delta}{\delta g_{\mu\nu}}\left(\tilde{G}_{3}^{(3)}\tilde{\mathcal{O}}_{3}\right) &\supset& -\tilde{G}_{3}^{(3)}\phi^{;\alpha}g^{\mu\nu}\mathring{\Box}\mathring{\nabla}_\alpha\phi \, ,\\
\frac{\delta}{\delta g_{\mu\nu}}\left(\tilde{G}_{3}^{(4)}\tilde{\mathcal{O}}_{4}\right) &\supset& -\tilde{G}_{3}^{(4)}\phi_{;\alpha}\mathring{\nabla}^{\mu}\mathring{\nabla}^\alpha\mathring{\nabla}^\nu\phi \, ,\\
\frac{\delta}{\delta g_{\mu\nu}}\left(\tilde{G}_{3}^{(5)}\tilde{\mathcal{O}}_{5}\right) &\supset& -\tilde{G}_{3}^{(5)}\phi^{;\mu}\mathring{\Box}\mathring{\nabla}^\nu\phi \, ,\\
\frac{\delta}{\delta g_{\mu\nu}}\left(\tilde{G}_{3}^{(6)}\tilde{\mathcal{O}}_{6}\right) &\supset& -\tilde{G}_{3}^{(6)}\phi_{;\alpha}\mathring{\nabla}^{\alpha}\mathring{\nabla}^\mu\mathring{\nabla}^\nu\phi \, ,\\
\frac{\delta}{\delta g_{\mu\nu}}\left(\tilde{G}_{3}^{(7)}\tilde{\mathcal{O}}_{7}\right) &\supset& -\tilde{G}_{3}^{(7)}\phi^{;\mu}\phi^{;\nu}\phi^{;\alpha}\mathring{\nabla}_{\alpha}\mathring{\Box}\phi \, ,\\
\frac{\delta}{\delta g_{\mu\nu}}\left(\tilde{G}_{3}^{(8)}\tilde{\mathcal{O}}_{8}\right) &\supset& -\tilde{G}_{3}^{(8)}\phi^{;\alpha}\phi^{;\beta}\phi^{;\rho}g^{\mu\nu}\mathring{\nabla}_{\rho}\mathring{\nabla}_\alpha\mathring{\nabla}_\beta\phi \, ,\\
\frac{\delta}{\delta g_{\mu\nu}}\left(\tilde{G}_{3}^{(9)}\tilde{\mathcal{O}}_{9}\right) &\supset& -\tilde{G}_{3}^{(9)}\phi^{;\rho}\phi^{;\nu}\phi^{;\alpha}g^{\beta\mu}\mathring{\nabla}_{\beta}\mathring{\nabla}_\alpha\mathring{\nabla}_\rho\phi \, ,\\
\frac{\delta}{\delta g_{\mu\nu}}\left(\tilde{G}_{3}^{(10)}\tilde{\mathcal{O}}_{10}\right) &\supset& -\tilde{G}_{3}^{(10)}\phi^{;\rho}\phi^{;\nu}\phi^{;\alpha}g^{\beta\mu}\mathring{\nabla}_{\rho}\mathring{\nabla}_\alpha\mathring{\nabla}_\beta\phi \, ,\\
\frac{\delta}{\delta g_{\mu\nu}}\left(\tilde{G}_{3}^{(11)}\tilde{\mathcal{O}}_{11}\right) &\supset& -\tilde{G}_{3}^{(11)}\phi^{;\mu}\phi^{;\nu}\phi^{;\alpha}\mathring{\Box}\mathring{\nabla}_{\alpha}\phi \, ,\\
\frac{\delta}{\delta g_{\mu\nu}}\left(\tilde{G}_{3}^{(12)}\tilde{\mathcal{O}}_{12}\right) &\supset& -\tilde{G}_{3}^{(12)}\phi^{;\mu}\phi^{;\nu}\phi^{;\alpha}\phi^{;\rho}\phi^{;\sigma}\mathring{\nabla}_{\alpha}\mathring{\nabla}_\rho\mathring{\nabla}_\sigma\phi \, ,
\end{eqnarray}
\end{subequations}
Variations with respect to the connection do not give rise to terms containing 3rd derivatives, since it enters in the scalars  without any derivatives acting on it.

\section{Invariants linear in second derivatives and in irreducible components of $Q_{\alpha\mu\nu}$} \label{app:operators-irrep}

The invariants in Eqs. \eqref{B-operators} can be broken into the irreducible components of nonmetricity
\begin{subequations} \label{B-operators-bis}
\begin{eqnarray}
\tilde{\mathcal{O}}_{1} =& \phi^{;\mu} Q_{\mu\nu}{}^{\nu} \, \mathring{\Box} \phi &= 4 \mathcal{O}_{W1} \, ,\\
\tilde{\mathcal{O}}_{2} =& \phi^{;\mu} Q_{\nu\mu}{}^{\nu} \, \mathring{\Box} \phi &= \frac{9}{4} \mathcal{O}_{\Lambda1} + \mathcal{O}_{W1} \, ,\\
\tilde{\mathcal{O}}_{3} =& \phi_{;\alpha} Q_{\beta\mu}{}^{\mu} \, \mathring{\nabla}^\alpha \mathring{\nabla}^\beta \phi &= 4 \mathcal{O}_{W2} \, ,\\
\tilde{\mathcal{O}}_{4} =& \phi_{;\alpha} Q_{\mu\beta}{}^{\mu} \, \mathring{\nabla}^\alpha \mathring{\nabla}^\beta \phi &= \frac{9}{4} \mathcal{O}_{\Lambda2} + \mathcal{O}_{W2} \, ,\\
\tilde{\mathcal{O}}_{5} =& \phi^{;\mu} Q_{\alpha\mu\beta} \, \mathring{\nabla}^\alpha \mathring{\nabla}^\beta \phi &= \mathcal{O}_{W2} + \frac{1}{4} \mathcal{O}_{\Lambda2} + \frac{1}{2} \mathcal{O}_{\Lambda1} + \frac{1}{6} \mathcal{O}_{\Omega1} + \mathcal{O}_{q1} \, ,\\
\tilde{\mathcal{O}}_{6} =& \phi^{;\mu} Q_{\mu\alpha\beta} \, \mathring{\nabla}^\alpha \mathring{\nabla}^\beta \phi &= \mathcal{O}_{W1} + \mathcal{O}_{\Lambda2} - \frac{1}{4} \mathcal{O}_{\Lambda1} - \frac{1}{3} \mathcal{O}_{\Omega1} + \mathcal{O}_{q1} \, ,\\
\tilde{\mathcal{O}}_{7} =& \phi^{;\mu} \phi^{;\nu} \phi^{;\alpha} Q_{\mu\nu\alpha} \, \mathring{\Box} \phi &= -2 X \, \mathcal{O}_{W1} - \frac{3}{2} X \, \mathcal{O}_{\Lambda1} + \mathcal{O}_{q2} \, ,\\
\tilde{\mathcal{O}}_{8} =& \phi^{;\mu} \phi^{;\nu} \phi^{;\alpha} Q_{\nu\beta}{}^{\beta} \, \mathring{\nabla}_\mu \mathring{\nabla}_\alpha \phi &= 4 \mathcal{O}_{W3}\, ,\\
\tilde{\mathcal{O}}_{9} =& \phi^{;\mu} \phi^{;\nu} \phi^{;\alpha} Q_{\beta\nu}{}^{\beta} \, \mathring{\nabla}_\mu \mathring{\nabla}_\alpha \phi &= \frac{9}{4} \mathcal{O}_{\Lambda3} + \mathcal{O}_{W3} \, ,\\
\tilde{\mathcal{O}}_{10} =& \phi^{;\mu} \phi^{;\nu} \phi^{;\alpha} Q_{\mu\nu}{}^{\beta} \, \mathring{\nabla}_\alpha \mathring{\nabla}_\beta \phi &= \mathcal{O}_{W3} - X \, \mathcal{O}_{\Lambda2} + \frac{1}{4} \mathcal{O}_{\Lambda3} + \frac{1}{6} \mathcal{O}_{\Omega2} + \mathcal{O}_{q3} \, ,\\
\tilde{\mathcal{O}}_{11} =& \phi^{;\mu} \phi^{;\nu} \phi^{;\alpha} Q^{\beta}{}_{\mu\nu}\, \mathring{\nabla}_\alpha \mathring{\nabla}_\beta \phi &= -2 X \, \mathcal{O}_{W2} + \mathcal{O}_{\Lambda3} + \frac{1}{2} X \, \mathcal{O}_{\Lambda2} - \frac{1}{3} \mathcal{O}_{\Omega2} + \mathcal{O}_{q3} \, ,\\
\tilde{\mathcal{O}}_{12} =& \phi^{;\mu} \phi^{;\nu} \phi^{;\alpha} \phi^{;\rho} \phi^{;\sigma} Q_{\mu\nu\alpha} \, \mathring{\nabla}_\rho  \mathring{\nabla}_\sigma \phi &= -2 X \, \mathcal{O}_{W3} - \frac{3}{2} X \, \mathcal{O}_{\Lambda3} + \mathcal{O}_{q4} \, ,
\end{eqnarray}
\end{subequations}
where the alternative basis of twelve invariants constructed  using the irreducible components of nonmetricity are defined as follows,
\begin{subequations} \label{B-operators-irrep}
\begin{eqnarray}
\mathcal{O}_{W1} &=& W_{\mu} \phi^{;\mu} \, \mathring{\Box} \phi \, , \\
\mathcal{O}_{W2} &=& W_{\alpha} \phi_{;\beta} \, \mathring{\nabla}^\alpha \mathring{\nabla}^\beta \phi \, , \\
\mathcal{O}_{W3} &=& W_{\mu} \phi^{;\mu} \phi^{;\alpha} \phi^{;\beta} \, \mathring{\nabla}_\alpha \mathring{\nabla}_\beta \phi \, , \\
\mathcal{O}_{\Lambda1} &=& \Lambda_\mu \phi^{;\mu} \, \mathring{\Box} \phi \, , \\
\mathcal{O}_{\Lambda2} &=& \Lambda_{\alpha} \phi_{;\beta} \, \mathring{\nabla}^\alpha \mathring{\nabla}^\beta \phi \, , \\
\mathcal{O}_{\Lambda3} &=& \Lambda_\mu \phi^{;\mu} \phi^{;\alpha} \phi^{;\beta}\, \mathring{\nabla}_\alpha \mathring{\nabla}_\beta \phi \, , \\
\mathcal{O}_{\Omega1} &=& {*\Omega}_{\alpha\beta\mu} \phi^{;\mu} \, \mathring{\nabla}^\alpha \mathring{\nabla}^\beta \phi \, , \\
\mathcal{O}_{\Omega2} &=& {*\Omega}_{\alpha\beta\mu} \phi^{;\alpha} \phi^{;\beta} \phi_{;\nu} \, \mathring{\nabla}^\mu \mathring{\nabla}^\nu \phi \, , \\
\mathcal{O}_{q1} &=& q_{\alpha\beta\mu} \phi^{;\mu} \mathring{\nabla}^\alpha \mathring{\nabla}^\beta \phi \, , \\
\mathcal{O}_{q2} &=& q_{\alpha\beta\mu} \phi^{;\mu} \phi^{;\alpha} \phi^{;\beta} \, \mathring{\Box} \phi \, , \\
\mathcal{O}_{q3} &=& q_{\alpha\beta\mu} \phi^{;\alpha} \phi^{;\beta} \phi_{;\nu} \, \mathring{\nabla}^\mu \mathring{\nabla}^\nu \phi \, , \\
\mathcal{O}_{q4} &=& q_{\alpha\beta\mu} \phi^{;\mu} \phi^{;\alpha} \phi^{;\beta} \phi^{;\rho} \phi^{;\sigma} \, \mathring{\nabla}_\rho \mathring{\nabla}_\sigma \phi \, . 
\end{eqnarray}
\end{subequations}

\section{FLRW Cosmological equations for $J_i$ and STKGB contributions }\label{appendixcosmo}
In Sec.~\ref{FLRW} we computed the FLRW for the contributions coming from the theory~\eqref{telehor}. However, the explicit form for the contributions from $J_i$ and STKG was omitted due to its cumbersome expressions. For completeness, we present them here. 

\subsection{First branch}\label{appendixa}
For the first branch (see Sec.~\ref{branch1}), the contributions from $J_i$ for the two flat FLRW equations are the following:
\begin{eqnarray}
{\cal E}_{\rm J_i}&=&2 K X \left(3 G_{\rm ST,J_1} H-4 H (-5 G_{\rm ST,J_{10}} X+G_{\rm ST,J_2}+2 G_{\rm ST,J_5})+3 \sqrt{2} G_{\rm ST,J_9} \sqrt{X}\right)\nonumber\\
&&-\frac{4}{3} H X \left(H (9 G_{\rm ST,J_1}+18 G_{\rm ST,J_{10}} X-4 G_{\rm ST,J_2}-8 G_{\rm ST,J_5})+6 \sqrt{2} G_{\rm ST,J_9} \sqrt{X}\right)\nonumber\\
&&+\frac{2}{3} K^2 X \Big[3 G_{\rm ST,J_1}+4 (-6 G_{\rm ST,J_{10}} X+G_{\rm ST,J_2}+2 G_{\rm ST,J_5})\Big]\,,\\
{\cal P}_{\rm J_i}&=&\frac{2}{9} X \Big[\dot{H} \Big(9 G_{\rm ST,J_1}-4 (-3 G_{\rm ST,J_{10}} X+G_{\rm ST,J_2}+2 G_{\rm ST,J_5})\Big)+3 H^2 \Big(9 G_{\rm ST,J_1}-4 (-3 G_{\rm ST,J_{10}} X\nonumber\\
&&+G_{\rm ST,J_2}+2 G_{\rm ST,J_5})\Big)+H \left(12 \dot{G}_{\rm ST,J_{10}} X+9 \dot{G}_{\rm ST,J_{1}}-4 \dot{G}_{\rm ST,J_{2}}-8 \dot{G}_{\rm ST,J_{5}}+9 \sqrt{2X} G_{\rm ST,J_9} \right)+3 \sqrt{2X} \dot{G}_{\rm ST,J_{9}}\Big]\nonumber\\&&
+K \Big[\frac{2}{9} X \Big(3 H (-3 G_{\rm ST,J_1}+4 G_{\rm ST,J_2}+8 G_{\rm ST,J_5})-12 X (3 G_{\rm ST,J_{10}} H+\dot{G}_{\rm ST,J_{10}})-3 \dot{G}_{\rm ST,J_{1}}+4 \dot{G}_{\rm ST,J_{2}}+8 \dot{G}_{\rm ST,J_{5}}\Big)\nonumber\\
&&+\frac{2}{9} \sqrt{2X}\ddot{\phi}\Big (4 (-6 G_{\rm ST,J_{10}} X+G_{\rm ST,J_2}+2 G_{\rm ST,J_5})-3 G_{\rm ST,J_1}\Big)\Big]\nonumber\\
&&+\ddot{\phi} \Big[\frac{2}{9} \sqrt{2X} H  \Big(9 G_{\rm ST,J_1}-4 (-6 G_{\rm ST,J_{10}} X+G_{\rm ST,J_2}+2 G_{\rm ST,J_5})\Big)+2 G_{\rm ST,J_9} X\Big]\nonumber\\
&&+\frac{2}{9} X \dot{K} \Big[4 (-3 G_{\rm ST,J_{10}} X+G_{\rm ST,J_2}+2 G_{\rm ST,J_5})-3 G_{\rm ST,J_1}\Big]\,,
\end{eqnarray}
and the contributions from the STKG (the two FLRW plus the scalar field contribution) are
\begin{eqnarray}
{\cal E}_{\rm G_{3}^{(i)}}&=&2 H K X \left[X (28 \bar{G}_{3,X}^{(2)}-3 G_{3,X}^{(1)})+28 \bar{G}_{3}^{(2)}-3 G_{3}^{(1)}\right]-H^2 X \left[2 X (28 \bar{G}_{3,X}^{(2)}+9 G_{3,X}^{(1)})+84 \bar{G}_{3}^{(2)}+27 G_{3}^{(1)}\right]\,,\\
{\cal P}_{\rm G_{3}^{(i)}}&=&\frac{1}{3} X \Big[\sqrt{2X} \Big(56 H \bar{G}_{3,\phi}^{(2)}-28 K \bar{G}_{3,\phi}^{(2)}+18 H G_{3,\phi}^{(1)}+3 K G_{3,\phi}^{(1)}\Big)+28 (2 \dot{H}+3 H^2-\dot{K}) \bar{G}_{3}^{(2)}+3 (6 \dot{H}+9 H^2+\dot{K}) G_{3}^{(1)}\Big]\nonumber\\
&&+\frac{1}{3} \sqrt{2X} \ddot{\phi} \Big[X (56 H \bar{G}_{3,X}^{(2)}-28 K \bar{G}_{3,X}^{(2)}+18 H G_{3,X}^{(1)}+3 K G_{3,X}^{(1)})+28 (2 H-K) \bar{G}_{3}^{(2)}+3 (6 H+K) G_{3}^{(1)}\Big]\,,\\
\mathcal{J}_{\rm 
G_{3}^{(i)}}&=&0\,,\\
P_{\rm G_{3}^{(i)}}&=&\mathcal{O}_1G_{3,\phi}^{(1)}+\mathcal{O}_2\bar{G}_{3,\phi}^{(2)}\,,\quad \mathcal{J}_{\rm G_{3}^{(i)}}=\frac{1}{\dot{\phi}}\mathcal{O}_1 G_{3,\phi}^{(1)}\Big[2G_{3}^{(1)}+\dot{\phi}^2G_{3,X}^{(1)}\Big]+\frac{1}{\dot{\phi}}\mathcal{O}_2\Big[2\bar{G}_{3}^{(2)}+\dot{\phi}^2\bar{G}_{3,X}^{(2)}\Big]\,.
\end{eqnarray}

\subsection{Second branch}\label{appendixsecondbranch}
For the second branch (see Sec.~\ref{branch2}), the contributions from $J_i$ for the two flat FLRW equations are
\begin{eqnarray}
{\cal E}_{\rm J_i}&=&\frac{8 X \dot{K} }{3 K}\Big[H (9 G_{\rm ST,J_1}-18 G_{\rm ST,J_{10}} X+4 G_{\rm ST,J_2}+8 G_{\rm ST,J_5})+3 \sqrt{2X} G_{\rm ST,J_9}\Big]+K X \Big[15 G_{\rm ST,J_1} H\nonumber\\
&&-4 H (10 G_{\rm ST,J_{10}} X+G_{\rm ST,J_2}-4 G_{\rm ST,J_5})+6 \sqrt{2X} G_{\rm ST,J_9} \Big]+\frac{4}{3} H^2 X \Big[15 G_{\rm ST,J_1}-18 G_{\rm ST,J_{10}} X\nonumber\\
&&+4 G_{\rm ST,J_2}+8 G_{\rm ST,J_5}\Big]+\frac{4 X \dot{K}^2 }{3 K^2}(3 G_{\rm ST,J_1}-18 G_{\rm ST,J_{10}} X+4 G_{\rm ST,J_2}+8 G_{\rm ST,J_5})+8 \sqrt{2} G_{\rm ST,J_9} H X^{3/2}\nonumber\\
&&+X \dot{K} \Big[3 G_{\rm ST,J_1}-4 (10 G_{\rm ST,J_{10}} X+G_{\rm ST,J_2}-4 G_{\rm ST,J_5})\Big]-\frac{16}{3} K^2 X (3 G_{\rm ST,J_{10}} X+G_{\rm ST,J_2}-G_{\rm ST,J_5})\,,\\
{\cal P}_{\rm J_i}&=&\frac{2}{9} X \Big[-H \Big(3 H (15 G_{\rm ST,J_1}+4 G_{\rm ST,J_2}+8 G_{\rm ST,J_5})+15 \dot{G}_{\rm ST,J_1}+4 \dot{G}_{\rm ST,J_2}+8 \dot{G}_{\rm ST,J_5}\Big)\nonumber\\
&&-\dot{H} (15 G_{\rm ST,J_1}+4 G_{\rm ST,J_2}+8 G_{\rm ST,J_5})+12 X \Big(G_{\rm ST,J_{10}} (\dot{H}+3 H^2)+\dot{G}_{\rm ST,J_{10}} H\Big)-3 \sqrt{2X} (3 G_{\rm ST,J_9} H+\dot{G}_{\rm ST,J_9})\Big]\nonumber\\
&&+\frac{1}{K}\Big[\dot{K} \Big(\frac{2}{9} X (-3 H (9 G_{\rm ST,J_1}+4 G_{\rm ST,J_2}+8 G_{\rm ST,J_5})+12 X (3 G_{\rm ST,J_{10}} H+\dot{G}_{\rm ST,J_{10}})-9 \dot{G}_{\rm ST,J_1}-4 \dot{G}_{\rm ST,J_2}-8 \dot{G}_{\rm ST,J_5})\nonumber\\
&&-\frac{2}{9} \sqrt{2X} \ddot{\phi} \Big\{9 G_{\rm ST,J_1}+4 (-6 G_{\rm ST,J_{10}} X+G_{\rm ST,J_2}+2 G_{\rm ST,J_5})\Big\}\Big)-\frac{2}{9} X \ddot{K} \Big(9 G_{\rm ST,J_1}+4 (-3 G_{\rm ST,J_{10}} X\nonumber\\
&&+G_{\rm ST,J_2}+2 G_{\rm ST,J_5})\Big)\Big]+K \Big[\frac{1}{9} X \Big(3 H (4 (6 G_{\rm ST,J_{10}} X+G_{\rm ST,J_2}-4 G_{\rm ST,J_5})-15 G_{\rm ST,J_1})+24 \dot{G}_{\rm ST,J_{10}} X-15 \dot{G}_{\rm ST,J_1}\nonumber\\
&&+4 \dot{G}_{\rm ST,J_2}-16 \dot{G}_{\rm ST,J_5}\Big)+\frac{1}{9} \sqrt{2} \sqrt{X} \ddot{\phi} (4 (12 G_{\rm ST,J_{10}} X+G_{\rm ST,J_2}-4 G_{\rm ST,J_5})-15 G_{\rm ST,J_1})\Big]\nonumber\\
&&+\ddot{\phi} \Big[-\frac{2}{9} \sqrt{2X} H  (15 G_{\rm ST,J_1}+4 G_{\rm ST,J_2}+8 G_{\rm ST,J_5})+\frac{16}{3} \sqrt{2} G_{\rm ST,J_{10}} H X^{3/2}-2 G_{\rm ST,J_9} X\Big]\nonumber\\
&&+\frac{2 X \dot{K}^2 }{9 K^2}(9 G_{\rm ST,J_1}+4 (-3 G_{\rm ST,J_{10}} X+G_{\rm ST,J_2}+2 G_{\rm ST,J_5}))\nonumber\\
&&+\frac{1}{9} X \dot{K} \Big(4 (6 G_{\rm ST,J_{10}} X+G_{\rm ST,J_2}-4 G_{\rm ST,J_5})-15 G_{\rm ST,J_1}\Big)\,,
\end{eqnarray}
and the contributions from $G_3^{(i)}$ for the two flat FLRW equations are
\begin{eqnarray}
{\cal E}_{\rm G_{3}^{(i)}}&=&\frac{1}{K}\Big[72 H X^3 \dot{K} (G_{3,X}^{(5)}-G_{3,X}^{(4)})+2 H X^2 \dot{K} \left(90 G_{3}^{(5)}-3 (30 G_{3}^{(4)}+G_{3,X}^{(1)}+4 G_{3,X}^{(3)})+28 G_{3,X}^{(2)}\right)\nonumber\\
&&-3 H X \dot{K} (3 G_{3}^{(1)}+12 G_{3}^{(3)}-28 G_{3}^{(2)})\Big]+K \Big[-24 \sqrt{2} X^{5/2} (G_{3,\phi}^{(5)}+2 G_{3,\phi}^{(4)})-8 H X^2 \Big(9 G_{3}^{(5)}+18 G_{3}^{(4)}\nonumber\\
&&-6 G_{3,X}^{(3)}+8 G_{3,X}^{(2)}\Big)+8 H X (6 G_{3}^{(3)}+G_{3}^{(2)})+24 \sqrt{2} X^{3/2} G_{3,\phi}^{(2)}\Big]+72 H^2 X^3 (G_{3,X}^{(5)}-G_{3,X}^{(4)})\nonumber\\
&&+2 X^2 \Big[6 (15 H^2-2 \dot{K}) G_{3}^{(5)}-6 (15 H^2+4 \dot{K}) G_{3}^{(4)}+H^2 (-15 G_{3,X}^{(1)}-12 G_{3,X}^{(3)}+28 G_{3,X}^{(2)})\Big]\nonumber\\
&&+X \Big[24 \dot{K} G_{3}^{(2)}-3 H^2 (15 G_{3}^{(1)}+12 G_{3}^{(3)}-28 G_{3}^{(2)})\Big]\,,
\end{eqnarray}
\begin{eqnarray}
{\cal P}_{\rm G_3^{(i)}}&=&\frac{1}{3K}\Big[\sqrt{2X}  \ddot{\phi} \Big(-4 G_{3}^{(2)} K^2+36 G_{3}^{(5)} X K^2+72 G_{3}^{(4)} X K^2-24 X G_{3,X}^{(3)} K^2+32 X G_{3,X}^{(2)} K^2-56 G_{3}^{(2)} H K\nonumber\\
&&-144 G_{3}^{(5)} H X K+144 G_{3}^{(4)} H X K-72 H X^2 G_{3,X}^{(5)} K+72 H X^2 G_{3,X}^{(4)} K+30 H X G_{3,X}^{(1)} K+24 H X G_{3,X}^{(3)} K\nonumber\\
&&-56 H X G_{3,X}^{(2)} K-28 G_{3}^{(2)} \dot{K}-72 G_{3}^{(5)} X \dot{K}+72 G_{3}^{(4)} X \dot{K}+3 G_{3}^{(1)} (10 H K+\dot{K})+12 G_{3}^{(3)} (-2 K^2+2 H K+\dot{K})\nonumber\\
&&-36 X^2 \dot{K} G_{3,X}^{(5)}+36 X^2 \dot{K} G_{3,X}^{(4)}+3 X \dot{K} G_{3,X}^{(1)}+12 X \dot{K} G_{3,X}^{(3)}-28 X \dot{K} G_{3,X}^{(2)}\Big)\Big]\nonumber\\
&&+\frac{X}{3 K^2}\Big[-72 \sqrt{2} H K^2 G_{3,\phi}^{(5)} X^{3/2}-36 \sqrt{2} K \dot{K} G_{3,\phi}^{(5)} X^{3/2}+72 H K^2 \sqrt{2} G_{3,\phi}^{(4)} X^{3/2}+36 K \sqrt{2} \dot{K} G_{3,\phi}^{(4)} X^{3/2}\nonumber\\
&&-108 G_{3}^{(5)} H^2 K^2 X+108 G_{3}^{(4)} H^2 K^2 X+36 G_{3}^{(5)} \dot{K}^2 X-36 G_{3}^{(4)} \dot{K}^2 X-72 G_{3}^{(5)} K^2 \dot{H} X+72 G_{3}^{(4)} K^2 \dot{H} X\nonumber\\
&&-36 G_{3}^{(5)} K \ddot{K} X+36 G_{3}^{(4)} K \ddot{K} X-84 G_{3}^{(2)} H^2 K^2+28 G_{3}^{(2)} \dot{K}^2-56 G_{3}^{(2)} K^2 \dot{H}+32 G_{3}^{(2)} K^2 \dot{K}-28 G_{3}^{(2)} K \ddot{K}\nonumber\\
&&+3 G_{3}^{(1)} \left(15 H^2 K^2+10 \dot{H} K^2+\ddot{K} K-\dot{K}^2\right)+12 G_{3}^{(3)} \left(3 H^2 K^2+2 (\dot{H}-\dot{K}) K^2+\ddot{K} K-\dot{K}^2\right)\nonumber\\
&&+30 H K^2 \sqrt{2} \sqrt{X} G_{3,\phi}^{(1)}+3 K \sqrt{2} \sqrt{X} \dot{K} G_{3,\phi}^{(1)}-24 \sqrt{2} K^3 \sqrt{X} G_{3,\phi}^{(3)}+24 H K^2 \sqrt{2} \sqrt{X} G_{3,\phi}^{(3)}+12 K \sqrt{2} \sqrt{X} \dot{K} G_{3,\phi}^{(3)}\nonumber\\
&&-56 \sqrt{2} H K^2 \sqrt{X} G_{3,\phi}^{(2)}+32 K^3 \sqrt{2} \sqrt{X} G_{3,\phi}^{(2)}-28 \sqrt{2} K \sqrt{X} \dot{K} G_{3,\phi}^{(2)}\Big]\,,
\end{eqnarray}
\begin{eqnarray}
\mathcal{J}_{\rm G_{3}^{(i)}}&=&\frac{1}{\dot{\phi}}\mathcal{O}_1 G_{3,\phi}^{(1)}\Big[2G_{3}^{(1)}+\dot{\phi}^2G_{3,X}^{(1)}\Big]+\frac{4 \dot{\phi} }{K}\Big[(7 H+3 K) \dot{K}+H K (7 H+K)\Big]G_{3}^{(2)}+12 K \dot{\phi}^2G_{3,\phi}^{(2)}\nonumber\\
&&+\frac{2 H \dot{\phi}^3 }{K}\Big[7 H K+7 \dot{K}-8 K^2\Big]G_{3,X}^{(2)}-\frac{12 \dot{\phi}^3 }{K}\Big[(3 H+K) \dot{K}+3 H K (H+K)\Big]G_{3}^{(4)}-12 K \dot{\phi}^4G_{3,\phi}^{(4)}\nonumber\\
&&-\frac{9 H \dot{\phi}^5 }{K}(H K+\dot{K})G_{3,X}^{(4)}+\frac{6 \dot{\phi}^3 }{K}G_{3}^{(5)}\Big[(6 H-K) \dot{K}+3 H K (2 H-K)\Big]-6 K \dot{\phi}^4G_{3,\phi}^{(5)}\nonumber\\
&&+\frac{9 H \dot{\phi}^5 }{K}\Big[H K+\dot{K}\Big]G_{3,X}^{(5)}+\frac{1}{\dot{\phi}}\mathcal{O}_3\Big[2G_{3}^{(3)}+\dot{\phi}^2G_{3,X}^{(3)}\Big]\,,\\
P_{\rm G_{3}^{(i)}}&=&\mathcal{O}_1 G_{3,\phi}^{(1)}+\mathcal{O}_2 G_{3,\phi}^{(2)}+\mathcal{O}_3 G_{3,\phi}^{(3)}+\mathcal{O}_4 G_{3,\phi}^{(4)}+\mathcal{O}_5 G_{3,\phi}^{(5)}\,.
\end{eqnarray}

\subsection{Third branch}\label{appendixthirdbranch}

For the third branch (see Sec.~\ref{branch3}), the contributions from $J_i$ for the two flat FLRW equations are
\begin{eqnarray}
{\cal E}_{\rm J_i}&=&-\frac{1}{3 K}\Big[8 X \dot{K} \left(H (3 G_{\rm ST,J_1}+18 G_{\rm ST,J_{10}} X-4 G_{\rm ST,J_2}-8 G_{\rm ST,J_5})+3 \sqrt{2} G_{\rm ST,J_9} \sqrt{X}\right)\Big]\nonumber\\
&&+K X \Big[21 G_{\rm ST,J_1} H+4 H (10 G_{\rm ST,J_{10}} X+G_{\rm ST,J_2}-4 G_{\rm ST,J_5})+6 \sqrt{2} G_{\rm ST,J_9} \sqrt{X}\Big]\nonumber\\
&&+\frac{4}{3} H^2 X \Big(-9 G_{\rm ST,J_1}-18 G_{\rm ST,J_{10}} X+4 G_{\rm ST,J_2}+8 G_{\rm ST,J_5}\Big)+\frac{4 X \dot{K}^2}{3 K^2}\Big( 3 G_{\rm ST,J_1}-18 G_{\rm ST,J_{10}} X+4 G_{\rm ST,J_2}+8 G_{\rm ST,J_5}\Big)\nonumber\\
&&+X \dot{K} (9 G_{\rm ST,J_1}+4 (10 G_{\rm ST,J_{10}} X+G_{\rm ST,J_2}-4 G_{\rm ST,J_5}))-\frac{8}{3} K^2 X (3 G_{\rm ST,J_1}+6 G_{\rm ST,J_{10}} X+2 G_{\rm ST,J_2}-2 G_{\rm ST,J_5})\nonumber\\
&&-8 \sqrt{2} G_{\rm ST,J_9} H X^{3/2}\,,
\end{eqnarray}
\begin{eqnarray}
{\cal P}_{\rm J_i}&=&\frac{2}{9} X \Big[\dot{H} \Big(9 G_{\rm ST,J_1}-4 (-3 G_{\rm ST,J_{10}} X+G_{\rm ST,J_2}+2 G_{\rm ST,J_5})\Big)+3 H^2 \Big(9 G_{\rm ST,J_1}-4 (-3 G_{\rm ST,J_{10}} X+G_{\rm ST,J_2}+2 G_{\rm ST,J_5})\Big)\nonumber\\
&&+H \left(12 \dot{G}_{\rm ST,J_{10}} X+9 \dot{G}_{\rm ST,J_{1}}-4 \dot{G}_{\rm ST,J_{2}}-8 \dot{G}_{\rm ST,J_{5}}+9 \sqrt{2} G_{\rm ST,J_9} \sqrt{X}\right)+3 \sqrt{2X} \dot{G}_{\rm ST,J_{9}} \Big]\nonumber\\
&&+\frac{1}{K}\Big[\dot{K} \Big\{\frac{2}{9} X \Big(3 H (3 G_{\rm ST,J_1}-4 G_{\rm ST,J_2}-8 G_{\rm ST,J_5})+12 X (3 G_{\rm ST,J_{10}} H+\dot{G}_{\rm ST,J_{10}})+3 \dot{G}_{\rm ST,J_{1}}-4 \dot{G}_{\rm ST,J_{2}}-8 \dot{G}_{\rm ST,J_{5}}\Big)\nonumber\\
&&+\frac{2}{9} \sqrt{2X}  \ddot{\phi} \Big(3 G_{\rm ST,J_1}-4 (-6 G_{\rm ST,J_{10}} X+G_{\rm ST,J_2}+2 G_{\rm ST,J_5})\Big)\Big\}+\frac{2}{9} X \ddot{K} \Big(3 G_{\rm ST,J_1}-4 (-3 G_{\rm ST,J_{10}} X\nonumber\\
&&+G_{\rm ST,J_2}+2 G_{\rm ST,J_5})\Big)\Big]+K \Big[-\frac{1}{9} X \Big(3 H (21 G_{\rm ST,J_1}+4 (6 G_{\rm ST,J_{10}} X+G_{\rm ST,J_2}-4 G_{\rm ST,J_5}))+24 \dot{G}_{\rm ST,J_{10}} X\nonumber\\
&&+21 \dot{G}_{\rm ST,J_{1}}+4 \dot{G}_{\rm ST,J_{2}}-16 \dot{G}_{\rm ST,J_{5}}\Big)-\frac{1}{9} \sqrt{2} \sqrt{X} \ddot{\phi} (21 G_{\rm ST,J_1}+4 (12 G_{\rm ST,J_{10}} X+G_{\rm ST,J_2}-4 G_{\rm ST,J_5}))\Big]\nonumber\\
&&+\ddot{\phi} \Big[\frac{2}{9} \sqrt{2} H \sqrt{X} (9 G_{\rm ST,J_1}-4 (-6 G_{\rm ST,J_{10}} X+G_{\rm ST,J_2}+2 G_{\rm ST,J_5}))+2 G_{\rm ST,J_9} X\Big]\nonumber\\
&&-\frac{2 X \dot{K}^2}{9 K^2}\Big[ 3 G_{\rm ST,J_1}-4 (-3 G_{\rm ST,J_{10}} X+G_{\rm ST,J_2}+2 G_{\rm ST,J_5})\Big]\nonumber\\
&&-\frac{1}{9} X \dot{K} \Big[21 G_{\rm ST,J_1}+4 (6 G_{\rm ST,J_{10}} X+G_{\rm ST,J_2}-4 G_{\rm ST,J_5})\Big]\,,
\end{eqnarray}
and the contributions from $G_3^{(i)}$ for the two flat FLRW equations are
\begin{eqnarray}
{\cal E}_{\rm G_{3}^{(i)}}&=&\frac{1}{K}\Big[72 H X^3 \dot{K} (G_{3,X}^{(4)}-G_{3,X}^{(5)})+2 H X^2 \dot{K} (-90 G_{3}^{(5)}+90 G_{3}^{(4)}+3 G_{3,X}^{(1)}+12 G_{3,X}^{(3)}-28 G_{3,X}^{(2)})+3 H X \dot{K} (3 G_{3}^{(1)}\nonumber\\
&&+12 G_{3}^{(3)}-28 G_{3}^{(2)})\Big]+K \Big[-24 \sqrt{2} X^{5/2} (G_{3,\phi}^{(5)}+2 G_{3,\phi}^{(4)})-8 H X^2 (9 G_{3}^{(5)}+18 G_{3}^{(4)}-3 G_{3,X}^{(1)}-6 G_{3,X}^{(3)}+8 G_{3,X}^{(2)})\nonumber\\
&&+8 H X (3 G_{3}^{(1)}+6 G_{3}^{(3)}+G_{3}^{(2)})+24 \sqrt{2} X^{3/2} G_{3,\phi}^{(2)}\Big]+72 H^2 X^3 (G_{3,X}^{(4)}-G_{3,X}^{(5)})\nonumber\\
&&-2 X^2 \Big[6 (15 H^2+2 \dot{K}) G_{3}^{(5)}+(24 \dot{K}-90 H^2) G_{3}^{(4)}+H^2 (9 G_{3,X}^{(1)}-12 G_{3,X}^{(3)}+28 G_{3,X}^{(2)})\Big]\nonumber\\
&&+X \Big[24 \dot{K} G_{3}^{(2)}-3 H^2 (9 G_{3}^{(1)}-12 G_{3}^{(3)}+28 G_{3}^{(2)})\Big]\,,
\end{eqnarray}
\begin{eqnarray}
{\cal P}_{\rm G_3^{(i)}}&=&\frac{\sqrt{2X}  \ddot{\phi} }{3 K}\Big[-4 G_{3}^{(2)} K^2+36 G_{3}^{(5)} X K^2+72 G_{3}^{(4)} X K^2-12 X G_{3,X}^{(1)} K^2-24 X G_{3,X}^{(3)} K^2+32 X G_{3,X}^{(2)} K^2+56 G_{3}^{(2)} H K\nonumber\\
&&+144 G_{3}^{(5)} H X K-144 G_{3}^{(4)} H X K+72 H X^2 G_{3,X}^{(5)} K-72 H X^2 G_{3,X}^{(4)} K+18 H X G_{3,X}^{(1)} K-24 H X G_{3,X}^{(3)} K\nonumber\\
&&+56 H X G_{3,X}^{(2)} K+3 G_{3}^{(1)} (-4 K^2+6 H K-\dot{K})+28 G_{3}^{(2)} \dot{K}+72 G_{3}^{(5)} X \dot{K}-72 G_{3}^{(4)} X \dot{K}\nonumber\\
&&-12 G_{3}^{(3)} (2 K^2+2 H K+\dot{K})+36 X^2 \dot{K} G_{3,X}^{(5)}-36 X^2 \dot{K} G_{3,X}^{(4)}-3 X \dot{K} G_{3,X}^{(1)}-12 X \dot{K} G_{3,X}^{(3)}+28 X \dot{K} G_{3,X}^{(2)}\Big]\nonumber\\
&&+\frac{X}{3 K^2}\Big[72 H K^2 \sqrt{2} G_{3,\phi}^{(5)} X^{3/2}+36 K \sqrt{2} \dot{K} G_{3,\phi}^{(5)} X^{3/2}-72 \sqrt{2} H K^2 G_{3,\phi}^{(4)} X^{3/2}-36 \sqrt{2} K \dot{K} G_{3,\phi}^{(4)} X^{3/2}\nonumber\\
&&+108 G_{3}^{(5)} H^2 K^2 X-108 G_{3}^{(4)} H^2 K^2 X-36 G_{3}^{(5)} \dot{K}^2 X+36 G_{3}^{(4)} \dot{K}^2 X+72 G_{3}^{(5)} K^2 \dot{H} X-72 G_{3}^{(4)} K^2 \dot{H} X\nonumber\\
&&+36 G_{3}^{(5)} K \ddot{K} X-36 G_{3}^{(4)} K \ddot{K} X+84 G_{3}^{(2)} H^2 K^2-28 G_{3}^{(2)} \dot{K}^2+56 G_{3}^{(2)} K^2 \dot{H}+32 G_{3}^{(2)} K^2 \dot{K}+28 G_{3}^{(2)} K \ddot{K}\nonumber\\
&&-12 G_{3}^{(3)} \Big(3 H^2 K^2+2(\dot{H}+\dot{K}) K^2+\ddot{K} K-\dot{K}^2\Big)+3 G_{3}^{(1)} \Big(9 H^2 K^2+(6 \dot{H}-4 \dot{K}) K^2-\ddot{K} K+\dot{K}^2\Big)\nonumber\\
&&-\sqrt{2X}\Big(12 K^3  G_{3,\phi}^{(1)}-18 H K^2  G_{3,\phi}^{(1)}+3  K \dot{K} G_{3,\phi}^{(1)}+24  K^3  G_{3,\phi}^{(3)}+24  H K^2  G_{3,\phi}^{(3)}\nonumber\\
&&+12  K  \dot{K} G_{3,\phi}^{(3)}-32 K^3   G_{3,\phi}^{(2)}-56 H K^2   G_{3,\phi}^{(2)}-28 K \dot{K} G_{3,\phi}^{(2)}\Big)\Big]\,,
\end{eqnarray}
\begin{eqnarray}
\mathcal{J}_{\rm G_{3}^{(i)}}&=&\frac{1}{\dot{\phi}}\mathcal{O}_1 G_{3,\phi}^{(1)}\Big[2G_{3}^{(1)}+\dot{\phi}^2G_{3,X}^{(1)}\Big]+\frac{4 \dot{\phi} }{K}\Big[H (K^2-7 \dot{K})-7 H^2 K+3 K \dot{K}\Big]G_{3}^{(2)}+12 K \dot{\phi}^2G_{3,\phi}^{(2)}\nonumber\\
&&-\frac{2 H \dot{\phi}^3 }{K}\Big[7 H K+7 \dot{K}+8 K^2\Big]G_{3,X}^{(2)}-\frac{12 \dot{\phi}^3 }{K}\Big[3 H (K^2-\dot{K})-3 H^2 K+K \dot{K}\Big]G_{3}^{(4)}-12 K \dot{\phi}^4G_{3,\phi}^{(4)}\nonumber\\
&&+\frac{9 H \dot{\phi}^5 }{K}(H K+\dot{K})G_{3,X}^{(4)}-\frac{6 \dot{\phi}^3 }{K}G_{3}^{(5)}\Big[3 H (2 \dot{K}+K^2)+6 H^2 K+K \dot{K}\Big]-6 K \dot{\phi}^4G_{3,\phi}^{(5)}\nonumber\\
&&-\frac{9 H \dot{\phi}^5 }{K}\Big[H K+\dot{K}\Big]G_{3,X}^{(5)}+\frac{1}{\dot{\phi}}\mathcal{O}_3\Big[2G_{3}^{(3)}+\dot{\phi}^2G_{3,X}^{(3)}\Big]\,,\\
P_{\rm G_{3}^{(i)}}&=&\mathcal{O}_1 G_{3,\phi}^{(1)}+\mathcal{O}_2 G_{3,\phi}^{(2)}+\mathcal{O}_3 G_{3,\phi}^{(3)}+\mathcal{O}_4 G_{3,\phi}^{(4)}+\mathcal{O}_5 G_{3,\phi}^{(5)}\,.
\end{eqnarray}

\end{document}